\documentclass{LMCS}

\def\dOi{11(2:1)2015}
\lmcsheading%
{\dOi}
{1--23}
{}
{}
{Dec.~11, 2013}
{Apr.~14, 2015}
{}

\ACMCCS{[{\bf Theory of computation}]: Semantics and
  reasoning---Program semantics; Models of
  computation---Concurrency---Process calculi}

\subjclass{Semantics of Programming Languages, Concurrency, Process
  models}

\pdfoutput=1
\usepackage{amssymb}

\usepackage{tikz}
\usetikzlibrary{external} 
\usepackage{proof}
\RequirePackage{xspace}
\RequirePackage{mathtools}
\RequirePackage{centernot}

\usepackage{Mutualt}
\usepackage{neweqnarray}
\usepackage{hyperref}

\usetikzlibrary{calc,arrows,shapes,decorations.pathmorphing,backgrounds,positioning,fit}

\tikzstyle{state} = [rectangle,rounded corners,draw=black,thick,
                     minimum size=5mm, node distance=10mm and 12mm,
                     inner sep=3pt]
 \tikzstyle{action} = [auto]
 \tikzstyle{from} = [<->, shorten <=1pt, >=stealth',semithick]
 \tikzstyle{timeto} = [->>, shorten >=1pt, >=stealth',semithick]
 \tikzstyle{to} = [->, shorten >=1pt, >=stealth',semithick]

\theoremstyle{plain}\newtheorem{cexample}[thm]{Counterexample}

\newcommand{\csucc}{client-successful\xspace}   

\newcommand{\testleqArb}{\testleq[ \ensuremath{\star}]}
\newcommand{\testplusArb}{\testplus[\ensuremath{\star}]}
\newcommand{\testpreArb}{\testpre[\ensuremath{\star}]}
\newcommand{\testpreS}{\testpre[\svr]}
\newcommand{\testeqC}{\testeq[\clt]}

\newcommand{\testleqn}[1][]{\mathrel{\sqsubsetsim^n_{\text{\scriptsize  #1}}}}
\newcommand{\testleqnC}{\testleqn[\clt]}

\renewcommand{\Conv}[2][]{ \Downarrow^{ #1} \hspace{-0.3em} {#2}}

\newcommand{\uar}[1]{\ar{#1}\kern-4pt_\ut}
\newcommand{\uder}[2]{\mathsf{D}\kern-4pt_\ut(#1, #2)}

\newcommand{\freshf}{ \mathsf{f}} 

\newcommand{\notok}{{ \mathop{\not \hspace{-0.3em}\ok}}} 
\newcommand{\pnf}{\ensuremath{\mathsf{pnf}}\xspace}
\newcommand{\pnfs}{\ensuremath{\mathsf{pnfs}}\xspace }

\newcommand{\cnf}{\ensuremath{\mathsf{cnf}}\xspace } 
\newcommand{\cnfs}{\ensuremath{\mathsf{cnfs}}\xspace}

\newcommand{\testeqPlus}[1][]{\mathrel{\eqsim^{+}_{\scriptsize #1}}}

\newcommand{\testplusP}{\testplus[\peer]}
\newcommand{\testleqP}{\testleq[\peer]}

\newcommand{\eqleqP}{ \eqleq[\Ename{P2P}]  } 
\newcommand{\eqleqC}{ \eqleq[\Ename{CLT}]  } 
\newcommand{\eqP}{ =_\Ename{P2P}  } 
\newcommand{\eqC}{ =_\Ename{CLT}  } 
\newcommand{\Soption}[1]{ {\ \sset{+\,  #1 }}} 
\newcommand{\calD}{ {\mathcal D}}

\newcommand{\ntickvar}[1]{{#1}_{\kern-3pt{\ut}}  }
\newcommand{\ntickx}{\ntickvar{x}}
\newcommand{\nticky}{\ntickvar{y}}

\renewcommand{\Conv}[2][]{ \Downarrow^{ #1} \hspace{-0.3em} {#2}}

\newcommand{\cupdot}{\mathbin{\mathaccent\cdot\cup}}

\newcommand{\closure}[1] {\ensuremath{\mathop{\mathsf{cl}}({#1}) }}

\newcommand{\calC} {{\mathcal C}}

\begin{document}

\title[Mutually testing processes]{Mutually testing processes}
\thanks{{\lsuper{a,b}}Research supported by SFI project SFI 06 IN.1 1898.}

\author[G.~Bernardi]{Giovanni Bernardi\rsuper a}
\address{{\lsuper a}IMDEA Software Institute, Madrid, Spain}
\email{bernargi@tcd.ie}

\author[M.~Hennessy]{Matthew Hennessy\rsuper b}
\address{{\lsuper b}School of Statistics and Computer Science, Trinity College, Dublin, Ireland}
\email{matthew.hennessy@scss.tcd.ie}



\begin{abstract}
In the  standard  testing theory of DeNicola-Hennessy one process
is considered to be a refinement of  another  if every
test guaranteed by the former is also guaranteed by the latter. In
the domain of web services this has been recast, with processes viewed
as \emph{servers} and tests as \emph{clients}. In this way the
standard refinement preorder between  servers is determined by their
ability to satisfy clients. 

But in this setting there is also a natural refinement preorder
between \emph{clients}, determined by their ability to be satisfied by 
\emph{servers}.  In more general settings where there is no
distinction between \emph{clients} and \emph{servers}, but all
processes are \emph{peers}, there is a further refinement preorder
based on the mutual satisfaction of \emph{peers}.

We give a uniform account of these three preorders. In particular we
give two characterisations. The first is behavioural, in terms of traces
and ready sets. The second, for finite processes, is equational.
\end{abstract}

\maketitle


\section{Introduction}

The DeNicola-Hennessy theory of testing \cite{NH1984,DBLP:conf/tapsoft/NicolaH87,Hennessy88a}
considers a process $p$ to be a refinement of process $q$ if every
test passed by $p$ is also passed by $q$. Recently, in papers
such as \cite{LP07,DBLP:conf/ppdp/Barbanerad10,DBLP:journals/toplas/CastagnaGP09,DBLP:journals/tcs/Padovani10}, this refinement
preorder has been recast with a view to providing theoretical
foundations for web services.  Here processes are viewed as
\emph{servers} and tests viewed as \emph{clients}.  In this
terminology the standard (must) testing preorder is a refinement
preorder between servers, which we denote by $p \testleqS q$; this is
determined by the ability of the servers $p,\,q$ to satisfy
clients. However in this framework there are many other natural
behavioural preorders between processes. In this paper we investigate
two; the first, $p \testleqC q$, is determined by the ability of the
clients $p,\, q$ to be satisfied by servers. For the second we drop
the distinction between clients and servers.  Instead all processes
are viewed as peers of each other and the purpose of interaction
between two peers is the mutual satisfaction of both. The resulting
refinement preorder is denoted by $p \testleqSC q$. We give a uniform
behavioural characterisation of all three refinement preorders in
terms of traces and \emph{acceptances sets}
\cite{NH1984,Hennessy88a}. We also give equational characterisations
for a finite process calculus for servers/clients/peers.

We use an infinitary version of $\CCS$ \cite{ccs} augmented by
a \emph{success} constant $\Unit$, to describe processes, be they
servers, clients or peers.  Thus $p = \tau.a.(b.\Cnil \extc c.\Cnil) \extc \tau.a.c.\Cnil$ is a server which offers the action $a$ followed by either $b$ and $c$ depending on how choices are made, and then terminates,
denoted by $\Cnil$. On the other hand $r=
\overline{a}.\overline{c}.\Unit$ is a test or a client which seeks a
synchronisation on $a$ followed by one on $c$; as usual \cite{ccs} communication
or cooperation consists of the simultaneous occurrence of an action
$a$ and its complement~$\overline{a}$. Thus when the server $p$ is
executed in parallel with the client $r$, the latter will always be
satisfied, in that it is guaranteed to reach the successful state
$\Unit$ regardless of how the various choices are made. But if the
client is executed with the alternative server $q = \tau.a.b.\Cnil \extc \tau.a.c.\Cnil$ there is a possibility of the client remaining unhappy; for
this reason $p \nottestleqS q$. However it turns out that $q
\testleqS p$ because every client satisfied by $q$ will also be
satisfied by $p$.

The client preorder $p \testleqC q$ compares the processes as clients,
and their ability to be satisfied by servers. This refinement preorder turns out to be 
incomparable with the server preorder. For example $a.\Unit \extc b.\Cnil \nottestleqS a.\Unit$
because of the client $\overline{b}.\Unit$. But $a.\Unit \extc b.\Cnil \testleqC a.\Unit$
because every server satisfying the former also satisfies $a.\Unit$;
intuitively the extra component of the client $b.\Cnil$ puts no further demands on servers, because the execution of
$b$ will never lead to satisfaction. Conversely 
$a.\Unit \testleqS a.\Cnil$ because $\Unit$ plays no role for processes acting as servers,
while $a.\Unit \nottestleqC a.\Cnil$; $a.\Unit$ as a client is satisfied by the server 
$\overline{a}.\Cnil$ while $a.\Cnil$ can never be satisfied as a client by any server. 
Behaviour relative to the client preorder $\testleqC$ is very sensitive to the presence
of $\Unit$ and $\Cnil$; for example $\Cnil$ is a least element, that is $\Cnil \testleqC r$ for
any process $r$.\footnote{Note in passing that this is not the case for the server preorder; 
$\Cnil$ as a server guarantees the client 
$\overline{b}.\Cnil \extc \tau.\Unit$ but
the server $b.\Cnil$ does not.} 
However in general the precise role these constants play is difficult to discern; for example,
rather surprisingly we have 
$a.(b.\Cnil \extc c.\Unit) \extc a.(b.\Unit \extc c.\Cnil) \testleqC
\Cnil$.

If we ignore the distinction between servers and clients then every process plays an
independent role as a \emph{peer} to all other processes in its environment. This point of view
leads to another behavioural preorder. Intuitively, we say that 
the process $p$ satisfies its  peer $q$ if whenever they are executed in parallel
both are guaranteed to be satisfied; in some sense both peers test 
their partner.  Then   $p_1 \testleqSC p_2$ means that every peer 
satisfied by $p_1$ is also satisfied  by $p_2$.

The peer\leaveout{This third refinement} preorder is different from the server and
client preorders. In fact we will show that $p_1 \testleqSC p_2$
implies $p_1 \testleqC p_2$; but the converse is not true in general. For example $\Unit \extc b.\Cnil \testleqC \Unit$ but $\Unit
\extc b.\Cnil \nottestleqSC \Unit$ because of the peer $\co{b}.\Unit$.  In
our formulation $\Unit \extc b.\Cnil$ and~$\co{b}.\Unit$ mutually satisfy
each other, whereas the peers $\Unit$ and $\co{b}.\Unit$ do not.

The aim of the paper is to show that the theory of the standard (must) testing preorder
\cite{NH1984,Hennessy88a}, here
formulated as the server refinement preorder $\testleqS$, can be extended to both 
the client and the peer refinement preorders. 
 
It is well-known that the behaviour of processes relative to $\testleqS$ can be 
characterised  in terms of the traces they can perform followed by 
\emph{ready} or \emph{acceptance} sets; intuitively each ready set $A$ after a trace $s$ captures a possibility for  the
process to deadlock when interacting with a client. For example the
process $q = \tau.a.b.\Cnil \extc \tau.a.c.\Cnil$ has the ready set~$\sset{b}$
after the (weak) sequence of actions $a$; this represents the possibility of $q$ deadlocking if servicing 
a client which requests an action $a$ but then is not subsequently interested in the 
action $b$. The process $p = a.(b.\Cnil \extc c.\Cnil) \extc a.c.\Cnil$, also discussed above,
has no comparable ready set and for this reason  $p \nottestleqS q$. 

The first main result of the paper is a similar behavioural characterisation of
both the client and the peer refinement preorders, in terms of certain
kinds of traces and ready sets.  However the details are intricate. It
turns out that \emph{unsuccessful} traces, those which can be performed
without reaching a successful state, play an essential role. We also
need to parametrise these concepts, relative to \emph{usable} actions
and \emph{usable} processes; the exact meaning of \emph{usable} will
depend on the particular refinement preorder being considered.

It is also well-known that the standard testing preorders over finite
processes can be characterised by a collection of (in-)equations over
the process operators, \cite{NH1984,Hennessy88a}.  \leaveout{ The set
  which characterises the must-testing preorder from , and therefore
  $\testleqS$, are given in Figure~\ref{fig:equations} and are referred
  to as the standard equations.  } The second main result of the paper
is a similar characterisation of the new refinement
preorders. In fact there is a complication here, as these preorders
are not in general preserved by the external operator $\extc$. A
similar complication occurred in Section 7.2 of \cite{ccs} in the
axiomatisation of \emph{weak bisimulation equivalence}, and in
  the axiomatisations of the \emph{must testing} preorder in
  \cite{NH1984}, and we adopt the same solution.  We give sound and
complete (in-)equational theories for the largest pre-congruences
$\testpreC,\,\testpreSC$ contained in the refinement preorders
$\testleqC,\, \testleqSC$ respectively, over a finite version of \CCS.
The presence of the success constant $\Unit$ in this language
complicates the axiomatisations considerably, as the behaviour of
clients and peers is very dependent on their ability to immediately
report success. For this reason we reformulate the axiomatisation of
\emph{must testing} preorder from \cite{NH1984}, which in this paper
coincides with the server preorder $\testpreS$, as a two-sorted
equational theory. The characterisation of the client and server
preorders, $\testpreC, \testpreS$ respectively, requires extra equations to
capture the behaviour of the special processes $\Unit$ and $\Cnil$. For
example one of the inequations required by the client preorder is $x
\leq \Unit$, while those for the peer preorder include 
$\mu.( \Unit + x ) \leq  \Unit \extc \mu.x$.

\MHc{
The remainder of the paper is organised as follows. 
Section~\ref{sec:testing} is devoted to definitions and notation. 
We  introduce a language for describing processes, an
infinitary version of the \CCS used in \cite{ccs}, and give the standard 
intensional interpretation of it as a labelled transition system, LTS.
For the remainder of the paper, processes will then be considered to be
states in the resulting LTS.
  We also formally define
the three different refinement preorders discussed informally in the
Introduction, by generalising the standard notion from \cite{NH1984}
of applying tests to processes. 

We begin \rsec{characterisations} by recalling the
well-known characterisation of the must preorder (\rthm{fb}) for
finite branching LTSs from \cite{NH1984} in terms of traces and ready
sets.  To adapt this for the client preorder
we need some extra technical notation. This is motivated by a series of 
examples, 
until we finally obtain a statement of the characterisation theorem (\rthm{c.completeness}).
The proof of this result is delegated to a separate subsequent
section, \rsec{clientproofs}. Meanwhile
\rsec{characterisations} continues by showing how
the notation used in this characterisation of the client preorder can be 
modified in a uniform manner to give an analogous characterisation of
the server preorder, (\rthm{s.completeness}), which applies even in
LTSs which are not finite-branching.
Finally by combining these we get an analogous characterisation 
(\rthm{sc.completeness}) for the peer preorder. 

\rsec{clientproofs}, which contains the details of the behavioural 
characterisation theorem for clients, is divided into three
sub-sections. The first is devoted to some technical  results
concerning the relations used in the characterisation. 
The \emph{soundness} of the characterisation is the topic of the
next sub-section, \rsec{client.soundness}, while the converse
\emph{completeness} is covered in the final sub-section.

\rsec{peerproofs} is similar in structure, but deals with the
behavioural characterisation of the peer preorder.

In \rsec{axiomatisations} we restrict our attention to a finite sub-language $\CCSf$
and address the question of equational characterisations. We first show why the client and
peer refinement preorders are not preserved by the external choice operator $\extc$, and 
give a simple behavioural characterisation of the associated pre-congruences $\testpreS,\ \testpreC$ and
$\testpreSC$; this simply involves taking into account the initial behaviour of processes. 
We then explain the equations which need to be added to the standard set in order to obtain 
an equational characterisation of the client and peer pre-congruences;
 These  are stated in Theorem~\ref{thm:axioms-client} and
 Theorem~\ref{thm:mutual.completeness} respectively. 
The proof of the soundness of the equations is straightforward and is
left to the reader. But the completeness is considerably more complex 
and the details are self-contained in a separate section,
\rsec{eq.completeness}.
This again is divided into three sub-sections. The first is devoted to
the exposition of \emph{normal-forms} which are crucial to the
completeness proofs. This is followed by two sub-sections, dealing
with the client preorder first, followed by the peer preorder. 

The paper ends with 
 \rsec{conclusions}, where we present a
summary of our results, a comparison with the existing work, and a series of open questions.
Most of the material described  in the paper, in particular the results
in \rsec{characterisations} to \rsec{peerproofs}, was originally
reported in \cite{gbthesis}.


}

\section{Testing processes}
\label{sec:testing}

Let $\Act$ be a set of actions, ranged over by $a, b, c, \ldots$ and
let $\tau,\ok$ be two distinct actions {\em not} in $\Act$; the first
will denote internal unobservable activity while the second will be
used to report the success of an experiment.
To emphasise
their distinctness we use \Actt\ to denote the set $\Act \cup
 \sset{\tau }$, and similarly for $\Acttt$; we use $\mu$ to range
over the former and~$\lambda$ to range over the latter. 
We assume $\Act$ has an idempotent
complementation function, with $\overline{a}$ being the complement to
$a$.
A labelled transition system, LTS, consists of a triple $\langle \, P, \,
\Acttt, \, \ar{} \, \rangle$, where $P$ is a set of states and
$\ar{} \,\subseteq\, P \times \Acttt \times P$ is a transition
relation between states decorated with labels drawn from the set $\Acttt$.
We use the infix notation $ p \ar{\lambda} q$ in place of $(p,\lambda,q) \in \, \ar{}$.
An LTS is finite-branching if for all $p \in P$ and for all $\lambda \in \Acttt$, the 
set $\setof{q}{p \ar{\lambda} q}$ is finite.
\leaveout{
A labelled transition system, LTS, consists of a triple $\langle \, P, \,
\Acttt, \, \ar{} \, \rangle$, where $P$ is a set of processes and
$\ar{} \,\subseteq\, P \times \Acttt \times P$ is a transition
relation between processes decorated with labels drawn from the set $\Acttt$.
We use the infix notation $ p \ar{\lambda} q$ in place of $(p,\lambda,q) \in \, \ar{}$.
An LTS is finite-branching if for all $p \in P$ and for all $\lambda \in \Acttt$, the 
set $\setof{q}{p \ar{\lambda} q}$ is finite. }
Single transitions $p \ar{\lambda} q$ are extended to sequences of transitions 
$p \ar{t} q$, where $t \in (\Acttt)^\star$, in the standard manner. For 
$s \in (\Acttick)^\star$ we also have the standard weak transitions, $p \wt{s} q$, defined
by ignoring the occurrences of $\tau$s. Somewhat nonstandard is the use of infinite weak transitions, $p \wt{u} $, for $u \in (\Act)^\infty$. 
Finally we lift in the obvious way the complementation function
to both finite and  infinite traces, so that, for example, $\overline{s}$ is the complement
of $s$.

\begin{figure}[t]
\hrulefill
\begin{eqnarray*}
  p,q,r &::=&  \Unit \BNFsep A \BNFsep \mu.p \BNFsep \extsum_{i \in I} p_i   
\end{eqnarray*}
where $I$ is a countable index set,  and 
$A$ ranges over  a set
of definitional constants each of which has an associated definition $A \eqdef p_A$.
\caption{Syntax of infinitary \CCS. \label{fig:syntax}}
\hrulefill
\end{figure}

It will be convenient to have a notation for describing LTSs; we use
an infinitary version of $\CCS$, \cite{ccs},
augmented with a \emph{success} operator, $\Unit$. The syntax of the
language is depicted in \rfig{syntax}.
We use $\Cnil$ to denote the empty external sum $\sum_{i \in
  \emptyset} p_i$ and $p_1 \extc p_2$ for the binary sum $\extsum_{i
  \in \sset{1,2}} p_i$. 
If $I$ is a non-empty set, we use $\intsum_{i \in I} p_i$ to denote the sum~$\extsum_{i \in I} \tau.p_i$.
For the remainder of the paper  we use the LTS whose states are
the terms in \CCS\ and where the relations $p \ar{\lambda} q$ are the
least ones determined by the (standard) rules in \rfig{opsem}.
We use \emph{finite branching \CCS\ } to refer to the LTS whose states
are terms from \CCS\ which generate finite branching
structures. These are the $p$s in \CCS such that the set $\setof{ q }{ p \ar{\lambda} q} $ is finite.
\leaveout{
We use \emph{finite branching \CCS\ } to refer to the LTS which
consists only of terms from \CCS\ which generate finite branching
structures.
}

\begin{figure}[t]
\hrulefill
\centering
  \[
  \begin{array}{l@{\hskip 2em}l}
    \infer[\rname{a-Ok}] {\Unit \ar{\ok} \Cnil}{}&
    \infer[\rname{a-Pre}]{\mu.p \ar{\mu} p}{}
    \\[1em]
    \infer[\rname{r-Ext-l}]
    {\raisebox{-1mm}{$p \extc q\ar{\lambda} p'$}}
    {p\ar{\lambda}p'}
    &
    \infer[\rname{r-Ext-r}]
    {\raisebox{-1mm}{$p \extc q \ar{\lambda} q'$}}
    {q\ar{\lambda}q'}
    \\[1em]

\multicolumn{2}{c}{
\infer[A \eqdef p; \; \rname{r-Const}]
{A \ar{\lambda} p'}
{p \ar{\lambda} p'}
}
\end{array}
\]
  \caption{The operational semantics of \CCS}
  \label{fig:opsem}
\hrulefill
\end{figure}

\begin{figure}[t]
\hrulefill
\centering
  \[
  \begin{array}{l@{\hskip 2em}l}
    \infer[\rname{p-Left}]
    { q\Par  p\ar{\lambda} q'\Par  p}
    { q\ar{\lambda} q'}
    &
    \infer[\rname{p-Right}]
    { q\Par p\ar{\lambda} q\Par p'}
    { p\ar{\lambda} p'}
    \\[.5em]
\multicolumn{2}{c}{
      \infer[\rname{p-Synch}]
      { q\Par  p\ar{\tau}  q'\Par p'}   
      { q\ar{ a } q'\quad p\ar{\co{ a }}  p'}
}
\end{array}
\]
  \caption{The operational semantics of contract composition}
  \label{fig:interactions}
\hrulefill
\end{figure}

To model the interactions that take place between the server and the
client contracts, we introduce a binary composition of contracts,
$p \Par r$, whose operational semantics is in
Figure~(\ref{fig:interactions}).

A \emph{computation} consists of series of $\tau$ actions of the form 
\begin{align}\label{eq:comp}
  p \Par r &= p_0 \Par r_0 \ar{\tau} p_1 \Par r_1 \ar{\tau} \ldots 
               \ar{\tau} p_k \Par r_k \ar{\tau} \ldots 
\end{align}
It is \emph{maximal} if it is infinite, or whenever  $p_n \Par r_n$ is the
last state then~$p_n \Par r_n \Nar{\tau} $.
A computation may be viewed as two
processes $p,\, r$, one a server and the other a client, co-operating
to achieve individual goals, which may or may not be independent. We
say that the computation in (\ref{eq:comp}) is \emph{\csucc} if there exists some
$k \geq 0$ such that $r_k \ar{\ok}$. It is \emph{successful} if it is
\emph{\csucc} and there exists an $l \geq 0$ such that $p_l
\ar{\ok}$. In a \emph{\csucc} computation the client can
report \emph{success} while in a \emph{successful} computation\leaveout{one} both the client
and the server can report success; note however that they are not required to 
do so at the same time.

\begin{defi}[ Passing tests ]
\label{def:passing}
  We write $p \Must r$ if every maximal computation from $p \Par r$  is
  \emph{\csucc}. We write $p \MustSC r$ if every such computation is \emph{successful}. 
\end{defi}

Intuitively, $p \Must r$ means that the client $r$ is satisfied by the
server $p$, as $r$ always reaches a state where it can report
success. On the other hand, $p \MustSC r$ means that $p$ passes $r$
{\em and} $r$ also passes $p$; so $p$ and $r$ {\em have to
  collaborate} in order to pass each other. This is why\leaveout{Thus,} when using the
testing relation $ \MustSC$ we think of $p$ and $r$ as two peers
rather than a server and a client.

\begin{defi}[ Testing preorders ]
\label{def:preorders}
  In an arbitrary LTS we write
  \begin{enumerate}
  \item $p_1 \testleqS p_2$ if for every $r$, $p_1 \Must r$ implies $p_2 \Must r$

  \item $r_1 \testleqC r_2$ if for every $p$, $p \Must r_1$ implies $p \Must r_2$
    
  \item $p_1 \testleqSC p_2$ if for every $r$, $p_1 \MustSC r$ implies $p_2 \MustSC r$.
  \end{enumerate}
We use the obvious notation for the kernel of these preorders; for
instance $p_1 \testeq[\peer] p_2$ means 
that  $p_1 \testleqSC p_2$ and  $p_2 \testleqSC p_1$.
\end{defi}
\noindent
The preorder $\testleqS$ is meant to compare servers, as $p_1
\testleqS p_2$ ensures that all the clients passed (wrt $\Must$) by
$p_1$ are passed also by $p_2$.
\leaveout{The relation $\testleqS$ is a mild
generalisation of the standard testing preorder of
\cite{NH1984,Hennessy88a}, in that it relates also terms that can
perform $\ok$, whereas the processes of \cite{Hennessy88a} cannot
perform $\ok$.}
The preorder $\testleqC$ relates processes seen as clients, because $r_1
\testleqC r_2$ means that all the servers that satisfy $r_1$ satisfy
also $r_2$.
\leaveout{ If two terms $r_1$ and $r_2$ cannot perform $\ok$, then
they are equivalent as clients ($ r_1 \testleqC r_2$ and $r_2
\testleqC r_1$) because no server can ever satisfy them; thus to compare
$\testleqS$ with $\testleqC$ in a meaningful way we have to use terms
that contain $\Unit$.}
The third preorder, $\testleqSC$, relates processes seen as {\em
  peers}; this follows from the fact that $p \MustSC r$ is true only
if $p$ and $r$ mutually satisfy each other.

\section{Semantic characterisations}
\label{sec:characterisations}

The standard (must) testing preorder from \cite{NH1984,Hennessy88a} has been
characterised for finite-branching LTSs using two behavioural
predicates.
The first, $p \Conv s$, 
says that $p$ can never come across a divergent residual while executing the
sequence of actions $s \in \Act^\star$. We use the notation $p \Conv {}$, $p$ \emph{converges},  
to mean that
there is no infinite sequence $$p \ar{\tau} p_1 \ar{\tau} \ldots 
\ar{\tau} p_k \ar{\tau} \ldots $$\noindent
Then the general convergence predicate is defined 
inductively as follows:
\begin{enumerate}[label=\({\alph*}] 
\item $p \Conv{\varepsilon}$ whenever $p \Conv{} $

\item $p \Conv {as}$ whenever 
\begin{enumerate}[label=(\arabic*)]
\item $p \Conv{}$ and 
\item if $p \wt{a} $ then $\intsum (p  \after a) \Conv{s}$
\end{enumerate}
\end{enumerate}
\noindent where  $(p \after s)$ denotes the set  $\setof{ p' }{ p \wt{s} p' }$. Note that 
$p \wt{a} $ ensures that $(p \after a)$  is non-empty; thus
$\intsum (p  \after a)$ consists of the choice between the
elements of the non-empty set $(p \after a)$, which may in general 
be infinite.\leaveout{
  $\intsum (p  \after a)$ represents a (well-formed) process
  consisting of the choice between the elements of the non-empty set
  $(p \after a)$, which may in general 
be infinite.}

The second predicate codifies the possible deadlocks which may occur when a process
$p$ attempts to execute the (weak) trace of  actions $s \in \Act^\star$:
\begin{align}\label{eq:acceptances}
  &\acc{p}{s} = \setof{S(q)}{p \wt{s} q \Nar{\tau} }
\end{align}
where $S(q) = \setof{ a \in \Act }{q \ar{ a } }$ is the set of
actions performed strongly by $q$.
The sets~$S(q)$ are called {\em ready sets}, while we say that
$\acc{p}{s}$ is the {\em acceptance set} of $p$ {\em after} a trace~$s$.
Ready sets  are essentially the complements of the \emph{refusal sets} used in
\cite{csp}.
The sets in $\acc{p}{s}$ describe the interactions that can lead $p$ out
of a possible deadlock,  reached by executing the trace $s$ of external
actions.

\begin{thm}
\label{thm:fb}\cite{DBLP:conf/tapsoft/NicolaH87,Hennessy88a}
  In finite branching $\CCS$, $p \testleqS q$
  if and only if, for every $s \in \Act^\star$, if $p \Conv s$ then 
\begin{enumerate}[label=(\roman*)]
  \item $q \Conv s$, 
  \item  for every $B \in \acc{q}{s}$ there exists
  some $A \in \acc{p}{s}$ such that $A \subseteq B$. 
\end{enumerate}
\end{thm}

\noindent
As might be expected, the behavioural properties used in \rthm{fb} do not
characterise the preorder $\testleqC$:
\begin{cexample}
\label{ex:B-in-acc-wrong}
We prove that $ r_1 \testleqC r_2 $ does not imply that $r_1$ and $r_2$ satisfy the requirements 
of \rthm{fb}.

One can prove that  $r_1 \testleqC r_2$, where $r_1,r_2$ denote 
$b.a.\Unit$ and  $b.(c.\Cnil \extc \Unit) $ respectively.
Now consider the singleton trace $b$; obviously $r_i \Conv b$ for $i=1,2$. 
However calculations show that $\set{c} \in \acc{r_2 }{ b }$ 
and that $\acc{r_1}{b} = \set{\set{a}}$.  
So there is no set $A$ in $\acc{r_1}{b}$ satisfying $A \subseteq \set{c}$.
\leaveout{
Calculations show that  $\acc{b.a.\Unit}{b} = \set{\set{a}}$.
But  $\set{c} \in \acc{ q }{ r_2 }$ and so  there is no set
$B$ in $\acc{b.a.\Unit}{b}$ satisfying $B \subseteq \set{c}$.}
\end{cexample}

\medskip
Our intention is to provide a behavioural characterisation of 
the client preorder $\testleqC$, and later the peer preorder,  
along the lines of the characterisation of the server preorder~$\testleqS$
given in \rthm{fb}. This will require the elaboration of new behavioural
predicates which capture behaviour relevant to clients. 
Or approach is incremental. To motivate the role of these new predicates we first
define  two tentative ``bad''
characterisations of $\testleqC$. In a series of examples we show the problems
which arise with these characterisations, and which \leaveout{hopefully} will motivate the necessity
for our new client oriented behavioural predicates. These are then collated in \rdef{test.sem}
to give the \leaveout{resulting}behavioural characterisation of 
$\testleqC$.

Some additional notation is in order. In \rexa{B-in-acc-wrong}
there is no need to require that the ready set  $\set{c} \in \acc{ q }{ b }$ 
be matched by one in $\acc{b.a.\Unit}{b}$,  because $q$ can report
success immediately after performing $b$.
Intuitively ready sets need only be matched when success has not yet been reported.
In order to capture this intuition we use the following predicate. 

For every  $s \in \Act^\star$ let $p \uwt{s} q$ be the least relation satisfying
\begin{enumerate}[label=\({\alph*}]
\item $p \Nar{\ok} $ implies $p \uwt{\varepsilon} p$ 
\item if $p' \uwt{s} q$ and   $p \Nar{\ok} $ then 
  \begin{itemize}
  \item   $p \ar{a} p'$ implies $p \uwt{as} q$
  \item   $p \ar{\tau} p'$ implies $p \uwt{s} q$
  \end{itemize}
\end{enumerate}
Intuitively, $p \uwt{s} q$ means that $p$ can perform the sequence of
external actions $s$ ending up in state $q$ without passing through
any state which can report success; in particular  neither $p$
nor $q$ can report success. This notation is extended to infinite
traces,  $ u \in \Act^\infty$,  by letting $p \uwt{u}$ whenever there 
exists a $t \in (\Actt)^\infty$ such that $t = \mu_1\mu_2\ldots$,
(a) $p = p_0 \ar{\mu_1} p_1 \ar{\mu_2} p_2 \ar{\mu_3}\ldots$ implies that 
$p_i \Nar{\ok}$ for every $p_i$, and
(b) for every $n \in \mathbb{N}$ there exists a $k \in
\mathbb{N}$ such that, $u_n = \wtrace{t_k}$; where $\wtrace{t}$
is the string obtained removing the $\tau$s from $t$.
\begin{defi}
  \label{def:acc-set-ut}
  For every process $p$ and trace $s \in \Act^\star$, let
  \begin{align*}
    &\accut{p}{s} = \setof{S(q)}{p \uwt{s} q \Nar{\tau}}
  \end{align*}
  We call the set $\accut{p}{s}$ the {\em unsuccessful} acceptance set of $p$ after $s$.
\end{defi}

Our first attempt at adapting the characterisation for servers in \rthm{fb} to clients
is as follows:
\begin{defi}
  \label{def:sbad}
  Let $ r_1 \sbad r_2$ if for every  $s \in \Act^\star$, 
  if $r_1 \Conv{s}$ then (i) $r_2 \Conv{s}$, and (ii) for every  $B
  \in \accut{r_2}{s}$, there exists some $A \in \accut{r_1}{s}$
  such that $A \subseteq B$.  
\end{defi}

\begin{cexample}
\label{ex:rs-not-sound}
In this counterexample we show that the relation $\sbad$ is not complete
  wrt. $\testleqC$, that is $ \testleqC \; \not\subseteq \; \sbad $.
One can show that $r \testleqC c.a.\Unit$ where $r$ denotes 
the client $ c.(a.\Unit + b.\Cnil)$. However $r$ and $c.a.\Unit$ are not related 
by the proposed $\sbad$ in \rdef{sbad}. Obviously 
$r \Conv{c}$ and $\set{a} \in \accut{c.a.\Unit}{c}$.
But there is no
$A \in \accut{r}{c}$ such that $A \subseteq \set{a}$;
this is because $\accut{r}{c} = \set{\set{a,b}}$; thus $ r
  \not \sbad c.a.\Unit$.

The problem is the presence of $b$ in the ready set of $a.\Unit \extc b.\Cnil$.
\end{cexample}
Intuitively, in the previous example the action $b$ is \emph{unusable} for $r$
after having performed the unsuccessful trace $c$; 
this is because performing $b$ leads to a client, $\Cnil$, which is 
\emph{unusable}, in the sense that it can never be satisfied by any
server. When comparing ready sets after unsuccessful traces in
\rdef{sbad} we should ignore occurrences of
\emph{unusable} actions.

Let $\Usable[\clt] = \setof{ r }{ p \MustP r, \,\,\mbox{for some
    server $p$}}$. The set $\Usable[\clt]$ contains the usable
clients, those satisfied by at least  one server.
We also need to consider the residuals of a client $r$ only after
unsuccessful traces (see \rcexa{after-not-sound}): for any process $r$ and $s \in \Act^\star$ let
\begin{equation}
\label{eq:afterut}
 (r \afterut s)  = \setof{q}{r \uwt{s} q  }
\end{equation}
\noindent
The usability of a client, then, is parametrised over traces:
for every $s \in \Act^\star$, the client
usability along an unsuccessful trace $s$, denoted $\usbut{s}$, 
is defined by induction on $s$:
\begin{enumerate}[label=\({\alph*}]
\item $r \usbut{\varepsilon}$ whenever $r \in \Usable[\clt]$
\item $r \usbut{as}$ whenever 
\begin{enumerate}[label=(\arabic*)]
\item $r \in \Usable[\clt]$ and
\item if $r \uwt{a}  $ then ${\intsum} (r \afterut a) \usbut{s}$
\end{enumerate}
\end{enumerate}
The predicate $\usbut{}$is extended to infinite traces $u \in \Act^\infty$ in the obvious manner. 
Intuitively  $r \usbut{s}$ means that any 
state reachable from $r$ by performing any subsequence of~$s$
is usable. Note that only unsuccessful traces have to be taken into
the account, and also that the definitions of the 
  predicates $\Conv{}$ and $\usbut{}$have the same structure.
\begin{defi}
\label{def:uaut}
The set of usable actions for a client $r$ after the trace $s$ is defined as
$$
  \uaut{r}{s} = \setof{a \in \Act}{ r \uwt{sa} \text{ implies } r \usbut{sa}}
$$
\end{defi}
Now we define a second tentative characterisation of $\testleqC$,
denoted $\sbad'$, by replacing in \rdef{sbad} the
set inclusion~$A \subseteq B$ with the more relaxed condition 
\begin{equation}
\label{eq:relaxed}
A \cap \uaut{r_1}{s}\subseteq B
\end{equation}
\begin{example}
\label{ex:rs-not-sound.again}
We revisit \rcexa{rs-not-sound}, and prove that $r \sbad'
  c.a.\Unit$, thereby correctly reflecting the fact that  $r \testleqC
c.a.\Unit$ and improving on $\sbad$.

Let us see why $r \sbad'
  c.a.\Unit$. Recall that $\accut{r}{\varepsilon} = \sset{ \sset{ c,
    b}}$, and observe that the action $b$ is not in $\uaut{r}{c}$ 
because $r \uwt{cb}$ and $r \not \usbut{cb}$. The last fact is true
because $(r \afterut cb)$ is the singleton set $\sset{
  \Cnil }$, and $\Cnil$ is not in $\Usable[\clt]$. Instead we have $\uaut{r}{c} =
\set{a}$, and so the
inclusion in \req{relaxed} used to define $\sbad'$ is satisfied
by the ready sets at hand, because
$$
\sset{ a, b } \cap \sset{ a } \subseteq \sset{ a }
$$
thus $r \sbad' c.a.\Unit$.
\end{example}

Before critiquing the second attempted chaaracterisation, $\sbad'$, let us
point out that in the definition of $\usbut{}$it is necessary to consider
only unsuccesful traces.
\begin{cexample}
\label{ex:after-not-sound}
 In this counter example we prove that if in the definition
of $\usbut{}$ above we consider all the traces rather than
only the unsuccessful ones, then the relation~$\sbad'$ is
not sound wrt $\testleqC$, that is $\sbad' \; \not\subseteq \;
\testleqC$.

Consider the client 
$r = b.(\tau.(\Unit \extc a.\Cnil) \extc \tau.a.\tau.\Unit)$. First note that
 $ \co{b}.\co{ a }.\Cnil \Must r$ while  $ \co{b}.\co{ a }.\Cnil \not
\Must b.\Cnil $ and therefore  $r \not \testleqC b.\Cnil$.

Suppose that $\usbut{}$ was defined using $\wt{}$ and $\after$ in
place of $\uwt{}$ and $\afterut$. We prove that $ r \sbad' b.\Cnil $.

The set $( r \after ba )$ is $\set{\Cnil,\Unit }$,
so $\intsum ( r \after ba )$ is the client $\tau.\Cnil \extc \tau.\Unit$, which is not in~$\Usable[\clt]$. 
This implies that $\uaut{r}{b}$ contains only actions not performed by
$r$ after $b$, therefore any ready set $B \in \accut{b.\Cnil}{b}$
can be matched according to \req{relaxed} by the ready set
$\sset{a} \in \accut{r}{b}$, because $ A \cap \uaut{r}{b} =
\emptyset$.
From this $r \sbad' b.\Cnil$ follows.

However with the correct definition of $\usbut{}$this reasoning
no longer works because $\uaut{r}{b}  = \set{a}$.
\end{cexample}

Unfortunately, as might be expected,  the relation $\sbad'$ is still not sufficient to obtain a
complete characterisation of the client preorder; one more adjustment is required.
\begin{cexample}
\label{ex:rel-convergence}
Here we prove that the relation $\sbad'$ is not
  complete wrt $\testleqC$.
Consider the clients $r_1 = a.(b.d.\Cnil + b.\Unit)$ and $r_2 = a.c.d.\Unit$.
As $r_1$ is not usable $r_1 \testleqC r_2$, although $r_1 \Nsbad' r_2$.
To see this first note $\set{d} \in \accut{r_2}{ac}$, and  $r_1 \Conv{ac}$,
although $r_1$ can not actually perform the sequence of actions $ac$;
$r_1 \Conv{ac}$ merely says that if $r_1$ can perform any prefix of
the sequence $ac$ to reach $r'$ then $r'$ must converge. 
Since $r_1$ can not perform the sequence of actions $ac$,
$\accut{r_1}{ac}$ is empty; thus no ready set $B$ can
be found to match the ready set $\set{d}$.
\end{cexample}

To fix the problem highlighted in \rexa{rel-convergence} 
we need to reconsider when ready sets are to be matched. In \rdef{sbad}
this matching is moderated by the 
predicate  $ \Conv{s}$; 
for example
$a.(\tau^\infty \extc b.\Unit) \sbad a.c.d.\Unit$, where $\tau^\infty$
denotes some process which does not converge. This is because
$a.(\tau^\infty \extc b.\Unit) \Conv{a}$ is false and therefore the
ready set $\set{c} \in \accut{a.c.d.\Unit}{a}$ does not have to be 
matched by  $a.(\tau^\infty \extc b.\Unit) $.
However the client preorder is largely impervious to 
convergence/divergence. For example 
$\Unit \testeqC (\Unit + \tau^\infty)$. 

It turns out that we have to moderate the matching of ready sets, not 
via the convergence predicate, but instead via \emph{usability}. 
One can show that if $r_1 \testleqC r_2$ and $r_1 \usbut{s}$ then $r_2 \usbut{s}$. 
In fact this predicate  describes precisely when we expect ready sets
of clients to be compared.

\begin{defi}[ Semantic client-preorder ]
  \label{def:test.sem}
  In any LTS, let $r_1 \stmpo r_2$ if 
\begin{enumerate}
 \item  for every $s \in \Act^\star$ such that   $r_1
  \usbut{s}$, 
   \begin{enumerate}[label=\({\alph*}]
   \item  $r_2 \usbut{s}$, 
   \item  for every $B \in
  \accut{r_2}{s} $ there exists some $A \in \accut{r_1}{s}$ such that
  $$A \cap  \uaut{r_1}{s} \subseteq B$$
  \end{enumerate}

 \item 
 for every $w \in
  \Act^\star \cup \Act^\infty$ such that $r_1
  \usbut{ w }$,  
  $r_2 \uwt{ w } $ implies $r_1 \uwt{ w }$.
\end{enumerate}
\end{defi}

\begin{example}
  Let us revisit the clients $r_1,\, r_2,$ in \rcexa{rel-convergence}. 
The client $b.d.\Cnil + b.\Unit$ is not usable, that is 
$b.d.\Cnil + b.\Unit \not\in \Usable[\clt]$, because it cannot be satisfied by
any server. Consequently $r_1 \usbut{ac}$ does not hold, and
therefore when checking whether $r_1 \stmpo r_2$ holds
the ready set  $\set{d} \in \accut{r_2}{ac}$ does not have to be matched by
$r_1$. 

Indeed it is now straightforward  to check that $r_1 \stmpo r_2$;
the only $s \in \Act^\star$ for which $\accut{r_2}{s}$ is non-empty and 
$r_1 \usbut{s}$ is the empty sequence $\varepsilon$. 
\leaveout{
The use of the predicate $\usbut{s}$ in Definition~\ref{def:test.sem} 
is very strong. As an example consider again the client 
$ r =  a.(b.\Cnil \extc c.\Unit) \extc a.(b.\Unit \extc c.\Cnil)$. Note that
$r \not\in \Usable[\clt]$ as there is no server which can satisfy it. 
Consequently $r \usbut{s}$ is false for every trace $s$,
from which it follows that $r \stmpo r'$ for any other client. 
}
\end{example}

\begin{figure}[t]
\hrulefill
\begin{center}
\begin{tikzpicture}[scale=0.5]
       \node[state,scale=0.7](t){$q$}; 
        \node[state,scale=0.7][below=  of t](p3){$q_2$};
      \node[state,scale=0.7][left=of p3](p2){$q_1$};
      \node[state,scale=0.7][left =of p2](p1){$q_0$};
      \node[state,scale=0.7][right=of p3](p4){$q_3$};  
      \node[][right=of p4](p5){\ldots};

\path (t) edge[to] node[action,swap] {$a$} (p1)
      (t) edge[to] node[action] {$a$} (p2)
      (t) edge[to] node[action] {$a$} (p3)
      (t) edge[to] node[action,swap] {$a$} (p4)
      (t) edge[to] node[action] {$a$} (p5)
      (t) edge[loop,to] node[action,swap] {$a$} (6);
\end{tikzpicture}
\end{center}
  \caption{Infinite traces}
  \label{fig:inf.tr}
\hrulefill
\end{figure}

In passing let us note that 
in general, and in particular in LTSs which are not
finite branching, the condition on the existence of infinite
computations in (2) of \rdef{test.sem} does not follow from the condition on finite
computations.

\begin{example}
  Consider the process $q$ from \rfig{inf.tr}, where $q_k$ denotes a
  process which performs a sequence of $k$ $a$ actions followed by $\Unit$. 
Let $p$ be a similar process, but without the self loop. Then  
$p \usbut{s}$ and  $q \usbut{s}$ for every $s$, and 
the pair $(p,q)$ satisfies condition (1) of $\stmpo$, and condition (2) on finite $w$s.
However condition (2) on infinite $w$s is not satisfied: if $u$ denotes the infinite sequence of $a$s then $q \uwt{u} $ but $p \Nuwt{u} $.

In fact $p \nottestleqC q$. For consider the process
$A \eqdef \overline{a}.A$. When $p$ is run as a test on
$A$, or as a \emph{client} using the \emph{server}
$A$, every computation is finite and successful; $A \Must p$.
However when $q$ is run as a test, there
is the possibility of an infinite computation, the indefinite
synchronisation on $a$, which is not successful; $A \NMust q$. 
\end{example}

\begin{thm}
\label{thm:c.completeness}
  In \CCS, $r_1 \testleqC r_2$ if and only if
  $r_1 \stmpo r_2$.   
\end{thm}

\proof
  It follows from \rthm{c.soundness} and \rthm{c.completeness2} of Section~\ref{sec:clientproofs}.
\qed

The server-preorder $\testleqS$ can be characterised
behaviourally  in manner dual to that of
\rdef{test.sem}, 
using the set of usable servers 
$\Usable[\svr] = \setof{p}{ p \MustP r, \,\,\mbox{for some
    client $r$}}$,
the usable actions
$$\uaSvr{p}{s} = \setof{a \in \Act}{ p \wt{s} \text{ implies }
p \usb{sa}} $$
and the \emph{server} convergence predicate   $p \svrConv{s}$,
defined as the conjunction of $p \Conv{s}$ and 
a server usability predicate $p
\usb{s}$. This latter predicate is defined inductively in
a manner similar to  $\usbut{s}$, but over all traces $s$, 
rather than  simply the unsuccessful ones.
\leaveout{
  \begin{enumerate}[label=\({\alph*}]
  \item $r \usb{\varepsilon}$ if $r \in \Usable[\clt]$
  \item $r \usb{as}$ if  $r  \in \Usable[\clt]$, and if $r \wt{a}  $ then ${\intsum} (r \after a) \usb{s}$
  \end{enumerate}
}
\leaveout{
The server-preorder $\testleqS$ can be characterised
behaviourally  in manner dual to that of
\rdef{test.sem}, 
using the set of usable servers 
$\Usable[\svr] = \setof{p}{ p \MustP r, \,\,\mbox{for some
    client $r$}}$,
the usable actions
$\uaSvr{p}{s} = \setof{a \in \Act}{ \intsum ( p \after sa) \in
  \Usable[\svr]}$,
and the \emph{server} convergence predicate   $p \svrConv{s}$,
defined as the conjunction of $p \Conv{s}$ and 
a server usability predicate $p
\usb{s}$. This latter predicate is defined inductively in
a manner similar to  $\usbut{s}$, but over all traces $s$, 
rather than  simply the unsuccessful ones.  
}
\begin{defi}[ Semantic server-preorder ]
  \label{def:proc.sem}
  In any LTS, let $p \spmpo q$ if 
\begin{enumerate}
\item for every $s \in \Act^\star$ such that $p \svrConv{s}$,  
\begin{enumerate}[label=\({\alph*}]
  \item $q \svrConv{s}$,
  \item for every $B \in \acc{q}{s} $ there
    exists some $A \in 
    \acc{p}{s}$ such that $$A \cap \uaSvr{p}{s}
    \subseteq B$$
\end{enumerate}
  \item for every $w \in \Act^\star \cup \Act^\infty$ such that   $p \svrConv{w}$, $q \wt{w} $ implies $p \wt{u}$. 
\end{enumerate}
\end{defi}
\begin{thm}
\label{thm:s.completeness}
  In \CCS, $p \testleqS q$ if and only if $p \spmpo q$. 
\end{thm}
\proof
The standard argument of \cite{Hennessy88a} suffices, but for the
condition on infinite traces, which we prove here.

Let $u = a_1a_2a_3 \ldots $, and $C_n \eqdef \tau.\Unit
\extc \overline{a}_n.C_{n+1}$ for every~$n \in \mathbb{N}$. No
$C_n$ is successful, so the infinite computation of $ C_0 \Par p_2 $
due to the trace $u$ proves that~$p_2 \not\Must C_0$.
The hypothesis~$p_1 \testleqS p_2$ implies that~$p_1 \not\Must C_0$. 
For a contradiction, suppose that $p_1 \Nwt{ u }$.
The assumption $p_1 \svrConv{u}$ implies $p_1 \Conv{ u }$, which in turn
lets us prove that all the maximal computations of~$C_0 \Par p_1$ are
\csucc. But this is not possible, as $p_1 \not\Must C_0$. It
follows that~$p_1 \wt{u}$.
\qed
\noindent
\rthm{s.completeness} is a generalisation of \rthm{fb}, as the server
usability predicate~$\Usable[\svr]$ is degenerate: it holds for every
process, since any process used as a server trivially satisfies the
degenerate client $\Unit$.
\leaveout{
This can be seen to be a generalisation of \rthm{fb}, as the server
usability predicate $\Usable[\svr]$ is degenerate; it holds for every
process, since any process used as a server trivially satisfies the
degenerate client $\Unit$.
}

\medskip
Let us now consider the peer preorder. The following result is hopeful:
\begin{prop}
\label{prop:sc-in-c}
In \CCS,
  $r_1 \testleqSC r_2$ implies $r_1 \testleqC r_2$.
\end{prop}
\proof
First note that using Theorem~(\ref{thm:s.completeness}) 
one can prove that
$\Unit \extc p \testleqS p$ and that $p \testleqS \Unit \extc p$.

Now suppose that $ p \Must r_1$; it follows that 
$\Unit \extc p \Must r_1$, and so $\Unit \extc p \MustSC
r_1$ because $\Unit \extc p$ is trivially satisfied.
The hypothesis imply that $\Unit \extc p \MustSC r_2$, thus
$\Unit \extc p \Must r_2$.  In turn this ensures that $p \Must r_2$.
\qed
Unfortunately, the peer preorder is \emph{not} contained in the server preorder:
\begin{example}
\label{ex:testleqSC-not-in-testleqS}
 It is easy to see that~$a.\Cnil \testleqSC b.\Cnil $. This is
 true because~$a.\Cnil$ can never be satisfied, for it offers
 no~$\ok$ at all. However,~$a.\Cnil \not \testleqS b.\Cnil$,
 as the client~$\co{a}.\Unit$ is satisfied by~$a.\Cnil$,
 whereas~$b.\Cnil \not \Must \co{a}.\Unit$.
\end{example}

Intuitively, the reason why $\testleqSC \; \not \subseteq \;
\testleqS$ is that the server preorder does 
not reflect the factthat servers should now act as peers; that is they should also
be satisfied by their interactions with clients. To take this into account we introduce
the usability of peers and amend the definition of $\spmpo$
accordingly.
In principle we should introduce the set of usable peers,
$
\Usable[\peer] = \setof{p}{ p \MustSC r \,\,\mbox{for some peer $r$}}
$.
However, since $\Usable[\peer]$ turns out to coincide with~$\Usable[\clt]$,
instead we define the peer convergence predicate by using the
usability predicate of {\em clients}. For every $w \in \Act^\star \cup
\Act^\infty$, let 
$
p \peerConv{w} \text{ whenever } p \Conv{w} \text{ and } p \usbut{ w}.
$

\begin{defi}
\label{def:Usmpo}
Let $p \Usmpo q$ whenever 
\begin{enumerate}
\item for every $s \in \Act^\star$, if $p \peerConv{s}$ then
\begin{enumerate}[label=\({\alph*}]
\item $q \Conv{s}$, 
\item for every $B \in \acc{q}{s} $ there
exists some $A \in \acc{p}{s}$ such that $$A  \cap \uaut{p}{s} \subseteq B$$
\end{enumerate}
\item for every $w \in \Act^\star \cup \Act^\infty$, if $p
\peerConv{w}$, and $q \wt{w} $, then $p \wt{w}$.
\end{enumerate}
\end{defi}


\begin{defi}[ Semantic peer-preorder ]
\label{def:mut.sem}
Let $p \sptmpo q$ if  $p \stmpo q$ and $p \Usmpo q $.
\end{defi}
\noindent
Note that the definition of $p \sptmpo q$ is \textbf{not} simply the
conjunction of the client and server preorders from \rdef{test.sem}
and \rdef{proc.sem}.  It is essential that the usable set
of peers $\Usable[\peer]$ be employed.

\begin{thm}\label{thm:sc.completeness}
  In \CCS, $p \testleqSC q$ if and only if $p \sptmpo q$.
\end{thm}
\proof
  See Section~\ref{sec:peerproofs}. 
\qed

\section{Characterising the client behaviour}
\label{sec:clientproofs}

This section is devoted to the proof of behavioural characterisation of 
the client preorder, Theorem~\ref{thm:c.completeness}. For convenience it is
divided into three subsections. The first gathers some preliminary technical 
properties of the various predicates used in the characterisation; the second is
devoted to \emph{soundness} and the final one to \emph{completeness}.

\subsection{Preliminaries}
Here we collect some technical results on the interplay between the testing
predicate $p \Must r$ and the client and action usability predicates. 
The two corollaries below are the main results of the section. 

\begin{lem}\label{lemma:one}\label{lem:one}
  Suppose $p \Must r$ where $p \wt{\co{s}} q$.
  Then $r \uwt{s} r' $ implies $q \Must r'$.
 \end{lem}
\proof
  Straightforward as any maximal computation from 
  $q \Par r'$ can be prefixed by an unsuccessful sequence of 
reduction steps to obtain a maximal computation from $p \Par r$. 
\qed

\begin{cor}
\label{cor:one}
    Suppose $p \Must r$ where $p \wt{\co{s}} q$.
    Then $r \usbut{s}$. 
\end{cor}
\proof
  By induction on $s$. If $s$ is the empty sequence $\varepsilon$
then the result is immediate, as $p \Must r$ ensures that $r \in
\Usable[\clt]$. 
So assume $s$ has the form $b.t$ and $r \uwt{b}$. 
We have to show that $\intsum   (r \afterut a) \usbut{t}$. 

Let $p \wt{\co{b}} p_b \wt{\co{t}} q$.
By \rlem{one} we know $p_b \Must r'$ whenever $r \uwt{b} r'$. 
This in turn means that  $p_a \Must \intsum   (r \afterut a) $. 
Now apply induction. 
\qed
\leaveout{
\begin{lem}
  Suppose $p \Must r$ and $p \wt{\co{a}} p_a$.
Then  $r \uwt{\varepsilon} r' \ar{a} r''$ implies  $p_a \Must r''$. 
\end{lem}
\proof
  Similar to the proof of Lemma~\ref{lemma:one};  as any maximal computation from 
  $p_a \Par r''$ can be prefixed by an unsuccessful sequence of 
reduction steps to obtain a maximal computation from $p \Par r$. 
\MHc{needs to be changed}
\qed
}

\begin{prop}
  Suppose $p \Must r$ and $p \wt{\co{sa}} q$.
  Then $r \uwt{s} r' \ar{a} r''$ implies  $q \MustP r''$. 
\end{prop}
\proof
Suppose that $p \wt{\co{sa}} q$ and that $r \uwt{s} \ar{a} r''$.
We prove $q \NMust r''$ implies $p \NMust r$.

Since $q \NMust r''$ there must exist a maximal unsuccessful computation from 
\begin{equation}
\label{eq:unsucces-max-comp-qr}
q \Par r'' = q_0 \Par r''_0 \ar{\tau} q_1 \Par r''_1 \ar{\tau}
q_2 \Par r''_2 \ar{\tau} \ldots
\end{equation}
such that $r''_k \Nar{\ok}$ for every $k \geq 0$. In particular $r'' \Nar{\ok}$.

The two derivations  $p \wt{\co{sa}} q$ and $r \uwt{s} r' \ar{a} r''$ can be zipped 
  together to obtain a computation
\begin{equation}
\label{eq:unsucc-max-comp}
  p \Par r = p_0 \Par r_0 \ar{\tau}
p_1 \Par r_1 \ar{\tau} \ldots 
p_n \Par r_n = q \Par r'' 
\end{equation}
Moreover here $r_i \Nar{\ok}$ for every $0 \leq i \leq n$.

Now the computation in (\ref{eq:unsucc-max-comp}) can be continued using
the one in (\ref{eq:unsucces-max-comp-qr}), leading to a maximal computation from $p \Par r$ which is unsuccessful. It follows that $p \NMust r$.
\leaveout{
\MHc{There is a simpler proof}
  By induction on the size of $s$. 
  When $s$ is the empty sequence the result is given by the previous
lemma.
So let us assume that $s$ has the form $b.t$. In this case we have
\begin{align*}
  &p \wt{\co{b}} p_b \wt{\co{t.a}} q \\
  & r \uwt{\varepsilon} \uar{b} r_b \uwt{t} r' \ar{a} r''
\end{align*}
An application of the lemma gives $p_b \Must r_b$ and by induction
we can conclude the required $q \Must r''$. }

\qed

\begin{cor}\label{cor:usable.actions}
  Suppose $p \Must r$ where  $p \wt{\co{sa}} q$
  and $r \wt{s} r' \ar{a} r''$.
  Then $a \in  \uaut{r}{s}$.
\end{cor}
\proof
If $r \Nuwt{sa}$, then by \rdef{uaut} $a \in \uaut{r}{s}$.
If $r \uwt{sa}$, then we have to prove that~$r \usbut{sa}$;
the argument relies on the previous proposition and induction on $s$.

If $s$ is empty then $\intsum (r \afterut a) \usbut{\varepsilon}$
because of the previous proposition.

If $s = bt$, then $p \wt{\co{b}} p' \wt{\co{ta}} q$. An application of
the previous proposition to $p \wt{\co{b}} p'$ ensures that
$p' \Must r''$ for every $r'' \in ( r \afterut b )$, so $p' \Must \intsum ( r \afterut b )$.
As $p' \wt{ta}$, induction implies that $\intsum (r \afterut b) \usbut
ta$.
\qed

\leaveout{
\begin{cor}\label{cor:usable.actions}
  Suppose $p \Must r$ where  $p \wt{\co{s.a}} q$
  and $r \uwt{s} r' \ar{a} r''$.
  Then $a \in   \uaut{r}{s}$.
\end{cor}
\proof
If $r \Nuwt{sa}$, then by \rdef{uaut} $a \in \uaut{r}{s}$.
If $r \uwt{sa}$, then we have to prove $r \usbut{sa}$;
the argument relies on the previous proposition and induction on $s$.

If $s$ is empty then $\intsum (r \afterut a) \usbut{\varepsilon}$
because of the previous proposition.

If $s = b.t$, then $p \wt{b} p' \wt{t} q$ and
the previous proposition ensures that
$p' \Must r''$ for every $r'' \in ( r \afterut b )$.
So $p' \intsum ( r \afterut b )$.
As $p' \wt{ta}$, induction implies that $\intsum (r \afterut b) \usbut
ta$.

We know that the set of clients 
$\uder{r}{s}$ is non-empty. 
The previous proposition ensures that 
$q \Must r''$ for every $r'' \in \uder{r}{s}$, from which it follows
immediately that $q \Must \intsum (r \afterut s)$. 
\qed
}

Usability ensures that even when a client diverges, it can report
success.
\begin{lem}
\label{lem:convergence-tick}
If $r \in \Usable[\clt]$ and $r \Uparrow$, then for every infinite
reduction sequence $r = r_0 \ar{ \tau } r_1  \ar{ \tau } r_2 \ar{ \tau
} r_3 \ar{ \tau } \ldots$, there exists an $n \in \mathbb{N}$ such that $r_n \ar{ \ok }$.
\end{lem}
\proof
  As~$ r \in\Usable[\clt]$ there exists a server $p$ such that~$p
  \Must r$. Fix a divergent computation of $r$, and zip it with~$p$,
  $$
  p \Par r = p \Par r_0 \ar{\tau} p \Par r_1 \ar{\tau} p \Par r_2
  \ar{\tau} \ldots
  $$
  The computation must be \csucc, so $r_n \ar{ \ok }$ for
  some $n \in \mathbb{N}$.
\qed

\subsection{Soundness}
\label{sec:client.soundness}

Here we prove that the behavioural preorder in Definition~\ref{def:test.sem} 
provides a sufficient set of conditions to capture the client-preorder. 
It is difficult to break the proof into a series of manageable independent results; 
instead we have one long monolithic proof.

\begin{thm}[Soundness client preorder]\label{thm:c.soundness}
  $r_1 \stmpo r_2$ implies $r_1 \testleqC r_2$.
\end{thm}
\proof
  Fix a pair~$ r_1 \stmpo r_2 $, and let~$ p \Must
  r_1$; we have to show that all the maximal computations of the
  composition~$ r_2 \Par p $ are \csucc.
  The argument is by contradiction, in that we show that if a maximal computation of $p \Par r_2$ is not \csucc, then
  also~$p \Par r_1$ performs a non \csucc computation, so $p \NMust r_1$.

  Fix a maximal computation from $p \Par r_2$,
  \begin{equation}
    \label{eq:scmpo-sound-comp}
p \Par r_2
    =
     p^0 \Par r^0_2 
    \ar{ \tau } 
    p^1 \Par  r^1_2 
    \ar{ \tau }
    p^3 \Par r^3_2 
    \ar{ \tau } 
    p^4 \Par  r^4_2 
    \ar{ \tau } \ldots
  \end{equation}
  The computation in (\ref{eq:scmpo-sound-comp}) is
  finite or infinite. We discuss the two cases separately.
  
  Suppose that the computation is finite, and unzip it; the resulting
  contributions of~$p$ and~$r_2$ are
  $$
  r_2 \wt{ s } r^k_2, \qquad
  p \wt{ \co{ s }} p^k
  $$
  for some~$s \in \Act^\star$, and stable $p^k \Par r^k_2$.
  The hypothesis~$p \Must r_1$,~$p \wt{ \co{ s }}$, and \rcor{one}
  imply that~$ r_1 \usbut{ s } $.
  Suppose that the computation in (\ref{eq:scmpo-sound-comp}) is not
  \csucc, 
so no state in the contribution of~$r_2$ reports success. It follows~$r_2 \uwt{ s }
  r^k_2$, and as~$r^k_2 \Nar{\tau}$, $ S( r^k_2 ) \in \accut{ r_2 }{ s } $; so 
  part (1b) of \rdef{test.sem} implies that~$ A \in \accut{ r_1 }{ s } $,
  for some~$A$ such that~$A \cap \uaut{r_1}{s} \subseteq
  S( r^k_2 )$. \rdef{acc-set-ut} implies that there exists a~$r'_1$ such that~$S(
  r'_1 ) = A$ and $ r_1 \uwt{ s } r'_1 \stable $.
Zipping  together the contributions along~$s$ of~$p$
and~$r_1$, the resulting computation reaches the state $
p_k \Par r'_1$; if this state is stable, then the computation is maximal and
{\em not} \csucc, so~$p \not\Must r_1$. This contradicts our
assumption that~$p \Must r_1$.

So it remains to show that $p_k \Par r_1'$ is stable. 
Suppose, for a contradiction, that $p_k \ar{\co{c}} $
and $r_1' \ar{c}$, that is  $c \in A$, for some action $c$. 
This situation matches the assumptions in
Corollary~\ref{cor:usable.actions} exactly, which gives that $c \in
\uaut{r_1}{s}$. Since $A \cap \uaut{r_1}{s} \subseteq S(r_2^k)$ this
in turn means that $r_2^k \ar{c} $, which contradicts the assumption that $p \Par r_2^k$ is stable.

We have discussed when the computation in (\ref{eq:scmpo-sound-comp}) above is finite. Now let us suppose that it is infinite.
As before unzip it.

Either~$ p $ and~$r_2$ perform infinite traces, or they perform
finite traces and then (at least) one of them diverge.

If we are in the first case, then
$$
r_2 \wt{ u }, \qquad
p \wt{ \overline{ u } }
$$
The assumption~$p \Must r_1$, the fact that~$p \wt{ \overline{ u } }$,
and \rcor{one} applied to every prefix of $u$, imply that~$ r_1 \usbut{ u }$.
The proof that there is a successful term in~$r_2 \wt{ u }$ is by
contradiction; for suppose that~$r_2 \uwt{ u }$; then part (2) of
\rdef{test.sem} implies that~$r_1 \uwt{ u }$. By zipping~$r_1 \uwt{ u
}$ with~$p \wt{ \overline{ u } }$ we obtain a maximal computation of~$
p \Par r_1$ which is not \csucc; this implies that~$ p \NMust
r_1$, which contradicts our original assumption on~$p$.

Suppose now that~$p$ and~$r_2$ engage in a finite trace and then there is a
divergence; by unzipping the computation in
(\ref{eq:scmpo-sound-comp}) we get the contributions
$$
r_2 \wt{ s } r^k_2, \qquad
p \wt{ \overline{ s } } p^k
$$
The assumption~$p \Must r_1$, the fact that~$p \wt{ \overline{ s } }$,
and \rcor{one} imply that~$ r_1 \usbut{ s }$.
Either~$p^k$ diverges or~$r^k_2$ diverges, or both diverge.

Suppose that~$p_k$ diverges. To prove that the computation in
(\ref{eq:scmpo-sound-comp}) is \csucc we reason by
contradiction: suppose that there is no successful state among~$ r_2,
\ldots, r^k_2 $; this implies that~$r_2$ performs the trace~$s$ unsuccessfully,
$$ r_2 \uwt{ s } $$
Part (2) of \rdef{test.sem} ensures that~$ r_1 \uwt{ s } r'_1$. We zip
the contribution of~$ p $ with the unsuccessful transition of~$r_1$; as
$p_k$ diverges the resulting computation is maximal,
\begin{equation}
\label{eq:testleqc-char-comp}
p \Par r_1 
\wt{ } 
p_k \Par r'_1
\wt{ }  
p_k  \Par r'_1 
\wt{ }  \ldots 
\end{equation}
All the derivatives of~$r_1$ in the maximal computation above are in
$r_1 \uwt{ s } r'_1$, so they are not successful. It follows that the
computation in (\ref{eq:testleqc-char-comp}) is not \csucc.
However this contradicts the assumption $p \Must r_1$.

Finally suppose that~$r^k_2$ diverges.
If there is a successful state in~$ r_2 \wt{ s } r^k_2$ then the maximal
computation we unzipped is \csucc. Therefore suppose that there is no successful state in the contribution of~$r_2$, that is~$ r_2
\uwt{ s } r^k_2$. As~$ r_1 \usbut{ s } $,
part (1a) of \rdef{test.sem} implies that~$ r_2 \usbut{ s }$. Now one
can show that this implies that $r^k_2 \usbut{ \varepsilon } $.
So an application of \rlem{convergence-tick} ensures that the unzipped computation is \csucc.

\leaveout{
We have to show that $r_1' \Par p_k$ is stable. 
Suppose, for a contradiction, that $p_k \ar{\co{c}} $
and $r_1' \ar{c}$, that is  $c \in A$, for some action $c$. 
This situation matches the assumptions in Corollary~\ref{cor:usable.actions} 
exactly, which gives that $c \in \uaut{r_1}{s}$. Since 
$A \cap \uaut{r_1}{s} \subseteq S(r_2^k)$ this in turn means that 
$r_2^k \ar{c} $, which contradicts the assumption that $p \Par r_2^k$ is stable. 
}
\qed

\subsection{Completeness}

Here we show the converse of \rthm{c.soundness}, which involves 
showing that the testing preorder $r_1 \testleqC r_2$ implies
the collection of properties gathered together in Definition~\ref{def:test.sem}.
These in turn are quantified over all sequences $s \in \Act^\star$ and $w \in \Act^\infty$; we handle this quantification
using induction over the length of $s$. 
First a technical lemma.
\begin{lem}
\label{lem:nil-not-must}
  Suppose $r \in \Usable[\clt]$. Then $\Cnil \NMust r$ if and only if
$r \uwt{\varepsilon} r' \Nar{\tau}$ for some client~$r'$.
\end{lem}
\proof
  One direction is straightforward. 
For the converse
suppose  $\Cnil \not\Must r$; we have to show that there exists some $r'$ satisfying
$r \uwt{\varepsilon} r' \Nar{\tau}$.

Since  $\Cnil \not\Must r$ there must exist some unsuccessful maximal computation
\begin{align}\label{eq:max.comp2}
  \Cnil \Par r = \Cnil \Par r_0  \ar{\tau} \ldots   \ar{\tau} \Cnil \Par r_k \ar{\tau} \ldots
\end{align}
Suppose this is infinite. Then $p \Must r$ can not be true for any server $p$, as
$\Cnil$ can be replaced in (\ref{eq:max.comp2}) by any $p$, to obtain an unsuccessful 
maximal computation from $p \Par r$. This contradicts the assumption that  $r \in \Usable[\clt]$.

So  (\ref{eq:max.comp2}) has to be finite, with terminal element $\Cnil \Par r_n$. 
The required $r'$ is $r_n$. 
\qed

\begin{prop}\label{prop:client.props}
  Suppose $r_1 \testleqC r_2$ where $r_1 \in \Usable[\clt]$.
  \begin{enumerate}
  \item \label{pt:action-sim}
If $r_2 \uwt{a} $ then $r_1 \uwt{a} $

\item
for every $B \in \accut{r_2}{\varepsilon}$ there exists some set 
 $A \in \accut{r_1}{\varepsilon}$ such that
  $$A \cap  \uaut{r_1}{\varepsilon} \subseteq B$$

\item 
if $r_2 \uwt{a} $ then 
$\intsum  (r_1 \afterut a) \testleqC  \intsum (r_2 \afterut a)$.
  \end{enumerate}
\end{prop}
\proof
  Throughout let $p_1$ be a server such that $p_1 \Must r_1$.
  \begin{enumerate}
  \item Let $p = p_1 \extc \co{a}.\tau^\infty$. As $p$ diverges after
    the interaction on $\co{a}$, and $r_2$ performs $a$ without reaching successful states, $p \not \Must r_2$. The hypothesis implies $p \not\Must r_1$.
In turn this ensures that $r_1 \uwt{a}$, for otherwise the assumption on $p_1$ would imply that $p \Must r_1$.
\leaveout{\footnote{
This follows from the fact that if $p \Must r$ then 
$r \not\uwt{a}$ if and only if $p \extc \co{a}.\Cnil \Must r$.
\GBc{
COUNTER EXAMPLE: let $r = \tau.\Unit \extc a.\tau.\Unit$. Then
$ \Cnil \Must r$,
$ \Cnil \extc \co{a}.\Cnil \Must r$,
and $ r \uwt{a}$.
}
}}

\item
  Let $\accut{r_1}{\varepsilon}$  be denoted by $\setof{A_i}{i \in I}$, for some index set $I$.
Note that \leaveout{the index set} $I$ may be empty, or indeed infinite.
For convenience we use $U$ to denote the set $\uaut{r_1}{\varepsilon}$. 
Suppose, for a contradiction, that there exists some $B \in \accut{r_2}{\varepsilon}$
such that 
\begin{equation}
\label{eq:absurd}
\text{for every }i \in I\text{ there exists some action }a_i \in (A_i \cap
U) \backslash B
\end{equation}

We will eventually show that this assumption contradicts the 
hypothesis that  $r_1 \testleqC r_2$. 

But first we show that it implies that the index set $I$ is 
non-empty. The existence of $B$ ensures that 
$ \accut{r_2}{\varepsilon}$ is not empty; that is  there exists a client $r'_2$
such that $r_2 \uwt{ \varepsilon } r'_2 \stable$. Thus \rlem{nil-not-must}
implies that $\Cnil \not\Must r_2$. As $r_1 \testleqC r_2$, $\Cnil
  \not\Must r_1$, and another  application of \rlem{nil-not-must}
  ensures  that 
  $I$ is not empty.

For each $i \in I$ let  $D_i$ denote the set $\setof{r'}{ r_1
  \uwt{\varepsilon} \ar{a_i} r'}$;
because $a_i \in U$\leaveout{ we may assume that} each of these sets are
non-empty. 
We also know, again because $a_i \in U$, 
that for every $i \in I$ there is some
server $p_i$ satisfying $p_i \Must r'$ for every $r' \in D_i$.
This is true because either $r' \ar{\ok}$ for every $r' \in D_i$,
or \rdef{uaut} ensures that there exists a $p_i$ such that 
$p_i \Must \bigoplus \setof{r' \in D_i}{ r' \Nar{\ok}}$. Plainly $p_i \Must r'$ for every $r' \in D_i$ such that $r' \ar{\ok}$, so the server $p_i$ indeed satisfies all the clients in $D_i$.

Let $J = \setof{ i \in I}{ r_1 \Nuwt{a_i} }$; 
this subset of $I$ contains the indices of all the actions $a_i$
which $r_1$ can perform weakly while passing through a successful state.
Now let $p$ denote the server
$$\sum_{i \in I \setminus J}
\co{a_i}. p_i \extc \sum_{j \in J} \co{a_j}.\Cnil$$

To establish the contradiction to   $r_1 \testleqC r_2$, it remains to
show that
$p \not\Must r_2$ and $p \Must r_1$.
\begin{enumerate}[label=\({\alph*}]
\item 
$p \not\Must r_2$. A finite  unsuccessful maximal computation is ensured by the
existence of $B$ in $\accut{r_2}{\varepsilon}$, and the assumption
(\ref{eq:absurd}) above.

\item 
To prove  that $p \Must r_1 $ is a little delicate. Consider a maximal 
computation
\begin{align}\label{eq:max.comp}
  p \Par r_1 = p^0 \Par r^0 \ar{\tau} \ldots  p^k \Par r^k \ldots
\end{align}
If the server $p$ remains untouched then the same sequence of clients can be
used to construct a maximal computation from $p_1 \Par r_1$; so some $r^k$ 
must report success. 
On the other hand suppose $p$ is touched. For example $p^k$ is $p$ while 
$p^{k+1}$ is $\Cnil$ or~$p_i$ for some $i \in I$. If no $r^j$, for $j \leq k$, reports success then $r^{k+1} \in D_i$, from which $p_i \Must r^{k+1}$ follows, and $p^{k+1}$ indeed equals $p_i$. This means again that
(\ref{eq:max.comp}) above is successful. 
\end{enumerate}

\item
For convenience let $\hat{r}_i$ denote $\intsum  (r_i \afterut a)$,
for $i = 1,2$.  
Suppose $p \Must \hat{r}_1$. 
Then one can argue that $p_1 \extc \co{a}.p \Must r_1$.
Since $r_1 \testleqC r_2$ this ensures that 
 $p_1 \extc \co{a}.p \Must r_2$. From this it is easy to see
that $p \Must r'$ for every $r'$ in the non-empty set 
$r_2 \afterut a$. Now the required result, $p \Must \hat{r}_2$,
follows. \qed
\end{enumerate}

\noindent
We note in passing that the first part of this  proposition  depends on the
possibility of processses diverging. In the absence of divergence, that
is if we confine our attention to both servers and clients which can
never diverge, one can prove that $a.\Unit \testleqC a.\tau.\Unit$; see Example~4.2.26 in
\cite{gbthesis}.
 This inequation would provide a counter example to part (1) of \rprop{client.props}, for $ a.\tau.\Unit
 \uwt{a}$, whereas $a.\Unit \Nuwt{a}$.

The property involving infinite sequences in \rdef{test.sem} 
does not follow from \rpt{action-sim} of \rprop{client.props},
and requires an additional argument. 
\begin{lem}
   Suppose $r_1 \testleqC r_2$. If $r_1 \usbut{u}$ and 
   $r_2 \uwt{u} $, where $u \in \Act^\infty$, then $r_1 \uwt{u} $.
\end{lem}
\proof
  Let~$u = a_1a_2a_3 \ldots $.
  To show that $r_1 \uwt{ u }$ we have to exhibit a $t \in
  \Actt^\infty$ such that $t = \mu_1\mu_2\mu_3 \ldots $ and
\begin{itemize}
  \item $ r_1 = r_1^0 \ar{\mu_1} r_1^1 \ar{\mu_2} r_1^2 \ar{\mu_3}
    \ldots $
  \item for every~$n \in \mathbb{N}$, $u_n = \wtrace{ t_k }$ for some~$k \in \mathbb{N}$
  \item for every~$n \in \mathbb{N}$, $r_1^n \Nar{ \ok }$
\end{itemize}

  The hypothesis~$r_1 \usbut{ u }$ ensures that for every 
  $u_k$ there is a $p_k$ such that $p_k \Must \intsum ( r_1 \afterut
  u_k )$.
  For every
  ~$k \in \mathbb{N}$, let~$A_k \eqdef p_k \extc \overline{ a
  }_{k+1}.A_{k+1}$.

  By zipping $r_2 \uwt{ u }$ with $A_0 \wt{\co{u}}$ one sees that
  $A_0 \not\Must r_2$, for the client $r_2$ does not report success.
  In turn $A_0 \not\Must r_1$, so there exists a maximal
  computation of~$r_1 \Par A_0$ which is not
  \csucc.
  Given the construction of the $A$'s and the $P_k$'s,
  this is possible only if the computation is due to the infinite
  trace $u$. So $r_1 \wt{ u }$ , which ensures the first two
  properties above.
  As the computation is unsuccessful, $r_1^i \Nar{\ok}$ for every $i
  \in \mathbb{N}$.
\qed

We have now gathered sufficient material to give the proof of completeness.

\begin{thm}[Completeness]\label{thm:c.completeness2}
  $r_1 \testleqC r_2$ implies $r_1 \stmpo r_2$.
\end{thm}
\proof
  We have to infer all the properties used in Definition~\ref{def:test.sem}. 
The property (2) for $w \in \Act^\infty$ follows directly from the 
preceding lemma. All other properties are parametrised on $s \in \Act^\star$;
they can be inferred using induction on the length of $s$, and Proposition~\ref{prop:client.props}.
Here we give one example, and the remaining ones can be established in a similar manner. 

We show that $r_1 \usbut{s}$ implies $r_2 \usbut{s}$. 
If $s$ is the empty string this follows immediately. 
So suppose it has the form $bt$ and $r_1 \usbut{b.t}$ ; we have to prove that 
 $r_2 \usbut{b.t}$ follows. This requires establishing 
(a) $r_2 \in \Usable[\clt]$, which is a consequence of $r_1$ being in $\Usable[\clt]$
and
(b) if $r_2 \uwt{b} $ then   ${\intsum} (r_2 \afterut b) \usbut{t}$.
So suppose  $r_2 \uwt{b} $. 
But by part (3) of Proposition~\ref{prop:client.props} we know that 
 ${\intsum} (r_1 \afterut b) \testleqC  {\intsum} (r_2 \afterut b)$. 
Moreover unravelling the assumption  $r_1 \usbut{s}$ gives 
that   ${\intsum} (r_1 \afterut b) \usbut{t}$.
The required result,  ${\intsum} (r_2 \afterut b) \usbut{t}$, now follows by
induction. 
\leaveout{
\MHc{MH: The other cases should also be checked, but not written}\\
\GBc{
We show that $r_2 \usbut{s}$ and $r_1 \usb{s}$ implies $r_1
  \uwt{s}$.
If $s$ is empty, then we apply Lemma~(\ref{lem:nil-not-must}),
the hypothesis $r_1 \testleqC r_2$, and then again the lemma.
If $s = bt$ then ${\intsum} (r_2 \afterut b) \uwt{t}$. 
Reasoning as above, we obtain the other two properties that
allow us to apply the inductive hypothesis, namely
${\intsum} (r_1 \afterut b) \testleqC  {\intsum} (r_2 \afterut b)$,
${\intsum} (r_1 \afterut b) \usbut{t}$.
Induction ensures that  ${\intsum} (r_1 \afterut b) \uwt{t}$,
which in turn let us prove $r_1 \uwt{bt}$.

We show that if $r_1 \usbut{t}$ then for every $B \in \accut{r_2}{s}$
there exists a $A \in \accut{r_1}{s}$ such that $A \cap \uaut{r_1}{s}
\subseteq B$.
If $s$ is empty, then this follows from the preceding proposition.
If $s = bt$, then we use again 
${\intsum} (r_1 \afterut b) \testleqC  {\intsum} (r_2 \afterut b)$, and
${\intsum} (r_1 \afterut b) \usbut{t}$.
These facts and induction implies that
for every $B \in \accut{ {\intsum} (r_2 \afterut b) }{ t }$ there
exists
a $A \in \accut{ {\intsum} (r_1 \afterut b) }{t}$ such that $A \cap
\uaut{ {\intsum} (r_1 \afterut b) }{t} \subseteq B$.
The property follows from an application of the equality
$ \accut{r_i}{s} = \accut{ {\intsum} (r_i \afterut b)}{t}$,
where $i \in \sset{ 1,2 }$.
}}
\qed

\section{Characterising the peer behaviour}
\label{sec:peerproofs}

In this section we are concerned with the behavioural characterisation
of the peer preorder, Theorem~(\ref{thm:sc.completeness}). The
material is organised in three subsections, where we respectively
gather ancillary results, we prove the {\em
  soundness} of the characterisation, and then prove its {\em
  completeness}.

\subsection{Preliminaries}

\begin{lem}
  \label{lem:exists-cool-server}
  If $ r \in\Usable[\clt]$, then there exists a $p$ such that $ p
  \Must r$, $ p \Nar{ \ok }$ and $ p \Nar{\tau}$.
\end{lem}
\proof

Suppose  $r \in\Usable[\clt]$. This means that there is some $p_a$ such that 
 $p_a \Must r$. As a first step in the proof of the lemma we show that
 \begin{align}\label{eq:nottick}
   p_n \Must r  &\qquad\text{for some $p_n$ satisfying $p_n \Nar{\tau}$}
 \end{align}
The argument proceeds on whether or not $p_a$ diverges.
\begin{enumerate}[label=\({\alph*}]
\item 
$p_a$ diverges: Here $p_a \Must r$ implies that $r \ar{
    \ok }$. It follows that $\Cnil \Must r$; as $\Cnil \Nar{\tau}$, so we
  can take the required $p_n$ to be  $\Cnil$.

\item
$p_a$ converges: Here let $p_n$ be any process satisfying $p_a \wt{} p_n$ and
$p_n \Nar{\tau}$; there must exist at least one. All maximal computations 
of $p_n \Par r$ are extensions of the initial computation 
$p_a \Par r$, which ensures that $p_n \Must r$. 
\end{enumerate}

Having established (\ref{eq:nottick}) above we now complete the proof of the
lemma by examining the structure of $p_n$. If $p_n \Nar{\ok}$ we are done. 
Otherwise, because of the possible structure of processes, see \rfig{syntax}, 
$p_n$ must take the form  
$p \extc \extsum_{i\in I}\Unit$ 
for some non-empty set $I$ and $p$ such that $p \Nar{\ok}$; moreover $p_n \Nar{\tau}$
ensures that $p \Nar{\tau}$ also. 

It is easy to use \rthm{s.completeness}  to prove that $ p_n \testeqS p$, since
adding $\Unit$ to  terms has no impact on their traces and acceptance sets.
It follows that $p \Must r$, and so $p$ enjoys all the required properties.
\qed
\leaveout{
Suppose  $r \in\Usable[\clt]$. This means that there is some $p_a$ such that 
 $p \Must r$. As a first step in the proof of the lemma we show that
 \begin{align}\label{eq:nottick}
   p_n \Must r  &\qquad\text{for some $p_n$ satisfying $p_n \Nar{\tau}$}
 \end{align}
If $p_a \Nar{\ok}$ then we can take the required $p_n$ to be this $p_a$. 
So we can assume $p_a \ar{\tau}$. The argument proceeds on whether or not 
$p_a$ diverges.
\begin{enumerate}[label=\({\alph*}]
\item 
$p_a$ diverges: Here $p_a \Must r$ implies that $r \ar{
    \ok }$. It follows that $\Cnil \Must r$; as $\Cnil \Nar{\tau}$,  and so we
  can take the required $p_n$ to be  $\Cnil$.

\item
$p_a$ converges: Here let $p_n$ be any process satisfying $p_a \wt{} p_n$ and
$p_n \Nar{\tau}$; there must exist at least one. All maximal computations 
of $p_n \Par r$ are extensions of the initial computation 
$p_a \Par r$, which ensures that $p_n \Must r$. 
\end{enumerate}

Having established (\ref{eq:nottick}) above we now complete the proof of the
lemma by examining the structure of $p_n$. If $p_n \Nar{\ok}$ we are done. 
Otherise, becuase of the possible structure of processes, see \rfig{syntax}, 
$p_n$ must take the form  
$p \extc \extsum_{i\in I}\Unit$ 
for some non-empty set $I$ and $p$ such that $p \Nar{\ok}$; moreover $p_n \Nar{\tau}$
ensures that $p \Nar{\tau}$ also. 

It is easy to use \rthm{s.completeness}  to prove that $ p_n \testeqS p$, since
adding $\Unit$ to  terms has no impact on their traces and acceptance sets.
It follows that $p \Must r$. 
\qed
}

The next lemma tells what it means for a process $r$ to be usable along an unsuccessful trace $s$.
\begin{lem}
\label{lem:meaning-usb}
  For every process $r$ and trace $s$, if $r \usbut{s}$ then
  for every $s'$ prefix of $s$ if $r \uwt{s'}$ there exists a server $p$ such that $ p \Must \bigoplus ( r \afterut s' )$.
\end{lem}
\proof
  As $r \usbut{s}$, $p \Must r$ for some $p$.

  We reason by induction on $s$.
  In the base case ($s = \varepsilon$) observe that $p \Must \intsum (r \after \varepsilon)$.

  In the inductive case let $s = a\hat{s}$.
  Fix a prefix $s'$ of $s$ such that $r \uwt{s'}$;
  we have to show a server $p$ which must pass $\intsum ( r \afterut
  s')$.

  If $s'$ is empty we reason as in the base case.
  If $s'$ is not empty, then $s' = as''$. Let $\hat{r} = \intsum
    ( r \afterut a )$, since $r \uwt{s}$, the hypothesis $r \usbut{s}$
  ensures that $ \hat{r} \usbut{\hat{s}}$.
  The inductive hypothesis ensures that for every $s'''$ prefix of 
  $\hat{s}$, if $ \hat{r} \uwt{s'''}$ then there exists
  a server $p$ such that $p \Must \intsum (  \hat{r} \after s''')$.
  Since $s''$ is a prefix of $\hat{s}$, 
  the equality $$ \intsum (  \hat{r} \after s'' ) = \intsum (r
  \afterut s')$$ implies that there exists a $p$ such that $p \Must
  \intsum (r \afterut s')$.
\qed

The next result gives a proof method for the predicate $\Conv{}$.
This proof method is based on the convergence of the residuals of
processes after traces.
\begin{lem}
  \label{lem:conv-traces}
  If for every $s'$
  prefix of $ s $, $p \wt{ s' } p'$ implies~$p' \Conv{}$, then $p \Conv{ s }$.
\leaveout{
  Let~$p$ be a process;~$p \Conv{ s }$ if and only if for every~$s'$
  prefix of~$ s $, if~$p \wt{ s' } p'$ then~$p' \Conv{}$
}
\end{lem}
\proof
\leaveout{
  We have to show two implications, namely
\begin{enumerate}[i)]
\item\label{Conv-impl1} if ~$p \Conv{ s }$ then for every~$s'$
  prefix of~$ s $, if~$p \wt{ s' } p'$ then~$p' \Conv{}$
\item\label{Conv-impl2} if for every~$s'$
  prefix of~$ s $,~$p \wt{ s' } p'$ implies~$p' \Conv{}$, then ~$p \Conv{ s }$.
\end{enumerate}
We prove them separately. We begin by showing (\ref{Conv-impl1}). The
assumption that~$p \Conv{ s }$ ensures that there exists a finite
derivation
\begin{equation}
\label{deriv-Conv}
\infer[]
{ p \Conv{ s } }
{ \vdots }
\end{equation}
We have to prove that
\begin{equation}
\label{Conv-traces-aim1}
\text{for every }s'\text{ prefix of }s, p \wt{ s' } p' \text{ implies
} p' \Conv{}.
\end{equation}
The argument is by induction on the derivation in (\ref{deriv-Conv}).

\paragraph{Base case.} In this case the whole derivation in 
(\ref{deriv-Conv}) amounts an application of the axiom
\rname{conv-ax}, or of the axiom \rname{conv-ax-not}.

If \rname{conv-ax} was applied then (\ref{deriv-Conv}) is
$$
\infer[ p \Conv{}; \; \rname{conv-ax} ]
{ p \Conv{ \varepsilon } }
{}
$$
It follows that~$s$ is the empty string, and so 
(\ref{Conv-traces-aim1}) requires us to prove that if~$ p
\wt{ \varepsilon } p'$ then~$p' \Conv{}$.  Fix such a~$p'$; the
definition of~$\wt{}$ ensures that~$ p \weakN{n} p'$ for some~$n \in
\N$. Reasoning by induction on~$n$, we can show that the side
condition~$p \Conv{}$ implies~$p' \Conv{}$.

If \rname{conv-ax-not} was applied, then (\ref{deriv-Conv}) is
$$
\infer[ p \Conv{}, p \Nwt{ \alpha}; \; \rname{conv-ax-not} ]
{ p \Conv{ \alpha s' } }
{ }
$$
It follows that~$s = \alpha s'$. The side condition~$p \Nwt{ \alpha}$
implies that the only prefix of~$s$ performed by~$p$ is the empty
string, so (\ref{Conv-traces-aim1}) requires us to prove that if~$ p
\wt{ \varepsilon } p'$ then~$p' \Conv{}$. To prove this we reason as
we did discussing the application the axiom \rname{conv-ax}.

\paragraph{Inductive case.}
In this case the last step in the derivation in (\ref{deriv-Conv}) is
rule \rname{conv-alpha}, so the derivation in (\ref{deriv-Conv}) is
$$
\infer[ p \Conv{}, p \wt{ \alpha}; \; \rname{conv-ax-not} ]
{ p \Conv{ \alpha s' } }
{ \infer[]{\intsum ( p \after \alpha ) \Conv{ s'}}{ \vdots } }
$$
The derivation of~$\intsum ( p \after \alpha ) \Conv{ s'}$ is
shorter than the derivation above, so we are allowed to use the
inductive hypothesis, for every~$s''$ prefix of~$s'$, if~$\intsum (
p \after \alpha ) \wt{ s''} p'$ then~$ p' \Conv{} $.

We prove (\ref{Conv-traces-aim1}). Fix a~$\hat{s}$ prefix of~$s$ such that
~$ p \wt{ s' } p'$. If~$\hat{s} = \varepsilon$, then we reason as we did in
the base case to prove that~$ p' \Conv{} $. If~$\hat{s} \neq \emptyset $
then~$\hat{s} = \alpha s''$ for some~$s''$ prefix of~$s'$. The
definition of~$\after$ ensures that~$ \intsum (
r_1 \after a ) \wt{ s'' } p' $, and so the inductive hypothesis
implies that~$ p' \Conv{ }$.

We have proven the first implication of the lemma,
namely \ref{Conv-impl1}. Now we prove the second implication,
\ref{Conv-impl2}.
}

Let us assume that for every~$s'$ prefix of~$ s $,~$p \wt{ s' } p'$ implies
~$p' \Conv{}$. The string~$\varepsilon$ is a prefix
of every string~$s$, so the assumption and~$ p \wt{
  \varepsilon} p $ imply that~$ p \Conv{}$.

We proceed by induction.
As $p \Conv{}$ the base case is true. In the inductive case $s = a.s'$
and either $ p \Nwt{ a }$ or $p \wt{ a }$. In the first case $ p
\Conv{ a.s' }$ follows, while in the second case $\intsum ( p \after
a) \wt{ s '}$. Induction on $s'$ implies that $\intsum (p \after a) \Conv{s'}$, and so $p \Conv{a.s'}$.


\qed

\subsection{Soundness}
Our aim in this section is to prove that the peer preorder contains
the  behavioural preorder of \rdef{mut.sem}.
Roughly speaking, the proof is
a combination of the standard arguments that show the soundness of the
server preorder \cite{Hennessy88a}, with the arguments on usability that we used
to prove the soundness of the client preorder, \rthm{c.soundness}.
Much in the same style of
\rsec{client.soundness}, the proof is monolithic. 

\begin{thm}[Soundness peer]\label{thm:sc.soundness}
  $ p \sptmpo q$ implies $ p \testleqSC q$.
\end{thm}
\proof
 Fix two processes $p$ and $q$ such that $ p \spmpo q$. We are
 required to show that $ p \testleqSC q$, that is 
 $ p \MustSC r$ implies $ q \MustSC r$ for every process $r$.
 Fix a process $r$ such that $ p \MustSC r$; we explain why 
 all the maximal computations of $ q \Par r$ are {\em successful}.
  
  The definition of $\sptmpo$ ensures that $p \stmpo q$, so
  \rthm{c.soundness} implies that $p \testleqC q$.
  The assumption $p \MustSC r$ ensures that $r \Must p$, thus $r \Must
  q$. It follows that all the maximal computations of $q \Par r$ are
  \csucc, that is $q$ reach a successful state.
  
  What is left to prove is that the maximal computations
  of $q \Par r$ contain a state $q' \Par r'$ wherein $r' \ar{\ok}$.

  Fix a maximal computation of $ q \Par r $,
  \begin{equation}
    \label{peer-max-comp}
    q \Par r = q_0 \Par r_0 \ar{ \tau } q_1 \Par r_1 \ar{ \tau } q_2 \Par
  r_2 \ar{ \tau } \ldots
\end{equation}
Unzip the computation above.  We obtain the contributions
  $$
  q \wt{ w }, \quad r \wt{ \co{w }}
  $$
for some possibly infinite $w$.

The argument now depends on $p$. 
Either $ p \not \Conv{w}$ or $p \Conv{w}$.

In the first case $p$ performs a prefix of $w$, say $s$, and
reaches a state $p'$ that diverges: $ p \wt{s} p' \ar{\tau} p'  \ar{\tau} \ldots $.
Zip this diverging trace of $p$ with a prefix of the trace $r \wt{ \overline{s
  }} $, and let $p'$ diverge. The result is an infinite (i.e. maximal) computation 
of $p \Par r$ that contains a successful derivative of $r$, because
$p \MustSC r$. The successful derivative of $r$ appears also
in~(\ref{peer-max-comp}) above. 

Suppose now that $p \Conv{ w }$.

The computation in~(\ref{peer-max-comp}) above is either finite or infinite. 
Suppose it is finite. Then the contributions are
  $$
  q \wt{ s } q_k, \quad r \wt{ \overline{s }} r_k
  $$
  where $q_k \Par r_k \Nar{\tau}$ and $ s = w$. The last fact ensures that $p \Conv{s}$.
  Since $ r \wt{ \overline{s}} r_k $, and $ p \MustSC r$ implies $r
  \Must p$, \rcor{one} guarantees that $ p \usbut{ s }$.

  Note that $q \wt{ s } q_k \Nar{\ok}$ ensures $S( q_k ) \in \acc{q}{s}$.
  As $ p \usbut{ s }$ and $p \Conv{s}$, we know $p \peerConv{s}$, so 
  part (1b) of  \rdef{Usmpo} implies that there exists a set $ A \in \acc{p}{s}$  
  such that $A \cap \uaut{p}{s} \subseteq S( q_k )$.
  In turn this means that there exists a stable\ $p'$ such that $S(p') =
  A$ and $p \wt{ s } p'$.
  Consider the computation $   p \Par r \wt{ } p' \Par r_k $.
  If the state $p' \Par r_k$ is stable, then the
  computation is maximal, thus $ p \MustSC r$ ensures that one of
  the derivatives of $r$ is successful. 
  This derivative appears also in~(\ref{peer-max-comp}) above.
   
  We have to prove that $ p' \Par r_k \Nar{\tau}$. The reasoning here
  is analogous to the one used in \rthm{c.soundness} to show that the
  state $r'_1 \Par p_k$ is stable, and relies on \rcor{one}.

   Thus far we have proven that if (\ref{peer-max-comp}) above is
   finite, then $r$ reaches a successful state. This is the case also
   if the computation is infinite. Let us see why.

   Either $q$ and $r$ engage in infinite traces, or (at least) one of
   them diverge.
   
   Suppose that the contributions obtained by by unzipping the computation in~(\ref{peer-max-comp}) are infinite
   \begin{equation}
     \label{eq:peer-inf-traces}
     q \wt{ u }, \quad r \wt{ \overline{ u }}
   \end{equation}
   with $ u = w$.
   We have to show that one of the derivatives of $r$ is successful.
   
   As $ r \wt{ \overline{ u }}$ and $ p \MustSC r$, \rcor{one} applied
   to every finite prefix of $u$ implies that $ p \usbut{ u }$. As 
   $p \Conv{u}$ it follows that $ p \peerConv{ u }$.
   Since $ q \wt{ u } $, part (2) of \rdef{Usmpo} implies
   that $ p \wt{ u }$. Zip this infinite trace of $p$ with $ r
   \wt{ \overline{u}}$. The resulting computation of $p \Par r$ is
   infinite as well, so the assumption $p \MustSC r$ ensures that $r$
   reaches a successful state. This state appears in (\ref{peer-max-comp})
   above.


   Now we discuss the case of  (\ref{peer-max-comp}) being due to
   finite traces and divergence of $q$ or $r$.
   To unzip (\ref{peer-max-comp}) gives the following
   contributions,
$$
     q \wt{ s } q_k, \quad r \wt{ \overline{s }} r_k
$$
with $w = s$. Note that $ p \Conv{s}$.
The fact that $r \wt{ \overline{s} } r_k$ implies $ p \usbut{
  s } $, so $ p \peerConv{s}$. Part  (1a) of
\rdef{Usmpo} implies that $q \Conv{ s }$, so the divergent process
must be $r_k$.


\leaveout{  
Suppose that $q_k$ diverges: $ q \not \Conv{ s }$. Part (1a) of
\rdef{Usmpo} implies that $ p \not \peerConv{ s }$.
Since $ p \usbut{ s } $, the fact that $p \not \peerConv{s}$
implies that $p \not \Conv{s}$. This implies that there exists a
prefix $s'$ of $s$ such that $ p \wt{ s' }p'$ and $p'$ diverges.
We obtain the result by reasoning as in the paragraph
``Divergence of $p$'' above.
   }

   Part (2) of \rdef{Usmpo}, $ q \wt{ s } $, and $p \peerConv{ s
   }$ imply that $p \wt{ s }$.
   Zipping this trace of~$p $ with the trace $r \wt{\co{s}} r_k$, and
   let $r_k$ diverge.
   The resulting computation of $ p \Par r$ is infinite, so one of the 
   derivatives of $r$ in it is successful; this is true because 
   of the assumption $p \MustSC r$.
   This successful derivative of $r$ appears also
   in~(\ref{peer-max-comp}) above.
\qed

\subsection{Completeness}
This section contains the proof that the behavioural
characterisation given in \rdef{mut.sem} is complete with respect to
the peer preorder. This result is the converse inclusion of
\rthm{sc.soundness}.

In view of \rprop{sc-in-c}, the bulk of the work is to prove that the peer
preorder is contained in the behavioural preorder $\Usmpo$, \rprop{sc-in-usmpo}.

In \rsec{clientproofs} we have proven a similar result for the client preorder
and its characterisation, \rthm{c.completeness}. Our reasoning there
is inductive, and relies on the property proven in part (3) of \rprop{client.props}.
That property is not true for the peer preorder and traces,
so here we will reason using techniques analogous to the standard ones
of \cite{Hennessy88a}.

\begin{example}
  It is not true that if $ p \testleqSC q$ and $ q \wt{ a }$ for some
  action $a$, then $ \bigoplus ( p \after a) \testleqSC \bigoplus ( q
  \after a) $.
  An example are the peers $ p = a.\Unit$ and~$q = \Unit \extc a.\Cnil$.
  First, the inequality $ p \testleqSC q$ is true,
  because the peers $p$ and $q$ engage exactly in the same
  interactions, and the latter is trivially satisfied (i.e. $ q \ar{\ok}$).
  Second, $ \bigoplus (p \after a) = \tau.\Unit$ and $ \intsum (q
  \after a) = \tau.\Cnil$. A peer that witnesses~$ \tau.\Unit \not
  \testleqSC \tau.\Cnil$ is $\Unit$, for $ \tau.\Unit \MustSC
  \Unit$, while $ \tau.\Cnil \not \MustSC \Unit$.
\end{example}

The remaining part of the section essentially shows what properties
typical of server behaviours are enjoyed by peers.
During {\em unsuccessful} execution of traces, peers behave at
the same time as clients and servers, whereas after reporting success
they behave only as servers. The tests that we will use in the
oncoming proofs witness this intuition, for they are
a combination of the tests used to reason on the client and on the
server behaviours.

\begin{lem}
\label{lem:testleqSC-conv}
if $ p \testleqSC q~$, $ p \peerConv{s}$, and $q \wt{s} q'$, then~$q'
\Conv{}$.
\end{lem}
\proof
It is enough to show a peer $C$ such that 
$ q \MustSC C$ and $ C \uwt{ s } \hat{C}$, for some $\hat{C}$.
These facts imply that if $ q \wt{s} q'$ then $q' \Conv{}$.
This is true for otherwise there exists a maximal computation of
$q \Par C$ which is not successful, namely
$$
q \Par C \wt{} 
q' \Par \hat{C} \ar{\tau} 
q^1 \Par \hat{C} \ar{\tau}
q^2 \Par \hat{C} \ar{\tau} \ldots
$$

As $q \MustSC C$ follows from $p \MustSC C$, we define $C$ and prove
the latter fact.
Let $s = a_1 a_2 \ldots a_n$ and let $s'$ be the longest prefix
of~$s$ such that~$ p \uwt{ s' }$.
The precise definition of $C$ depends on the existence of
$s'$, so we treat the two cases separately.

\paragraph{Suppose $s'$ does not exist}
In this case $ p \ar{ \ok }$.
For every~$ 0 \leq k \leq n$, let
$$
C_k \eqdef
\begin{cases}
  ( \, \tau.\Unit \,)\extc \overline{ a }_{k+1}.C_{k+1} & \text{if }0
  \leq k < n\\
  \tau.\Unit & \text{if }k = n+1
\end{cases}
$$

The reason why $p \Must C_0$, is that $p \Conv{s}$. This follows from $p
\peerConv{s}$, and ensures that all the maximal computations of
$C_0 \Par p$ contain a stable state $C' \Par p'$. As $ C_0  \wt{
  \co{s'} } C'$ for some $s'$ is a prefix of~$s$, the definition of
the $C_i$'s ensures that $C' \ar{\ok}$.

Since~$p \ar{ \ok }$ we also know that~$ C \Must p$, and so $p
\MustSC C$. The hypothesis~$p \testleqSC q$ implies that~$ q \MustSC C$.

\leaveout{
To prove $q \wt{s} q'$ implies $q' \Conv{}$ we reason by
contradiction. We suppose that $q'$ diverges and prove $q \NMust C$,
which contradicts $q \MustSC C$.

Suppose that (\ref{eq:server-conv-aim}) is false: there exists a~$k$
such that~$p_2 \wt{ s_k } p^0_2$ and~$p^0_2$ diverges; then the following maximal computation is not \csucc, 
$$
C_0 \Par p_2 \wt{}   C_{k+1} \Par p^0_2 \ar{ \tau } 
C_{k+1} \Par p^1_2 \ar{\tau}
C_{k+1} \Par p^2_2 \ar{\tau} \ldots
$$
It follows that~$ p_2 \not\Must C_0$, which contradicts~$p_2 \Must
C_0$. In view of this contradiction, (\ref{eq:server-conv-aim}) is
true.
}

\paragraph{Suppose $s'$ exists}
In this case~$p \uwt{s'}$ for some~$s'$; let~$s' = a_1 a _2
\dots a_m$, with~$m \leq n$.
For every~$0 \leq j \leq m$ the assumption~$p \uwt{ s' }$ ensures
that~$ p \uwt{ s_j }$. \rlem{meaning-usb} ensures that for every
$0 \leq j \leq m$ there exists a~$\hat{r}_j$ such that 
\begin{equation}
\label{eq:r.must.pafter}
\hat{r}_j \Must \intsum ( p \afterut s_j )
\end{equation}
For every~$ 0 \leq k \leq n+1$ let
$$
C_k \eqdef 
\begin{cases}
  ( \, \tau.(\hat{r}_k \extc \Unit) \, ) \extc
  \overline{ a }_{k+1}.C_{k+1} &\text{if }0 \leq k \leq m\\
  ( \, \tau.\Unit \, ) \extc \overline{ a }_{k+1}.C_{k+1} &\text{if } m
  < k \leq n\\
  \tau.(\hat{r}_n \extc \Unit) & \text{if }k = n + 1,\,  m = n\\
  \tau.\Unit & \text{if }k = n + 1, \,  m < n
\end{cases}
$$

We prove that~$p \MustSC C_0$.
Fix a maximal computation of~$p \Par C_0$,
\begin{equation}
\label{eq:testleqSC-conv-comp}
  p \Par C_0 = p^0 \Par C^0_0 \ar{\tau} p^1 \Par C^1_0 \ar{\tau}
  p^2 \Par C^2_0 \ar{\tau} \ldots
\end{equation}
Intuitively, if one of the $\hat{r}_k \extc \Unit$'s appears in the
computation, then there is a state $p^k \Par C^k_0$ with $ C^k_0 = \hat{r}_i \extc \Unit $. The peer $C_0$ reaches a successful state (namely $C^k_0$ itself).
As for $p$, either $p^j \ar{\ok}$ for some $j \leq k$, or (\ref{eq:r.must.pafter}) above ensures that in the computation of $p^k \Par C^k_0$ the peer $p^k$ reaches a successful state.

If no $\hat{r}_k \extc \Unit$ appears in the computation then the
convergence of $p$, $p \Conv{s}$, ensures that~$C_0$ reaches
$\Unit$. Moreover the construction of the $C_k$'s and the $\hat{r}_k$'s imply that $p$ reaches a successful state in the computation.

\leaveout{
Fix a maximal computation of~$p \Par C_0$,
\begin{equation}
\label{eq:testleqSC-conv-comp}
  p \Par C_0 = p^0 \Par C^0_0 \ar{\tau} p^1 \Par C^1_0 \ar{\tau}
  p^2 \Par C^2_0 \ar{\tau} \ldots
\end{equation}
Note that the computation in~(\ref{eq:testleqSC-conv-comp}) may be
infinite.

Because of the construction of~$C_0$, the computation
in~(\ref{eq:testleqSC-conv-comp}) begins with moves due to a (possibly
empty) prefix of~$s$, say~$s_j$ for some~$0 \leq j \leq n$, that leads
$C_0$ to~$C^j_0$ (which is~$C_0$ itself is~$s_j$ is empty). Let us consider the longest~$s_j$ that satisfies the condition just given.
By unzipping the computation at hand we obtain the contributions of
$p~$ and~$C_0$, which begins with the ensuing prefixes,
$$
p \wt{ s_j } p', \qquad C_0 \wt{ \overline{s}_j } C^j_0
$$
Our assumption on~$s_j$ ensures that~$C^j_0$ and~$p'$ cannot interact,
because the only action to synchronise is~$a_{j+1}$, and
$s_ja_{j+1}$ cannot appear in the computation.

We explain why in the computation there is a successful derivative of
$C_0$. If in the prefix of contribution of~$C_0$ there is an internal
move, then that contribution contain a successful state.
In the opposite case, observe that the hypothesis~$p\peerConv{s}$
ensures that~$p\Conv{s}$, and so~$p' \Conv{}$. This fact,~$C^j_0
\ar{\tau} \ar{\ok}$, and the fact that the computation
in~(\ref{eq:testleqSC-conv-comp}) is maximal imply that 
the computation contains a derivative of~$C^j_0$ which is successful.

Now we discuss why~$p$ reaches a successful state.
If the contribution~$p \wt{s_j} p'$ contains a successful state, then
we have nothing more to discuss. In the opposite case,~$j \leq m$, and
$p' \in ( p \afterut s_j)$. Since~$j \leq m$,~$C^j_0 = \tau.(\hat{r}_j \extc \Unit) \extc \co{a}_{j+1}.C_{k+1}$,
and by construction~$ \hat{r}_j \Must \intsum ( p \afterut s_j)$. It
follows that~$\hat{r}_j \Must p'$. Since~$p'$ and~$C^j_0$ cannot interact,
the computation in~(\ref{eq:testleqSC-conv-comp}) after the state
$p' \Par C^j_0$ contains a maximal computation of~$p' \Par \hat{r}_j$.
Since~$\hat{r}_j \Must p'$ it follows that~$p'$ reaches a successful
state.
}
\qed
\begin{cor}
  \label{cor:testleqSC-conv}
  if $ p \testleqSC q$ and $ p \peerConv{s}$, then $q \Conv{s}$.
\end{cor}
\proof
  For every $s'$ prefix of $s$, the hypothesis imply that $p
  \peerConv{s'}$ and that $q \wt{s'} q'$, so \rlem{testleqSC-conv}
  ensures that $q' \Conv{}$.
  \rlem{conv-traces} implies that $q \Conv{s}$. 
\qed

\begin{lem}
  \label{lem:testleqSC-finite-traces}
  Let $ p \testleqSC q $. For every $s \in \Act^\star$, if $p
  \peerConv{s}$ and $q \wt{s}$, then $p \wt{ s }$.
\end{lem}
\proof
  It suffices to define a peer $C$ such that
  $p \not \Must C$, $C \uwt{ \co{ s }} \hat{C} \Nar{\tau}$, and for
  every~$s'$ proper prefix of $s$, $C \uwt{ \co{s' }} C' $
  implies $C' \ar{\tau}\ar{\ok}$.
  These three conditions and $p \Conv{s}$ ensure that $ p \wt{ s } $, 
  for otherwise all the maximal
  computations of $ p \Par C$ would be \csucc, thereby
  contradicting $ p \NMust C $.
\leaveout{
We explain this fact.
  Let~$ k < n$ be such that~$s_k$ is the longest prefix of~$s$
  performed by~$p_1$. Every maximal computation of~$C_0 \Par p_1$ must
  be finite, because~$p \Conv{s}$ and the longest trace that~$C_0$
  performs is~$\overline{s}$. Every maximal computation of~$C_0 \Par
  p_1$ must contain the contributions
  $$
  C_0 \wt{ \overline{ s}_k } C_k, \qquad p_1 \wt{ \overline{s}_k} p'_1
  $$
  for otherwise the computations can be extended with one further
  interaction due an~$\alpha$.
  Since~$k \leq i < n$, by construction we know that~$C_k = (
  \tau.\Unit ) \extc \overline{ \alpha }_{k+1}.C_{k+1}$.
  As the computation at hand is maximal and~$C_i \ar{ \tau }$, the
  computation can be extended. Since~$p'_1 \Nwt{ \alpha_{k+1}}$, 
  the only way to extend the computation is an internal move of~$C_k$.
  The move reduces the client to a successful state, namely~$ \Unit
  \extc \overline{ \alpha }_{k+1}.C_{k+1}$, thus the computation is
  \csucc.
 } 
 To prove that $p \not \Must C$, it suffices to show that 
 $ q \not \MustSC C$ and $ C \Must p$. We show the first fact.
The hypothesis imply that there exists a $q'$ such that~$q \wt{ s } q'$.
If~$q'$ diverges we infer the maximal computation
$$
C_0 \Par q \wt{} 
\hat{C} \Par q'  \ar{\tau} 
\hat{C} \Par q^1 \ar{ \tau }
\hat{C} \Par q^2 \ar{ \tau} 
\ldots
$$
If~$q'$ does not diverge, then there exists a~$q''$ such that~$q'
\wt{ \varepsilon } q'' \Nar{\tau} $, and we infer the maximal
computation 
$C_0 \Par q \wt{} 
\hat{C} \Par q'' \Nar{\tau} $.
In both cases we have shown non \csucc computation
of~$C_0 \Par q$, so we have proven that~$q \not\Must C_0 $. This
ensures that $ q \not\MustSC C_0 $.

Now we define a suitable $ C $, and prove that $C \Must p$. 
Let $n$ be the length of $s$, $s = a_1a_2
\ldots a_n$, and let $s'$ be the longest prefix of $s$ such that $ p
\uwt{ s' }$. The construction of~$C$ depends on the existence of
$s'$, and so rest of the proof is divided in two parts.

\paragraph{Suppose $s'$ does not exist}
In this case $ p \ar{ \ok }$. Let for every $ 0 \leq i \leq n + 1$,
$$
C_i \eqdef 
\begin{cases}
  ( \,  \tau.\Unit \, ) \extc \overline{ a }_{i+1}.C_{i+1} & \text{if } 0 \leq i < n\\
  \Cnil & \text{if } i = n+1
\end{cases}
$$

We prove that~$ q \not \Must C_0$. The hypothesis imply that there
exists a $q'$ such that~$q \wt{ s } q'$.
If~$q'$ diverges we infer the maximal computation
$$
C_0 \Par q \wt{} \Cnil \Par q'  \ar{\tau} \Cnil \Par q^1 \ar{ \tau }
\Cnil \Par q^2 \ar{ \tau} \ldots
$$
If~$q'$ does not diverge, then there exists a~$q''$ such that~$q'
\wt{ \varepsilon } q'' \Nar{\tau} $, and we infer the maximal
computation 
$C_0 \Par q \wt{} \Cnil \Par q'' \Nar{\tau} $.
In both cases we have shown non \csucc computation
of~$C_0 \Par q$, so we have proven that~$q \not\Must C_0 $. This
ensures that $ q \not\MustSC C_0 $.
Since $p \ar{\ok}$, it is clear that $ C \Must p $.

  \paragraph{Suppose $s'$ exists}
  Let $ s' = a_0a_1 \ldots a_m$, with $m \leq n$.
  For every $0 \leq k \leq m$, the assumption $ p \uwt{s}$ ensures
  that $ p \uwt{ s_k }$, and so \rlem{meaning-usb} and
  \rlem{exists-cool-server} implies that
  there exists a $\hat{r}_k$ such that $\hat{r}_k \, \Must \, \intsum
  ( p \afterut s_k) $, $\hat{r}_k \Nar{\tau}$ and $\hat{r}_k \Nar{\ok}$.
  For every $0 \leq i \leq n+1 $, let
  $$
  C_i \eqdef
  \begin{cases}
    ( \tau.(\hat{r}_i
    \extc \Unit) ) \extc \overline{ a}_i
    .C_{i+1} &\text{if } 0 \leq i \leq m\\
    ( \tau.\Unit ) \extc \overline{ a }_{i+1}.C_{i+1} &\text{if } m < i < n\\
    \hat{r}_n & \text{if } i = n+1,\,  m = n\\
    \Cnil & \text{if } i = n+1,\,  m < n\\
  \end{cases}
  $$

\leaveout{
  We prove that $ q \not\MustSC C_0 $. By hypothesis $\acc{q}{s}
  \neq \emptyset$, so there exists a $q'$ such that $ q \wt{s} q'$.
  If $q'$ diverges then we infer the ensuing maximal computation
  $$
  q \Par C_0 \wt{} 
  q' \Par C_n \ar{ \tau } 
  q'_1 \Par C_n \ar{ \tau } 
  q'_2 \Par C_n \ar{ \tau }
  \ldots
  $$
  The computation above is not successful, because no $C_i$ is
  successful. If $q'$ converges, then there exists a stable $q''$ such
  that we can infer the following maximal computation
  $$
  q \Par C_0  \wt{} q'' \Par C_n  \Nar{\tau} 
  $$
  The finite computation above is not successful.
  What we have argued so far proves that $ q \not\MustSC C_0$.

}

  We prove that $C_0 \Must p$. Fix a maximal computation
$$
C_0 \Par p \ar{\tau} 
C^1_0 \Par p^1 \ar{\tau} 
C^2_0 \Par p^2 \ar{\tau} 
C^3_0 \Par p^3 \ar{\tau}
\ldots 
$$

  Either one of the $\hat{r}_i \extc \Unit$ (or $\hat{r}_n$) appears in
  the computation, or none does.
  Suppose $C^k_0 = \hat{r}_i \extc \Unit$ or $C^k_0 = \hat{r}_n$ for some
  state $ C^k_0 \Par p^k$. If some $p^j$ with $j \leq k$ is 
  successful the computation is \csucc. If no $p^j$ is
  successful, then the part of the computation starting at $C^k_0 \Par p^k$
  is \csucc, for $ \hat{r}_i \Must p^k$.
  If neither an $\hat{r}_i \extc \Unit$ nor $\hat{r}_n$ appear in the
  computation then there must a state $ C^k_0 \Par p^k$ with $C^k_0 =
  C_{n+1}$, and $ C^k_0 \in \sset{ \Cnil, \Unit } $, because neither 
  $\hat{r}_n$ not $\hat{r}_i \extc \Unit$ appear.
  This and the construction of $C$ ensure that $p$ reaches a successful
  state in the computation above, for otherwise $ C^k_0 = \hat{r}_n$
  or $ C^k_0 = \hat{r}_i \extc \Unit$ for some~$i$.
\leaveout{
  Fix a maximal computation of $C_0 \Par p$ and unzip it; the
  contributions that we obtain begin with the following prefixes of
  $s$,
  $$
  p \wt{ s_j } p', \qquad 
  C_0 \wt{ \overline{s}_j } C_j
  $$
  where we let $0 \leq j \leq m$ be greatest $j$ such that a state
  $C_j$ appears in the computation.
  If the action sequence $ p \wt{ s_j } p'$ contains a successful state, then the whole computation  is \csucc.
  In the opposite case, $p \uwt{s_j} p'$, and so $j \leq m$. 
  It follows that $C_j $ contains a term $\hat{r}_j$ such that $\hat{r}_j
  \Must \intsum (p \after s_j)$; this implies that $\hat{r}_j \Must p'$.
  Our assumption on $s_j$ and $C_j$ ensure that the remaining part of the computation we unzipped contain a maximal computation of $p' \Par \hat{r}_j$; it follows that the maximal computation at hand is \csucc.
  }

\qed

\begin{lem}
\label{lem:testleqSC-acc-set-2}
If $ p \testleqSC q $ and $p \peerConv{s}$, then for every $B \in
\acc{q}{s} $ there exists a set $A$ such that $A \in \acc{ p }{ s }$
and $A \cap \uaut{p}{s} \; \subseteq \; B$.
\end{lem}
\proof
Let $s = b_1b_2 \ldots b_n$.
We reason by contradiction. Suppose that there exists a set $A \in \acc{ p }{ s }$ such that $A \cap \uaut{p}{s} \;\not \subseteq \; B$. We use this assumption to define a  peer $C$ such that $ p \MustSC C$ and that $q \not \MustSC C$, thereby proving that $  p \not \testleqSC q$.

In particular, we build a $C$ such that $C \uwt{s} C'$, where $C'$
is similar to the external sum we used to prove part (2) of \rprop{client.props}. This allows us to prove that $ q \not \MustSC C $.

Let $I$ be the index set of $\acc{p}{s}$, let $J$ be the subset of $I$
which ranges over the ready sets of $p$ after successful executions of
$s$, and let $\co{a_i}$ an action in $A_i \cap \uaut{p}{s}$.

The construction of $C$ depends on the longest $s'$ prefix of $s$ that is performed unsuccessfully by $p$.
\paragraph{Suppose $s'$ does not exist}
In this case for every $0 \leq k \leq n+1$ we let
$$
C_k \eqdef 
\begin{cases}
  ( \,  \tau.\Unit \, ) \extc \overline{ b }_{k+1}.C_{k+1} & \text{if } 0 \leq k \leq n\\
  \sum_{j \in J} \co{a_j}.\Unit & \text{if } k = n+1
\end{cases}
$$

All the maximal computations of $ p \Par C_0$ are successful.
This is true because the assumption that $s'$ does not exist implies
that $p \ar{\ok}$, and because the hypothesis $p \peerConv{s}$ ensures
that $p \Conv{s}$. In turn this let us prove that $C_0$ reaches a
successful state in the maximal computations of $p \Par C_0$.
We have proven that $p \MustSC C_0$

To obtain an unsuccessful maximal computation of $q \Par C_0$
zip together $C_0 \wt{\co{s}} C_n$ with the execution of $s$
that leads $q$ to the state $q'$ with ready set $B$.
The state $q' \Par C'$ is stable, so in the computation $C_0$ does not
report success. This shows that $ q \not \MustSC C_0$, in turn leading
to the mentioned contradiction, $ p \not \testleqSC q$.

\paragraph{Suppose $s'$ exists} Let $ s' = a_0a_1 \ldots a_m$, with $m \leq n$.
The construction of $C$ in this case is more involved then the
previous case.
For every $0 \leq k \leq m$, $ p \uwt{ s_k }$ so the hypothesis $p \peerConv{s}$ implies that
  there exists a $\hat{r}_k$ such that $\hat{r}_k \, \Must \, \intsum
  ( p \afterut s_k) $. For every $0 \leq k \leq n+1$ we let
$$
  C_k \eqdef
  \begin{cases}
    (\, \tau.(\hat{r}_k
    \extc \Unit) \,) \extc \overline{ b }_k
    .C_{k+1} &\text{if } 0 \leq i \leq m\\
    (\, \tau.\Unit \,) \extc \overline{ b }_{k+1}.C_{k+1} &\text{if } m < k < n\\
    \sum_{i \in I \setminus J} \co{a_i}. p_i \extc \sum_{j \in J} \co{a_j}.\Cnil & \text{if } k = n+1,\,  m = n\\
    \sum_{j \in J} \co{a_j}.\Cnil & \text{if } k = n+1,\,  m < n\\
  \end{cases}
$$

To prove that $p \MustSC C_0$ we show that $ C_0 \Must p $ and that $p \Must C_0$.
The reason why $C_0 \Must p$ is same we used in \rlem{testleqSC-finite-traces}.
The symmetric statement, $p \Must C_0$, follows from the convergence of
$p$, which is ensures by the hypothesis $p \peerConv{s}$, and 
the fact that the stable states reached by $C_0$ are successful,
except $C_n$. This state, though, can interact with the derivative of
$p$ at hand, and reduce to a successful state.

To prove that $q \not \MustSC C_0$ we proceed as we did in the case that $s'$ does not exist.\qed

\leaveout{
We need one more result, which deals with infinite traces.
\begin{lem}
  \label{lem:testleqSC-inf-traces}
  If $ p \testleqSC q$,  $p
  \peerConv{ u }$ and $q \wt{ u }$, then $ p \wt{ u } $.
\end{lem}
\proof
  The hypothesis $p \peerConv{ u }$ implies that $p \Conv{ u }$.
  If $p \ar{ \ok } $, then the argument is the same we used in \rthm{s.completeness}.
  If $p \Nar{\ok}$, then either $p \uwt{ u }$ or $p \Nuwt{u}$.
  In the first case $p \wt{u}$ follows immediately. 
  In the second case, thanks to $p \Nar{\ok}$ there exists the
  greatest $m \in \mathbb{N}$ such that $p \uwt{u_m}$.
  The hypothesis $p \peerConv{u}$ ensures that $p \usbut{u}$, so for
  every $0 \leq i \leq m$ there exists a process $r_i$ such that $r_i
  \Must \intsum ( p \afterut u_i)$.
  For every $k \in N$ let
  $$
  C_k \eqdef
  \begin{cases}
    \tau.( r_k \extc \Unit ) \extc \overline{ a }_k.C_{k+1}&\text{if }k \leq m\\
    \tau.\Unit \extc \overline{a}_k.C_{k+1}&\text{otherwise}
  \end{cases}
$$
The remaining part of the proof is analogous to the one of
\rthm{s.completeness}, and relies on the fact that $p
\Conv{u}$ and that $r_k \extc \Unit \Must  ( p \afterut u_k)$.
\qed
}

\begin{prop}
\label{prop:sc-in-usmpo}
  $p \testleqSC q$ implies $ p \Usmpo q$.
\end{prop}
\proof
  It is a consequence of \rcor{testleqSC-conv},
  \rlem{testleqSC-acc-set-2}, \rlem{testleqSC-finite-traces} and
  of a fourth property of $\testleqSC$ that we prove here.

  We have to show that
  $ p \testleqSC q$,  $p
  \peerConv{ u }$ and $q \wt{ u }$, then $ p \wt{ u } $.
  Fix a pair $ p \testleqSC q$ that satisfies the first three
  conditions. $p \Conv{ u }$ is true because $p \peerConv{ u }$.
  If $p \ar{ \ok } $, then the argument is the same we used in \rthm{s.completeness}.
  If $p \Nar{\ok}$, then either $p \uwt{ u }$ or $p \Nuwt{u}$.
  In the first case $p \wt{u}$ follows immediately. 
  In the second case, thanks to $p \Nar{\ok}$ there exists the
  greatest $m \in \mathbb{N}$ such that $p \uwt{u_m}$.
  The hypothesis $p \peerConv{u}$ ensures that $p \usbut{u}$, so for
  every $0 \leq i \leq m$ there exists a process $r_i$ such that $r_i
  \Must \intsum ( p \after u_i)$.
  For every $n \in N$ let
  $$
  C_k \eqdef
  \begin{cases}
    \tau.( r_k \extc \Unit ) \extc \overline{ a }_k.C_{k+1}&\text{if }i \leq m\\
    \tau.\Unit \extc \overline{a}_k.C_{k+1}&\text{otherwise}
  \end{cases}
$$
The remaining part of the proof is analogous to the one of
\rthm{s.completeness}, and relies on the fact that $p
\Conv{u}$ and that $r_k \extc \Unit \Must  ( p \afterut u_k)$.
\qed

Now the proof of completeness is straightforward.
\begin{thm}[Completeness: peers]
  $ p \testleqSC q$ implies $ p \sptmpo q$.
\end{thm}
\proof
Fix a pair $ p \testleqSC q$. We have to show that
$ p \stmpo q$ and $ p \Usmpo q$.
The first fact follows from \rprop{sc-in-c} and \rthm{c.completeness}.
The second fact is \rprop{sc-in-usmpo}.
\qed

\section{Equational characterisation}
\label{sec:axiomatisations}

\begin{figure*}[t]
\hrulefill

\begin{alignat*}{2}
&
\begin{array}{lrcl}
    (\Ename{S1a})&     \mu.\ntickx + \mu.y  &=& \mu.(\tau.\ntickx + \tau.y)  \\
    (\Ename{S1b})&     \tau.x   &\leq& \tau.\tau.x   \\
      (\Ename{S2}) &   \ntickx + \tau.y   &=& \tau.(\ntickx + y) +
      \tau.y 
    \end{array}
&\quad
\begin{array}{lrcl}
       (\Ename{S3}) &  \mu.x + \tau.(\mu.y + z)  &=&  \tau.(\mu.x +
        \mu.y + z)\\
     (\Ename{S4})&      \tau.x + \tau.y  &\leq& x \\
     (\Ename{S5}) &  \tau^{\infty} &\leq& x 
    \end{array}
\end{alignat*}
  \caption{Standard inequations}
  \label{fig:equations}
\hrulefill
\end{figure*}

We use $\CCSf$ to denote the finite sub-language of $\CCS$; this
consists of all finite terms constructed from the operators
$\Cnil,\,\Unit, \extc, \, \mu._{-}$ for each $\mu \in \Actt$, together
with the special operator ${\tau^\infty}$; \leaveout{this last denotes the term
$\tau^\infty$ from \CCS and }its inclusion enables us to consider the
algebraic properties of divergent processes. Our intention is to use
equations, or more generally inequations, to characterise the three
behavioural preorders $p \testleqArb q$ over this finite algebra,
where $\star$ ranges over $\svr,\, \clt$ and $\peer$. 
\leaveout{ 
\MHf{This is wrong as the client preorder is not preserved by parallel. 
See addition to conclusion.} 
\MHc{Standard 
equations, \cite{ccs}, can be used for the other operators, such as
parallel and hiding. }
For a given set
of inequations $E$ we will use $p \eqleq[ E] q$ denote the fact that
the inequation $p \leq q$ can be derived from $E$ using standard
equational reasoning, while $t =_E u$ means that both $t \eqleq[E] u$
and $u \eqleq[E] t$ can be derived.
}
For a given set
of inequations $E$ we will use $p \eqleq[ E] q$ to denote that
the inequation $p \leq q$ can be derived from $E$ using standard
equational reasoning, while $t =_E u$ means that both $t \eqleq[E] u$
and $u \eqleq[E] t$ can be derived.

There are two immediate obstacles. 
The first is that none of these preorders are  pre-congruences for
the language \CCSf; specifically they are not preserved by the choice
operator~$+$. 

\begin{cexample}\label{ex:extchoice}
  Using the behavioural characterisation 
it is easy to 
  check that $\Cnil \testleqSC b.\Cnil$; in fact this is trivial because 
$\Cnil \not\in \Usable[\peer]$.  However  
$a.\Unit \extc \Cnil  \nottestleqSC a.\Unit + b.\Cnil$ because
$\overline{a}.\Unit \extc \overline{b}.\Cnil  \,\MustSC\,   a.\Unit \extc \Cnil$
while 
$\overline{a}.\Unit \extc \overline{b}.\Cnil  \,\not\MustSC\,   a.\Unit \extc b.\Cnil$; the latter
follows because of the possible communication on $b$.  

The same counter-example also shows the other preorders are also not
preserved by $\extc$. 

\end{cexample}

So in order to discuss equational reasoning   we focus on the
largest \CCSf pre-congruence contained in $\testleqArb$ which we denote by 
$\testpreArb$; by definition this is preserved by all the operators. But it is convenient 
to have an alternative more amenable characterisation. To this end we let 
$p \testplusArb q$ mean that $\freshf.\Unit \extc p \testleqArb \freshf.\Unit \extc q$ for some 
fresh action  $\freshf$. 
\begin{prop}\label{lemma:preSC}
  In an arbitrary LTS,  $p \testpreArb q$ if and only if  $p \testplusArb q$. 
\end{prop}
\proof
  One direction is immediate, namely  $p \testpreArb q$ implies  $p \testplusArb q$. 
To prove the converse it is sufficient to prove that each preorder $\testplusArb$ is 
preserved by the two operators $ - \extc - $ and~$\mu. -$. The details are straightforward,
and left to the reader. 
\qed
\noindent
Note that this is similar to the characterisation of observation-congruence in Section 7.2 of
\cite{ccs}; the same technique is also used in   \cite{NH1984}. 

Proposition~\ref{lemma:preSC} gives a convenient characterisation of the 
behavioural precongruences  $p \testpreArb q$ which we will use in the sequel.
One useful property of this characterisation is the following:
\begin{lem}\label{lemma:plus}
  Suppose $p \ar{\tau} $. Then $p \testleqArb q$ implies $p \testplusArb q$.
\end{lem}
\proof
Suppose $p \ar{\tau} $. Then for any client $r$, $\freshf.\Unit \extc
p \Must r$ implies  
$p \Must r$. This is sufficient to prove that $p \testleqS q$ implies 
$\freshf.\Unit \extc p \testleqS \freshf.\Unit \extc q$. 

Minor variations on this argument will show that the result also holds for 
the client and peer preorders. 
\qed

The second obstacle to the equational characterisation of the
behavioural preorders is that they are very sensitive to the ability
of processes to immediately report success, with the result that many
of the expected equations are not in general valid. For example the
innocuous
\begin{align}\label{eq:innocious}
  a.\tau.x  &= a.x,
\end{align}
valid in the theories of \cite{ccs,NH1984}, is not in general
satisfied by two of our behavioural theories. For example $a.\Unit
\not\testplusSC a.\tau.\Unit$ because of the peer
$\overline{a}.(\Unit \extc \tau^\infty)$.

Accordingly in order to have a more elegant presentation of the inequational
theory we will use two sorts of variables, the standard $x,y,\ldots$
which may be instantiated with any process from \CCSf, and $\ntickx,
\nticky, \ldots$ which may only be instantiated by a process $p$
satisfying $p \not\ar{\ok}$; in \CCSf such processes $p$ in fact have
a simple syntactic characterisation. With this convention in mind
consider the five \emph{standard} inequations given in
Figure~\ref{fig:equations},
which are satisfied by all three behavioural orders $\testplusArb$. We
also assume the standard equations for $(\CCSf,\extc,\Cnil)$ being a
commutative monoid.  Let $\Ename{SVR}$ denote the set of inequations
obtained by adding 
\begin{eqnarray*}
\Unit &=& \Cnil   &\Ename{(SVR1)}
\end{eqnarray*}
Intuitively $\Unit$ has no significance for server behaviour; this
extra equation captures this intuition and is sufficient to
characterise the server preorder:
\begin{thm}[Soundness and completeness for server-testing]
   In $\CCSf$,  $p \testpreS q $ if and only 
$p \eqleq[ \Ename{ SVR}] q$. 
\end{thm}
\proof [Outline]
  The equation \Ename{(SVR1)} means that every term can be reduced to
  one which does not contain any occurrence of the unit $\Unit$. This
  means that all terms can now match the special variables $\ntickx,\nticky, \ldots$
and therefore the equations  \Ename{(S1)} - \Ename{(S5)} can be rewritten
with them replaced with the standard variables $x,y, \ldots$. The resulting 
inequational theory coincides with that from \cite{NH1984} which characterises
the \emph{must} testing preorder over finite terms;\footnote{This is referred to as
$\eqleq_2$ in \cite{NH1984}.} 
this we know coincides with our server preorder $\testpreS$. 
\qed

All the standard inequations in Figure~\ref{fig:equations} are also valid
for the client and peer pre-congruences. Indeed the reason for introducing 
the two sorts of variables was to ensure that they remain valid for these
new pre-congruences. 
\begin{example}[The need for two sorted inequations]
  We have already seen why \Ename{(S1a)} would no longer hold, for the
  client and peer pre-congruences, if the meta-variable $\ntickx$ were
  replaced by the standard variable $x$. The innocuous equation 
(\ref{eq:innocious}) above would be a derived equation from 
this altered version of  \Ename{(S1a)}, because of the idempotency of $\extc$. 

Let $r_1,\,r_2$ denote the clients $\Unit \extc \tau.a.\Unit$ and 
$\tau.(\Unit \extc a.\Unit) \extc \tau.a.\Unit$ respectively. 
Then $(\Unit \extc \tau^\infty) \Must r_1$ whereas  
$(\Unit \extc \tau^\infty) \not\Must r_2$, and thus 
$r_1 \nottestleqC r_2$. This shows the need for the meta-variable $\ntickx$ in 
\Ename{(S2)}, for otherwise $r_1 = r_2$ would be an instantiation. 

The same example can be used to show that this restriction is also necessary
for the peer pre-congruence. 
\end{example}

Despite the use of two sorts of variables, much of the standard equational reasoning,
for example from \cite{NH1984} remains valid. Here is a typical example,
where we use $\Ename{ST}$ to denote the inequational theory generated by the 
standard inequations. 
\begin{lem}\label{lemma.tau.tau}
In the equational  theory \Ename{ST}, 
$\tau.x \extc \tau.y =_{\Ename{ST}} \tau.(\tau.x \extc \tau.y)$
is a derived equation.  
\end{lem}
\proof

Using \Ename{(S1b)} and \Ename{(S4)}, together with the idempotency of $\extc$, 
we have the derived equation $\tau.\tau.x = \tau.x$. 
Then applying this twice we obtain:
\begin{eqnarray*}
  \tau.x \extc \tau.y  =_{\Ename{ST}}  \tau.\tau. x \extc \tau.y
          & =_{\Ename{ST}}& \tau.(\tau.\tau.x  \extc \tau.y )  & \Ename{(S1a)}\\
         & =_{\Ename{ST}}& \tau.(\tau.x  \extc \tau.y )  & 
\end{eqnarray*}

Note that this derived equation is closely related to the standard
equation \Ename{(S1a)}. It is a restriction in that the $\mu$ is only
allowed to be $\tau$, but is a generalisation in that neither of the 
meta-variables $x,\,y$ need satisfy  the predicate $\notok$.
\qed

The equation \Ename{(SVR1)} is obviously not satisfied by either the
client or the peer pre-congruence. In order to characterise them we
need to replace it with inequations which capture the significance of the
operators $\Unit$ and $\Cnil$ for client and peers respectively.
First we consider the client case.  It is easy to see that $\Unit$ is
a maximal element for the preorder~$\testleqC$, and also for the
contextual preorder $\testpreC$: $p \Must \freshf.\Unit \extc \Unit$
for every server $p$, from which $r \testpreC \Unit$ follows for every
client~$r$.  We also have that $r \extc \Unit \testeqC \Unit$ for
every client $r$; intuitively once a client can report success
immediately then it does not matter what other behaviour it has. This
client behaviour of $\Unit$ is adequately captured by the two
inequations \Ename{(CLT1a)} and \Ename{(CLT1b)} in
Figure~\ref{fig:c.and.p.equations}. 
More specifically 
the equation 
\begin{align}
  \label{eq:clUnit}
   x \extc \Unit &= \Unit
\end{align}
is easily derivable from this pair of inequations.

Another property of $\Unit$ stems from the fact that for every client $r$,
\begin{displaymath}
 r \;\testleqC\; r \extc \mu.\Unit
\end{displaymath}
for every $\mu \in \Actt$; adding the capability $\mu.\Unit$ to a client does not decrease its 
ability to satisfy servers. This property is captured by the inequation
\Ename{(CLT1c)}. In a dual manner, adding the capability $\mu.\Cnil$ to a client does not increase 
its ability to satisfy; for every client~$r$
\begin{displaymath}
   r \extc \mu.\Cnil  \;\testleqC\; r
\end{displaymath}
This is captured by \Ename{(Zb)}.

Let us now look in more detail at the zero  $\Cnil$. Since $p \Must \Cnil$ for no server
$p$ it follows that $\Cnil \testleqC r$ for every client $r$. But $\Cnil \testpreC r$
does not in general hold. For example 
$\co{\freshf}.\Cnil \extc \co{b}.\Cnil  \Must \freshf.\Unit \extc \Cnil$ but 
$\co{\freshf}.\Cnil \extc \co{b}.\Cnil \, \not\Must\,  \freshf.\Unit \extc b.\Cnil$. 
In the latter, the  synchronisation on $b$ leads to the possibility of the client not being
satisfied. It follows that $\Cnil \not\testpreC b.\Cnil$.

However $p \,\Must\, \freshf.\Unit \extc \tau.\Cnil$ for no server
$p$, with the result that $\tau.\Cnil \testpreC r$ for every client
$r$; it follows that $\tau.\Cnil$ is a minimal element in the client
theory. Recall that ${\tau^\infty}$ is also a minimal element, and therefore
to capture this property of $\Cnil$ it is sufficient to add the
inequation \Ename{(Za)}.

Note that an application of \Ename{(S1a)}, together with the idempotency of $\extc$,
gives the derived equation $\mu.\ntickx = \mu.\tau.\ntickx$; this combined with \Ename{(Za),(S5)} 
gives the useful derived inequation 
\begin{eqnarray*}
    \mu.\Cnil &\leq& \mu.x            
\end{eqnarray*} 
in the client theory.
In Figure~\ref{fig:derived.equations} this is refered to as 
\Ename{(DZ1)} and will be used  extensively
in the sequel; intuitively this means that $\Cnil$ acts like a 
minimal element  underneath a prefix.

\begin{figure*}[t]
\hrulefill

\begin{alignat*}{2}
  & \begin{array}{lrcl}
    (\Ename{Za}) &  \tau.\Cnil &\leq& {\tau^\infty} \\ 
    \\
    (\Ename{CLT1a}) &x &\leq& \Unit \\
    (\Ename{CLT1b})  &\Unit  &\leq& x + \Unit   \\
    (\Ename{CLT1c})  &\Cnil  &\leq& \mu.\Unit
  \end{array}
&\;\;\quad
& \begin{array}{lrcl}
  (\Ename{Zb}) & \mu.\Cnil &\leq& \Cnil \\ 
  \\
  (\Ename{P2P1})\qquad &\Cnil &\,\leq\,& \Unit \\
  (\Ename{P2P2})\qquad &\mu.( \Unit + x ) &\leq&  \Unit \extc \mu.x  \\
  (\Ename{P2P3})\qquad &  \mu.(\Unit + x) + \mu.(\Unit + y)  &\leq&
  \mu.(\Unit +   \tau.x + \tau.y) \\
\end{array}
\end{alignat*}
\caption{Client and peer inequations }
\label{fig:c.and.p.equations}
\hrulefill
\end{figure*}

 Let $\Ename{CLT}$ denote the  set of
inequations obtained by adding to the standard one, the client inequations we have just discussed, 
\Ename{(Za)}, \Ename{(Zb)} and \Ename{(CLT1a)} - \Ename{(CLT1c)}.
\begin{thm}[Soundness and Completeness for client-testing]
\label{thm:axioms-client}
     In $\CCSf$,  $p \testpreC q $ if and only $p \eqleq[\Ename{CLT}] q$. 
\end{thm}
\proof
To prove soundness, again it is sufficient to show that $\testplusC$ satisfies
of the inequations concerned. 
Completeness requires the development of \emph{normal forms} for clients.   
This is the topic of Section~\ref{sec:completeness},
and the result is actually proved in Theorem~\ref{thm:comp.clients}.
\qed

Both the inequations \Ename{(Za)} and \Ename{(Zb)} remain valid for the peer preorder, 
but none of the unit inequations \Ename{(CLT1a)} -  \Ename{(CLT1c)} are. 
\begin{cexample}
  First consider \Ename{(CLT1a)}. It is easy to see that 
  $\co{a}.\Unit \MustP \freshf.\Unit \extc a.\Unit$ as both peers always evolve to success states. 
  However $\co{a}.\Unit \NMust[\peer]  \freshf.\Unit \extc \Unit$, 
because the peer~$\Unit$ can not help
the partner   $\co{a}.\Unit$ achieve success.  It follows that 
$ a.\Unit \not\testplusP \Unit$. 

Moving to \Ename{(CLT1b)},  $\Unit \NMust[\peer] \tau.\Cnil \extc \Unit$ because the activation
of the internal action can preempt one of the peers achieving success. However trivially 
$\Unit \MustP \Unit$, with the result that $\Unit \not\testplusP \tau.\Cnil \extc \Unit$; this
is a counterexample to  \Ename{(CLT1b)} for the peer pre-congruence. 

For the final counter-example note that $\co{a}.\Cnil \extc
\co{\freshf}.\Unit \MustP \freshf.\Unit \extc \Cnil$ because the
co-action $\co{a}$ is never activated. However $\co{a}.\Cnil \extc
\co{\freshf}.\Unit \NMust[\peer] \freshf.\Unit \extc a.\Unit$ because
the co-action $\co{a}$ here is activated, and the activation prevents one of the peers
from achieving success.  Thus \Ename{(CLT1c)} does not hold for the
pre-congruence $\testplusP$ for any external $\mu$; a minor variation
demonstrates that it also does not hold when $\mu$ is $\tau$.  
\end{cexample}

The unit inequations \Ename{(CLT1a)} -  \Ename{(CLT1c)}
need to be replaced by  unit inequations appropriate to peers. 
There are various possibilities we could add; we justify our particular
choice by considering properties we would like of the inequational theory; 
in total we add three new inequations.

In both the server and the client theory we know that for every action 
$\mu$ and processes $p,\,q$ there is another process $r$ satisfying
\begin{align}\label{eq:test.law}
  \mu.p \extc \mu.q  &= \mu.r
\end{align}
Indeed this is one of the most important laws which delineates
behavioural theories based on testing, rather than say
\emph{bisimulation equivalence} \cite{ccs}. It is derivable in the
theory of servers, where $r$ can be taken to be $\tau.p \extc \tau.q$.
It is also derivable in the theory of clients, although the form $r$
takes depends on whether both $p,q$ can immediately report success. If
at least one of $p,q$ can not report success immediately this is an
instance of \Ename{(S1a)} in Figure~\ref{fig:equations}. If this is
not the case then \Ename{(S1a)} can not be employed. But it turns out
we can still find an $r$ which satisfies (\ref{eq:test.law}) in
the algebraic theory of clients, namely $\Unit$.

We also require (\ref{eq:test.law}) to be derivable in the algebraic
theory of peers. Again if either $p$ or $q$ can not immediately report success
then this will be an instance of \Ename{(S1a)}.  One can also check that
\begin{align*}
   \mu.(\Unit + p) + \mu.(\Unit + q)  &\,\testeqPlus[\peer]\,
             \mu.(\Unit +   \tau.p + \tau.p)
\end{align*}
for all $p, q$. In order to make these derivable  in the algebraic theory it is sufficient to add the 
inequation \Ename{(P2P3)}, given in Figure~\ref{fig:c.and.p.equations}. From 
 \Ename{(P2P3)} and \Ename{(S4)} one then obtains the derived equation
\begin{align}\label{eq:P2P3}
   \mu.(\Unit + x) + \mu.(\Unit + y)  &\,=\,
             \mu.(\Unit +   \tau.x + \tau.y)
\end{align}
Another intrinsic property of extensional behavioural theories is the ability 
to abstract from internal activity. One equation capturing this has already been 
discussed in (\ref{eq:innocious}) above. This is valid in the server theory, and enables
us to forget about the intermediate internal action $\tau$. We have also seen that it does not
hold in the client theory; nor does it hold in the peer theory. However we are still able 
to abstract from intermediate internal actions in certain circumstances. 
For example 
\begin{eqnarray*}
  \mu.\tau.\ntickx &=& \mu.\ntickx
\end{eqnarray*}
is easily derivable from \Ename{(S1a)}. Other circumstances, the presence of $\Unit$, are 
summed up by
\begin{equation}
\label{eq:unit.tau}
   \mu.(\Unit \extc \tau.x)  \testeqPlus[\peer]  \mu.(\Unit \extc x)
\end{equation}
\noindent
This is immediately  derivable in the peer theory  from (\ref{eq:P2P3}) above.

Our other two  additions are  motivated by the requirement for both peers to always report
success. So adding \leaveout{this possibility to a peer}to a process the ability to report success
can only improve its behaviour as a peer.
This
is summed up by the inequation
\begin{eqnarray*}
  p \testplusP p \extc \Unit
\end{eqnarray*}
for every peer $p$.  This is captured as a derived equation if we add the inequation 
\Ename{(P2P1)} in Figure~\ref{fig:derived.equations} to the theory. 

Success does not have to be reported simultaneously by interacting pairs of peers; in particular
the ability of a peer is not damaged by bringing forward the reporting of success.  
This motivates the use of \Ename{(P2P2)} in
Figure~\ref{fig:c.and.p.equations}. 
An interesting consequence is the derived equation:
\begin{eqnarray*}
  \Unit \extc \mu.x &=& \Unit \extc \mu.(x \extc \Unit)  & 
\end{eqnarray*}
which is refered to as \Ename{(DP1)} in
  Figure~\ref{fig:derived.equations}.

Our inequational theory for peers is taken to consist of the standard inequations from 
Figure~\ref{fig:equations}, together with \Ename{(Za)},  \Ename{(Za)} and 
\Ename{(P2P1)} - \Ename{(P2P3)}.

\begin{thm}[Soundness and Completeness for peer-testing]\label{thm:mutual.completeness}
   In $\CCSf$,  $p \testpreSC q $ if and only $p \eqleq[\Ename{P2P}] q$. 
\end{thm}
\proof
  Again to prove soundness it is sufficient to show that all of the inequations
are valid for the preorder $\testplus[\peer]$. Completeness is proved in 
Theorem~\ref{thm:comp.peers}.
\qed

\section{Completeness proofs}
\label{sec:eq.completeness}

\begin{figure*}[t]
\hrulefill

\begin{displaymath}
   \begin{array}{l@{\hskip -1em}rcl}
    (\Ename{D1})&  \sum_{1 \leq i \leq n} \tau.x_i   &=& 
           \tau. (\sum_{1 \leq i \leq n} \tau.x_i)   \\
      (\Ename{D2}) &   \ntickx + \tau.(\ntickx \extc y)   &=& \tau.(\ntickx \extc y) \\
      (\Ename{D3}) &  \mu.x + {\tau^\infty}  &=& {\tau^\infty} \\
    (\Ename{D4a}) & \tau.\ntickx \extc \tau.y &=&\tau.\ntickx \extc \tau.y 
                        \extc \tau.(\ntickx \extc y) \\
  (\Ename{D5a}) & \tau.x \extc \tau.(x \extc \nticky \extc z) 
                 &=&
      \tau.x \extc \tau.(x \extc \nticky)  \extc \tau.(x \extc \nticky \extc z) \\
\\
\hline
\\
(\Ename{DZ1})  &    \mu.\Cnil &\leq& \mu.x        \\
\\
     (\Ename{DP1})  & \Unit \extc \mu.x &=& \Unit \extc \mu.(x \extc \Unit)  \\
    (\Ename{DP2})&   \tau.(\Unit \extc \sum_{1 \leq i \leq n} \tau.x_i)  &=& 
                       \sum_{1 \leq i \leq n}\tau.( \Unit \extc  x_i) \\
    (\Ename{DP3})\qquad &\mu.x &\,\leq\,& \mu.(\Unit \extc \tau.x)  \\
\\
    (\Ename{D4b}) & \tau.(\ntickx \extc \Unit) \extc \tau.(\nticky \extc \Unit)
                 &=& \tau.(\ntickx \extc \Unit) \extc \tau.(\nticky \extc \Unit)
                        \extc \tau.(\ntickx \extc \nticky \extc \Unit) \\
%
  (\Ename{D5b}) & \tau.x \extc \tau.(x \extc (\nticky \extc \Unit) \extc z) 
                 &=&
      \tau.x \extc \tau.(x \extc (\nticky \extc \Unit))  
            \extc \tau.(x \extc (\nticky \extc \Unit)  \extc z) \\
    \end{array}
\end{displaymath}

  \caption{Some derived equations}
  \label{fig:derived.equations}
\hrulefill
\end{figure*}

In this section we use a number of derived (in)equations, gathered in 
Figure~\ref{fig:derived.equations}. 
The first collection, (\Ename{D1}) - (\Ename{D5}), are derivable from
the standard equations, while the second follow from the peer
equations; see Appendix~\ref{app:derived.laws}.
Note however that the three peer inequations 
(\Ename{P2P1}) - (\Ename{P2P3}) are easily derivable in the client 
theory, using (\ref{eq:clUnit}) above.  So the second collection is also
available for reasoning about clients.

\subsection{Normal forms}

It will be notationally convenient to consider $\Unit$ as a 
prefix term, say~$\ok.\Cnil$, thus including $\ok$ as a possible
prefix action. We also use $p \ok$ to denote that $p$ can perform 
the success action, $p \ar{\ok}$, and $p \notok$ for the converse.   

The normal forms we use are an extension of those in \cite{NH1984}; considerable 
complications arise because of the presence of the unit operator $\Unit$. 
The central idea is that of \emph{saturated} collections of sets. 
Let $\calA$ be a collection of finite subsets of $\Acttick$. It is said to be
saturated if whenever  $X,\,Y \in \calA$,
\begin{enumerate}[label=(\roman*)]
\item $X \cup Y \in \calA$

\item $Z \in \calA$ whenever $X \subseteq Z \subseteq Y$
\end{enumerate}
\begin{lem}
  For every collection $\calA$ of finite subsets of $\Acttick$ there exists
a least collection $\closure{\calA}$ containing $\calA$ which is saturated. 
\end{lem}
\proof
  Straightforward. The existence of $\closure{\calA}$ can be shown from general
principles, but we can also give a constructive definition. Let 
\begin{align*}
  \calB =& \setof{Z}{ X  \subseteq Z \subseteq \,\cup\calA, \,
         \text{for some  $X \in \calA$}  }
\end{align*}
By definition $\calA \subseteq \calB$ and one can check that $\calB$ is 
saturated, that is it satisfies (i), (ii) above. 

Now let $\calC$ be any other saturated set containing $\calB$. Since it is closed 
under set theoretic union it must contain the set $\cup \calA$. Therefore, since
it satisfies (ii) above it must also contain all sets in $\calB$;
that is $\calB \subseteq \calC$. So we can set $\closure{\calA}$ to be $\calB$. 
\qed
\begin{defi}[Peer-normal forms, \pnfs]\label{def:pnfs}\mbox{}
  \begin{enumerate}
  \item ${\tau^\infty}  \Soption{\Unit} $ is a \pnf.

  \item        $n = (\sum_{a \in A} a.n_a)  \Soption{\Unit} $, for 
       $A \subseteq  \Act$,
        is a \pnf, provided $n \ok$ implies  $n_a \ok$ 

\item 
      Let $\calA$ be a non-empty saturated set of non-empty finite subsets of $\Acttick$.
      Suppose that for each $\lambda \in \cup \calA,\  n_\lambda$
      is a \pnf. Then 
      $
       \sum_{A \in \calA} \tau.n_A  \Soption{\Unit}  
      $
      is a \pnf, where 
      $n_A$ denotes the term $\sum_{\lambda \in A} n_\lambda$,
     provided $n \ok$ implies  $n_a \ok$.
  \end{enumerate}
\end{defi}
\noindent
Here we use the notation $p \Soption{\Unit}$ to indicate that the 
presence of $+ \Unit$  is optional. Thus by (1), both ${\tau^\infty}$ and ${\tau^\infty} \extc \Unit$ 
are \pnfs; by (2) $\Cnil$ is a \pnf, as are $a.\Cnil \extc \Unit$ and $a.\Cnil$.\footnote{unfortunately so 
is $\Cnil \extc \Unit$; this will be treated as $\Unit$.}

Before showing that all finite terms can be transformed into normal forms 
we need to develop some syntactic machinery for manipulating terms. We continue
to use the notation introduced in Definition~\ref{def:pnfs}, using $n_A$,
where $A \subseteq \Acttick$, to denote the term 
$\sum_{\lambda \in A} n_\lambda$, for some (assumed) collection of terms $n_\lambda$, and 
$n_\calA$ for the term  $\sum_{A \in \calA} \tau.n_A$. 

\begin{prop}[Saturation]\label{prop:saturation}
  Let $\calB = \closure{\calA}$. Then 
  $n_\calA \eqP n_\calB$. 
\end{prop}
\proof
  This relies on two auxiliary results. Suppose $A,\, B,\,C  \subseteq \Acttick$,
where $A \subseteq C \subseteq B$.  
  Then 
  \begin{eqnarray*}
    \tau.n_A + \tau.n_B &\eqP&  \tau.n_A +  \tau.n_B +  \tau.n_{A \cup B} & (Union)\\
     \tau.n_A +  \tau.n_B &\eqP&  \tau.n_A +  \tau.n_C +  \tau.n_B        &  (Sub)
  \end{eqnarray*}
By systematically employing both equalities, from left to right, we can 
transform $n_\calA$ into~$n_\calB$. So we concentrate on proving these properties. 

First we consider (Union). Suppose $\ok \not\in A$. Then the equality follows 
from an application of the derived equation \Ename{(D4a)} in 
Figure~\ref{fig:derived.equations}. This can also be applied if $\ok \not\in B$. 
Finally if $\ok \in A \cap B$ then we can use  \Ename{(D4b)}.

The proof of (Sub) above is similar, depending on whether $\ok \in
C$. If it is
then 
\Ename{(D5b)} is employed in the derivation; if not \Ename{(D5a)} is
required. 
\qed
\noindent
The next property has already been alluded to in (\ref{eq:test.law}). 
\begin{prop}[Uniqueness of derivatives]\label{prop:uniqueness}
  For all $p,\,q \in \CCSf$ and all actions $\mu \in \Actt$,  
   there exists some term $r$ such that 
     $\mu.p \extc \mu.q  \eqP \mu.r.$
\end{prop}
\proof
  If $p \notok$, or $q \notok$ then we can apply \Ename{(S1a)} directly,
obtaining  $r = \tau.p \extc \tau.q$. Otherwise we have both 
 $p \ok$ and $q \ok$ and the required $r$ is $\Unit \extc \tau.p \extc \tau.q$.
In one direction this is an application of \Ename{(P2P3)}. The reverse follows
from two applications of $\tau.x \leq x$, which is derivable from 
\Ename{(S4)}. 
\qed

We now show how to transform all terms into \pnfs. The main work is done 
in the following two lemmas. 
\begin{lem}
  If $n_1,\, n_2$ are \pnfs then there exists a \pnf $m$ such that 
   $\tau.n_1 \extc \tau.n_2 \eqP m$.
\end{lem}
\proof
  By induction on the combined size of $n_1,\, n_2$ and an analysis of
their structure. There are many cases to consider. We omit the cases when 
either take the form ${\tau^\infty} \Soption{ \Unit}$ as the result follows from
the derived rule \Ename{(D3)}. 
\begin{enumerate}[label=\({\alph*}]
\item Suppose $n_1 = \sum_{A \in \calA} \tau.n_A $ and  
$n_2 = \sum_{B \in \calB} \tau.m_B $.  An application of 
the derived rule \Ename{(D1)} from Figure~\ref{fig:derived.equations} gives
\begin{eqnarray*}
  \tau.n_1 + \tau.n_2 &\eqP& \sum_{A \in \calA} \tau.n_A \,+\, \sum_{B \in \calB} \tau.m_B
\end{eqnarray*}
Now suppose there exists some external action $c \in A \cap B$, where $A \in \calA$ and 
$B \in \calB$ such that $n_c \not= m_c$. We isolate the subterm $\tau.n_A + \tau.m_B$
so as to unify the $c$-derivatives. This subterm has the form 
$  \tau.(p \extc c.n_c) \extc \tau.(q \extc c.m_c)$ which 
can be rewritten to 
\begin{eqnarray*}
  &\eqP& 
  c.n_c \extc \tau.(p \extc c.n_c) \extc  c.m_c \extc \tau.(q \extc c.m_c)
  &by \Ename{(D2)}\\
&\eqP& 
  \tau.(p  \extc c.n_c  \extc c.m_c) \extc    \tau.(q  \extc c.n_c  \extc c.m_c)
  &by \Ename{(S3)}
\end{eqnarray*}
Now suppose at least one of $n_c,\,m_c$ satisfies $\notok$. Then we can use \Ename{(S1a)}
to proceed thus:
\begin{eqnarray*}
  &\eqP& 
  \tau.(p  \extc c.(\tau.n_c  \extc \tau.m_c) \extc    
   \tau.(q   \extc c.(\tau.n_c  \extc \tau.m_c)
\end{eqnarray*}
By induction there exists a normal form $o_c \eqP \tau.n_c  \extc \tau.m_c$
and so we may transform the subterm to 
\begin{eqnarray*}
  &\eqP& \tau.(p \extc c.o_c) \extc \tau.(q \extc c.o_c)
\end{eqnarray*}
On the other hand if both $n_c \ok$ and $m_c \ok$, we can imitate the above sequence
of steps, this time using \Ename{(P2P3)}, or rather its derived version (\ref{eq:P2P3}) 
above, to obtain 
\begin{eqnarray*}
  &\eqP& \tau.(\Unit \extc p \extc c.(\Unit \extc o_c)) 
                 \extc \tau.(\Unit \extc p \extc c.(\Unit \extc o_c))
\end{eqnarray*}

By systematically applying the \emph{derivative unification
  transformation} we can now assume that $ \tau.n_1 + \tau.n_2 $ has
the form $\sum_{C \in \calC} \tau. s_C$, where each $s_c$ is a \pnf.
Moreover by Proposition~\ref{prop:saturation} this can be transformed
into $\sum_{D \in {\mathcal D}} \tau.s_D$ where $\mathcal D$ is
saturated; this is the required \pnf.

\item 
 Suppose $n_1 = \Unit \extc  \sum_{A \in \calA} \tau. n_A $  and  
$n_2 = \sum_{B \in \calB} \tau.m_B $.

Applying the derived equation \Ename{(DP2)} we \leaveout{can} rewrite $\tau.n_1$ to the form
$$\sum_{A \in \calA} \tau.(\Unit \extc n_A)$$ which by 
\Ename{(D1)} can be transformed into  $\tau.\sum_{A \in \calA} \tau.(\Unit \extc n_A)$.
We \leaveout{can} now proceed as in the previous case. The same holds if $m$ has $\Unit$ as a summand. 

\item 
Suppose $n_1 = \sum_{a \in A}  a.n_a \Soption{\Unit}$ and $n_2$ is as in the previous case. 
Then by \Ename{(D1)} $$\tau.n_1 \eqP \tau. (\tau. (\sum_{A \in \sset{A}} n_A   \Soption{\Unit})$$ 
and we 
proceed as in case (a).  

Finally if $n_2$ contains $\Unit$ as a summand we \leaveout{can} proceed in much the same way, 
but using case (b). 
\end{enumerate}
\qed

\begin{lem}\label{lemma:plus.pnf}
  If $n_1,\, n_2$ are \pnfs then there exists a \pnf $m$ such that 
   $n_1 \extc n_2 \eqP m$.
\end{lem}
\proof
  Again the proof proceeds by induction on the combined sizes of $n_1,\, n_2$ and 
a case analysis of their form. 
\begin{enumerate}[label=\({\alph*}]
\item 
Suppose $n_1 = \sum_{a \in A} a.n_a$ and $n_2 = \sum_{B \in \calB } \tau.m_B$; this is the central
case. 

We know that $\calB$ is not empty. So using \Ename{(D1)},\Ename{(S2)} we have
$n_1 \extc n_2 \eqP \tau.(n_1 + m_{B1}) \extc \tau.n_2$, for some 
$B1 \in \calB$.  By induction 
$n_1 \extc m_{B1}$ has a \pnf. The required result now follows from the previous lemma. 

\item 
Suppose $n_1 = \Unit \extc n'_1$, where $n'_1 = \sum_{a \in A} a.n_a$ and $n_2$ is as in
the previous case. Then we can construct, as in case (a), a \pnf for 
$n'_1 \extc n_2$, which takes the form $\sum_{D \in \calD} \tau.o_D$. 
Then the required \pnf is 
\begin{align*}
  \Unit \extc \sum_{D \in \calD} \tau. (\Unit \extc \sum_{d \in D} d.(\Unit \extc o_d))
\end{align*}
This requires the repeated application of the derived rule \Ename{(DP1)}. 

The case where $n_2$ has $\Unit$ as an additional summand is handled in a similar 
manner. 

\item 
Suppose $n_1 = \sum_{A \in \calA} a.n_A$ and $n_2 = \sum_{B \in \calB} \tau.m_B$. 
Using \Ename{(D1)} we obtain $n_1 \extc n_2 =  \tau.n_1 \extc \tau.n_2$ and the result 
now follows by the previous lemma. 

If either $n_1$ or $n_2$, or both, have $\Unit$ as an additional summand we can proceed 
in the same manner. We may then have to apply \Ename{(DP1)} to ensure that 
the resulting \pnf~$\Unit \extc \sum_{D \in \calD} o_d$ is such that 
$o_d \ok$ for every $d \in \cup \calD$. 

\item Suppose $n_1 =  \sum_{a \in A} a.n_a \Soption{\Unit}$  and 
$n_2 = \sum_{b \in B} b.m_b \Soption{\Unit}$. Then using 
Proposition~\ref{prop:uniqueness} and induction we can construct 
a \pnf of the form $\sum_{d \in A \cup B} d.o_d \Soption{\Unit}$.

\item 
The final possibility, when either $n_1$ or $n_2$ is ${\tau^\infty}$ is straightforward, 
using the derived equation \Ename{(D3)}. 
\end{enumerate}
\qed

\begin{thm}[Peer normal forms]\label{thm:pnf}
  For every $p \in \CCSf$ there exists a \pnf $n$ such that 
  $p \eqP n$.
\end{thm}
\proof
  By structural induction on $p$. The main case is covered by 
Lemma~\ref{lemma:plus.pnf}. 
\qed

One consequence of the completeness theorem will be that 
$p \eqleqP  \Cnil$, whenever~$p \not\in \Usable[\peer]$, because 
for such $p$ we know $p \testplusP \Cnil$.\footnote{
In general $\Cnil \testplusP p$ is not true, even if $p \not\in \Usable[\peer]$.} 
However it is useful to already have this result when proving 
completeness. A direct proof of this fact is not obvious. 
For example consider 
$p = a.(b.\Cnil \extc c.\Unit)  \;\extc  a.(b.\Unit \extc c.\Cnil)$ which we
know not to be in $\Usable[\peer]$. The derivation of 
$p \eqleqP \Cnil$ is  not straightforward. But it becomes so 
 if we first convert $p$ to a \pnf. This turns out to 
be
\begin{align*}
   n_p &= a.\sum_{A \in \calA} n_A \qquad
\text{where}\  \calA = \sset{ \sset{b},\, \sset{c},\, \sset{b,c} }
\ \text{and}\ n_b = n_c = \tau.\Unit \extc \tau.\Cnil 
\end{align*}
Now \Ename{(S4)} gives $n_b = n_c \eqleqP \Cnil$ 
and $n_p \eqleq \Cnil $ then follows by  applications of the 
rule \Ename{(Zb)}. 
This technique is quite general, and powerful,
and is the basis of the proof of the following lemma.
\begin{lem}\label{lemma:notusable}
  Suppose $p \not\in \Usable[\peer]$.
   Then
   \begin{enumerate}
   \item 
   $p \testplusP q$ implies  $p \eqleqP q$ 
   \item 
   $p \eqleqP  \Cnil$. 
   \end{enumerate}

\end{lem}
\proof
Part (2) is an immediate consequence of part (1) and the observation
that   $p \not\in \Usable[\peer]$ implies $p \testplusP \Cnil$. 
So we concentrate on the part (1), and 
we may assume that $p$ is a \pnf. We now proceed by induction on
its size and a case analysis of its form. However we know that 
$p \notok$ because otherwise we would have $\Unit \MustP p$; this eliminates 
many of  possible forms. Also if $p$ is ${\tau^\infty}$ then the result is immediate,
since ${\tau^\infty}$ is a least element. So in effect we are left with two possibilities.
\begin{enumerate}[label=\({\alph*}]
\item Suppose $p$ has the form $\sum_{a \in A} a.n_a$ for some $A
  \subseteq \Act$.

If $n_a \in \Usable[\peer]$ for some $a \in A$ then it would follow that 
$p \in \Usable[\peer]$
for $p  \MustP \co{a}.q_a$ for any $q_a$ satisfying $n_a \MustP q_a$. So by induction we
have $n_a \eqleqP \Cnil$ for every $a \in A$. 

Because $p \testplusP q$ we also know that $q$ has essentially only one possible 
form, namely $m_B \Soption{\Unit}$ for some $B \subseteq \Act$. 
Moreover since $\freshf.\Unit \extc p \MustP \co{\freshf}.\Unit  \extc \co{c}.\Cnil$
for any $c \in \Act \backslash A$ we have that $B \subseteq A$. 
We can now reason as follows:
\begin{eqnarray*}
  \sum_{a \in A} a.n_a &\eqleqP& \sum_{a \in A} a.\Cnil   &Induction \\
                      &\eqleqP& \sum_{b \in B} b.\Cnil     & \Ename{(Zb)}\\
                      &\eqleqP& \sum_{b \in B} b.m_b     & \Ename{(DZ1)}\\
\end{eqnarray*}
Finally suppose $q$ has the summand $\Unit$. 
If $A$ is empty we can use \Ename{(P2P1)}. 
Otherwise  the extra summand $\Unit$ can be added
by an initial application of  the derived equation \Ename{(DP3)}.

\item 
Suppose $p$ has the form $\sum_{A \in \calA} \tau.n_A$ for some saturated set $\calA$. 

Now suppose the empty set is in $\calA$, that is $\tau.\Cnil$ is a summand. Then using 
\Ename{(S4)}  we obtain 
$p \eqleqP \tau.\Cnil$ and the required result now follows since $\tau.\Cnil$ is a 
least element. 

So we can assume $\emptyset \not\in \calA$. 
As a preliminary argument suppose that for all $A \in \calA$, either $\Unit \in A$ 
or there exists some $a_A \in A$ such that $n_{a_A} \in \Usable[\peer]$. Then let $p'$
denote the peer 
$$
\Unit \extc \sum_{A \in \calA, \Unit \not\in A} \co{a}.p'_{a_A}
$$
where $p'_{a_A}$ is chosen so that $p'_{a_A} \MustP n_{a_A}$. 
Then one can check that $p'\MustP p$, contradicting the fact that
$p \not\in \Usable[\peer]$. 

So we can assume that there is some $A_0 \in \calA$ such that  $\Unit \not\in A_0$ and 
$n_a \not\in \Usable[\peer]$ for every $a \in A_0$. By induction $n_a  \eqleqP \Cnil$ and
by \Ename{(Zb)}, $a.n_a \eqleqP \Cnil$ for every $a \in A_0$. As a result $n_{A_0} \eqleqP \Cnil$.
Now an application of \Ename{(S4)} allows us to conclude  $p \eqleqP \tau.\Cnil$,
from which again the required result follows. 
\qed
\end{enumerate}

Client normal forms are simplifications  of  their peer counterparts.
We have already remarked that the three extra peer inequations \Ename{(P2P1)} -
\Ename{(P2P3)} are derivable in the client theory, and so we will
obtain the client normal forms by using client inequations to simplify
peer normal forms.

\begin{defi}[Client normal forms, \cnfs]\label{def:cnf}\hfill
  \begin{enumerate} 
  \item Both ${\tau^\infty},\,\Unit$ and $\tau.\Unit$ are \cnfs. 

  \item For any $A \subseteq \Act$ the sum $\sum_{a \in A} a.n_a$ is 
        a \cnf, provided  each $n_a$ is a \cnf.
  \item   Let $\calA$ be a  non-empty saturated set of non-empty subsets of $\Act$.
      Suppose that for each $a \in \cup \calA,\  n_a$
      is a \cnf. Then 
      $
       n =   ( \sum_{A \in \calA} \tau.n_A ) \Soption{\tau. \Unit}  
      $
      is a \cnf. 
  \end{enumerate}
\end{defi}

\begin{thm}[Client normal forms]\label{thm:cnf}
   For every $p \in \CCSf$ there exists a \cnf $n$ such that 
  $p \eqC n$.
\end{thm}
\proof
  First recall that the usability sets $\Usable[\clt]$ and
  $\Usable[\peer]$ are identical.  Using Theorem~\ref{thm:pnf}, we can
  assume that $p$ can be transformed into a \pnf $m$, that is $p \eqC
  m$. Then by induction, and a systematic application of 
the derived unit equation $x + \Unit = \Unit$,
discussed in (\ref{eq:clUnit}) above,
$m$ can then be transformed
  into a \cnf.  For example if $m$ has the form~$\Unit + m'$ then the
  resulting \cnf is $\Unit$.  Suppose it has the form $( \sum_{A \in
    \calA} \tau.m_A )$, where $\calA$ is a saturated set of non-empty
  subsets of $\Actt$.  Let $\calB = \setof{A \in \calA}{\Unit \not \in
    A}$; $\calB$ is still saturated. If it is empty the required \cnf is 
   $\tau.\Unit$. Otherwise it is
\begin{align*}
  \sum_{B \in \calB} \tau.m'_B    \Soption{ \tau.\Unit \ \mbox{if $\Unit \in \cup \calA$} }
\end{align*}
where for each $b \in \cup \calB$, $m'_b$ is the \cnf obtained from $m_b$ by induction. Here again 
the derived equation  $x + \Unit  = \Unit$ is used to transform $\tau.m_A$ into $\tau.\Unit$
for any $A$ containing $\ok$. 
\qed

\subsection{Completeness for clients}
\label{sec:completeness}

We first tackle the more straightforward case, the client preorder. 
For convenience we isolate a particularly significant case in the following 
lemma.

\begin{lem}[Stable state]\label{lemma:tau.to.stable}
  Suppose $n =   ( \sum_{A \in \calA} \tau.n_A ) \Soption{\tau. \Unit} $
          and $m_B$ are both \cnfs such that $n \testleqC m_B$, and $n \in \Usable[\clt]$. 
Let 
$N = \setof{a \in \cup\calA}{n_a \in \Usable[\clt]}$
and  
$B_0 = \setof{b \in B}{m_b \,\text{is different from}\ \Unit}$. 
Then
\begin{enumerate}
\item there exists some  $A \in \calA$ such that $B_0 \subseteq A$ and $A \cap N \subseteq B$ 
\item $\tau.n_b \testleqC m_b$, for every $b \in B \cap  ( \,
    \cup \calA \,)$. 
\item $n_b \ok$ implies $m_b \ok$ for every  $b \in B \cap  (
    \, \cup \calA \, )$. 
\end{enumerate}
\end{lem}
\proof

 \begin{enumerate}
 \item 
Since $n \in \Usable[\clt]$ we know that there exists some server $p_n$
such that $p_n \Must n$. Now suppose  there is  some $b \in 
B_0 \backslash ( \cup\calA)$. Then  $p_n \extc \co{b}.\tau^\infty   \Must n$,
from which 
it follows $p_n \extc \co{b}.\tau^\infty   \Must m$.
 But $m_b$ is a
\cnf which is different from $\Unit$. By examining the other possibilities for 
$m_b$ we see that $\tau^\infty \Must m_b$ is not possible, which contradicts
 $p_n \extc \co{b}.\tau^\infty   \Must m$.   So we can conclude that 
$B_0 \subseteq \cup \calA$. 

Now suppose, for another  contradiction, that for every $A \in \calA$ there exists some
$a_A \in (A \cap N) \backslash B $. Let $p$ denote the server 
$\sum_{A \in \calA} a_A.p_a$, where the servers $p_a$ are chosen so that 
$p_a \Must n_a$. Then because $\calA$ is not empty $p \Must n$.
This  would imply $p \Must m_B$, which is
clearly not possible. What this means is that there is some $A_1 \in \calA$ such 
that $(A_1 \cap N) \subseteq B$. Let $A = A_1 \cup B_0$. Then since 
$B_0 \subseteq \cup \calA$ and $\calA$ is saturated we know $A \in \calA$,
and by construction it has the required properties.

\item 
Suppose $p \Must \tau.n_b$, where $b \in B$; 
we have to show that $p \Must m_b$ follows. Let~$p_n$ be the server 
used in part (1); it satisfies $p_n \Must n$. 
Then one can show that $p_n \extc \co{b}.p \Must
n$, from which $p_n \extc \co{b}.p \Must m$ follows. 
But this is only possible if
$p \Must m_b$.

\item 

Let $b \in B \cap \cup \calA$ be such that $n_b \ok$. Then 
$p_n + b.\tau^\infty \Must n$ from which $p_n + b.\tau^\infty \Must m_B$
follows. But this will only be possible if $m_b \ok$. 
\qed

 \end{enumerate}

\begin{thm}[Completeness: clients]\label{thm:comp.clients}
  In \CCSf, $p \testplusC q$ implies $p \eqleqC q$. 
\end{thm}
\proof
  Let $n,\ m$ be the \cnfs for $p,\,q$ respectively; we know  that 
$n \testplusC m$. The proof proceeds by induction on the combined size
of $n,\ m$, and an exhaustive analysis of their possible structure,
dictated by Definition~\ref{def:cnf}. Note that because of 
 Lemma~\ref{lemma:notusable} we can assume $n \in \Usable[\clt]$. 
\begin{enumerate}[label=\({\alph*}]

\item If $n$ is ${\tau^\infty}$ the result is obvious, since ${\tau^\infty}$ is a least
element in the client equational theory. If it is $\Unit$, the argument is
also straightforward. We have $\Unit \extc \tau^\infty \Must m$, since
this server is guaranteed by $n = \Unit$. But this is only possible if $m \ok$;
looking at the possible forms of \cnfs in Definition~\ref{def:cnf} we see that
$m$ also has to be $\Unit$. 

A similar argument, using the server $\Cnil$, gives the result when $n$ is 
$\tau.\Cnil$. 

\item Now suppose $n$ has the form $n_A$. Let us first look at the
  possible forms for the \cnf $m$.  Because $\freshf.\Unit \extc n_A
  \testleqC \freshf.\Unit \extc m$ where $\freshf$ is fresh, $m$ can not perform a
  $\tau$ action. So the only remaining possibility is that $m = m_B$
  for some set of actions $B$.

Now suppose that there exists some $b \in B \backslash A$. Then since
$\co{\freshf}.\Cnil \extc \co{b}.\tau^\infty  \Must \freshf.\Unit + n_A$ 
we must have that $\tau^\infty \Must m_b$, for this is the only way
to ensure that 
$\co{\freshf}.\Cnil \extc \co{b}.\tau^\infty  \Must \freshf.\Unit + m_B$.
But $m_b$ is a \cnf and so it must be precisely $\Unit$. 

If $A$ is the empty set then
  the result now is immediate, since then $n = \Cnil$, and we can 
apply \Ename{(CLT1c)} repeatedly to obtain $\Cnil \eqleqC m_B$.

At this stage we can use information available from
Lemma~\ref{lemma:tau.to.stable} because 
$$\sum_{A \in \sset{A}} \tau.n_A   \eqleqC  n_A \testplusC m_B $$
Part (1) gives that $ B_0 \subseteq A$, and $N \subseteq B$
where $B_0$ and $N$ are as    defined 
 in the statement of the lemma. So from part (2) 
we have that $\tau.n_a \testleqC m_a$ for every $a \in A \cap B$. 
From Lemma~\ref{lemma:plus} this gives $\tau.n_a \testplusC m_a$; now 
using induction, which recall is on the combined size of the terms, 
we can assume $\tau.n_a \eqleqC m_a$, and therefore $a.\tau.n_a \eqleqC a.m_a$. 
If~$n_a \notok$ an application of the standard equation \Ename{(S1a)}, and the idempotence
of $\extc$ we obtain $a.n_a \eqleqC a.m_a$.
On the other hand if $n_a \ok$ then from part (3) of Lemma~\ref{lemma:tau.to.stable}
we also have $m_a \ok$. But both are \cnfs and therefore both must coincide with 
$\Unit$. So 
for every $a \in A \cap B$ we have established  $a.n_a \eqleqC a.m_a$.

The argument is now completed as follows:
\begin{eqnarray*}
  n_A  &=& \sum_{a \in  N} a.n_a  \;\extc  \sum_{a \in A \backslash N} a.n_a&\\
       &\eqleqC& \sum_{a \in N} a.m_a   \;\extc  \sum_{a \in A \backslash N} a.n_a &as argued above\\
       &\eqleqC& \sum_{a \in N} a.m_a   \;\extc  \sum_{a \in  (A\backslash N) \cap B } a.n_a 
                \;\extc  \sum_{a \in (A \backslash N) \backslash B)}  a.n_a &\\
       &\eqleqC& \sum_{a \in  N} a.m_a  \;\extc  \sum_{a \in (A\backslash N) \cap B} a.n_a   
                                      & Lemma~\ref{lemma:notusable}, \Ename{(Zb)} \\  
       &\eqleqC& \sum_{a \in  N} a.m_a  \;\extc  \sum_{a \in (A\backslash N) \cap B} a.m_a   
                                      & Lemma~\ref{lemma:notusable}, \Ename{(DZ1)} \\   
       &\eqleqC& \sum_{a \in  N} a.m_a  \;\extc  \sum_{a \in w} a.m_a   
                      \;\extc  \sum_{b \in B\backslash A} b.\Unit & \Ename{(CLT1c)} \\      
       &=& \sum_{b \in B} b.m_b   
\end{eqnarray*}
The last line follows because 
\begin{itemize}
\item $B$ can be decomposed into the three disjoint sets 
$N,\, (A\backslash N) \cap B$ and $B \backslash A$

\item 
if $b \in B \backslash A$ then $m_b$ is $\Unit$; this follows because $B_0 \subseteq A$. 
\end{itemize}

\item
There is one remaining case for the structure of $n$, namely 
$ ( \sum_{A \in \calA} \tau.n_A ) \Soption{\tau. \Unit} $. 
Here again we have to look at the possible structure of $m$. 
There are only two interesting cases. 

The first is when $m$ has the form $m_B$ for some set $B \subseteq
\Act$. This case fits the statement of  Lemma~\ref{lemma:tau.to.stable}
precisely. There must be some $A \in \calA$ such that $B_0 \subseteq A$
and $A \cap N \subseteq B$, where again $B_0$ and $N$ are as defined in the
lemma. 

Now using the fact that
\begin{eqnarray*}
  A  &=& (A \cap N)  \ \cupdot\ (A\backslash N) \cap B  \ \cupdot\  (A\backslash N) \backslash B
\end{eqnarray*}
we can proceed as in case (b) to show
\begin{align}\label{eq:a}
  n_A &\eqleqC \sum_{a \in A \cap N} a.m_a  \;\extc  \sum_{a \in (A\backslash N) \cap B} a.m_a   
\end{align}
The set $B$ can also be decomposed as
\begin{eqnarray*}
  B  &=& (A \cap N)  \ \cupdot\ (A\backslash N) \cap B  \ \cupdot\  B \cap (N\backslash A) 
\end{eqnarray*}
Moreover since $B_0 \subseteq A$ for every $b \in  B \cap (N\backslash A) $ the residual $m_b$ must
be $\Unit$. Therefore using applications of \Ename{(CLT1c)}  to (\ref{eq:a}) we can obtain 
$n_A \eqleqC n_B$. The required result, $n \eqleqC m_B$  now follows by \Ename{(S4)}. 

The other interesting case is when $m$ has either the form $ \sum_{B \in
  \calB} \tau.m_B $ or the form
 $ ( \sum_{B \in  \calB} \tau.m_B ) + \tau. \Unit $.\leaveout{
The other interesting case is when $m$ has the form $ ( \sum_{B \in
  \calB} \tau.m_B ) \Soption{\tau. \Unit} $.} Here $m \testplusC m_B$ for
every $B \in \calB$, from which $n \testplusC m_B$ follows.  Again we
can proceed as in (b) to show $n \eqleqC m_B$.

To complete we use the fact that 
$n \eqC \sum_{B \in \calB} \tau.n$. 
This follows from the derived 
law \Ename{(D1)} and the  idempotency of $\extc$.  
\end{enumerate}
\qed

\subsection{Completeness for peers}

The completeness result for peers follows the same structure as that
for clients. But it is complicated by the more intricate form of
\pnfs; in particular \pnfs of the form $\Unit \extc n$, where $n$ is
non-trivial. 
We need a generalisation of Lemma~\ref{lemma:tau.to.stable} for peers, which in turn 
requires a preliminary result.

\begin{lem}\label{lemma:usable.nottick}
  Suppose $p \in \Usable[\peer]$ and $p \notok$. 
Then there exists some $q$ such that $q \notok$ and $q \MustP p$.
\end{lem}
\proof
  We may assume that $p$ is a \pnf. If it has the form 
$\sum_{a \in A} a.n_a$ there must exist some $a \in A$ such that $n_a \in \Usable[\peer]$.
From this we get some $q_a$ satisfying $q_a \MustP n_a$, and the required 
$q$ is $\co{a}.q_a$. 

Otherwise $p$ must have the form $\sum_{A \in \calA} \tau.n_A$. 
From the analysis carried out in the proof of Lemma~\ref{lemma:notusable} 
we know that for every $A \in \calA$ either $\Unit \in A$ or there exists some 
$a_A \in A$ such that $n_a \in \Usable[\peer]$. The required $q$ is then 
$\tau.\Unit + \sum_{A \in \calA, \Unit \notin A} \co{a}.q_{a_A}$, where the peer 
$q_{a_A}$ is chosen so that $n_a \MustP q_{a_A}$.

\qed

\begin{lem}[Stable state]\label{lemma:tau.to.stable.peers}
  Suppose $n =   ( \sum_{A \in \calA} \tau.n_A ) \Soption{ \Unit} $
          and $m_B$, where $B \subseteq \Acttick$, 
are both \pnfs such that $n \testleqP m_B$ and 
 $n \in \Usable[\clt]$. 
Let 
$N = \setof{a \in \cup\calA}{n_a \in \Usable[\peer]}$
and 
$B_0 = B \backslash{\ok}$. 
Then
\begin{enumerate}
\item there exists some  $A \in \calA$ such that $B_0 \subseteq A$ and 
$A \cap N \subseteq B_0$ 


\item $\tau.n_b \testleqP m_b$, for all $b \in B \;\cap\;  \cup \calA $

\item 
 $n_b \ok$ implies $m_b \ok$, for all $b \in B \;\cap\;  \cup \calA $.
\end{enumerate}
\end{lem}
\proof
 Let $p_n$ be any peer satisfying $p_n \MustP n$. We  know at least 
one exists and because of the previous lemma we may assume that $p_n \notok$. 
\begin{enumerate}
\item  This is similar to the proof of part (1) of
Lemma~\ref{lemma:tau.to.stable} although
here we are dealing with peers rather than clients. 
Suppose there is some $b \in B$ such that $b \not\in \cup\calA$.
Then  $p_n \extc \co{b}.\Cnil  \Must n$. This contradicts
the fact that $n \testleqP m_B$ since $m_B$ can not guarantee the success
of the peer  $p_n \extc \co{b}.\Cnil$.  So we have established
$B_0 \subseteq \cup \calA$. 

We can continue as in part (1) of Lemma~\ref{lemma:tau.to.stable} to show that
there exists some $A_1 \in \calA$ such that $A_1 \cap N \subseteq B_0$. The required $A$ 
can now be taken to be $A_1 \cup B_0$.


\item 
Suppose $p \MustP \tau.n_b$. This means that 
$p_n \extc \co{b}.p  \MustP n$, from which 
$p_n \extc \co{b}.p \MustP m_B$ follows. By construction $p_n \notok$ 
and so if $\ok \not \in B$ this implies that $p \MustP m_b$. On the other hand
if  $\ok \in B$ we can only deduce that $m_b \Must  p$. But by the construction
of \pnfs, if $\ok \in B$ then we also know that $m_b \ok$. The required
$p \MustP m_b$ now follows. 

\item
For an arbitrary  $b \in B \;\cap\;  \cup \calA $ suppose  
 $n_b \ok$. Then 
$p_n \extc \co{b}.(\Unit \extc \tau^\infty)  \MustP n$, and so this must also 
be true of $m_B$. But, since $p_n \notok$, this is only possible if 
$m_b \ok$. 
\qed
\end{enumerate}

\noindent Before embarking on the main proof of completeness 
it is convenient to isolate one particular case. 
\begin{lem}\label{lemma:unitcase}
  $n \testplusP \Unit$ implies $n \eqleqP \Unit$.
\end{lem}

\proof

We may assume that $n$ is a \pnf, and we use a case analysis on its structure. 
When it has the form ${\tau^\infty} \Soption{\Unit}$ the result is obvious. 
\begin{enumerate}
\item 
So consider the case when it has the form
$\sum_{a \in A} a.n_a \Soption{\Unit}$
for some $A \subseteq \Act$. 
Now suppose there is some $a \in A$ such that $n_a \in \Usable[\peer]$; so there is a peer
$p_a$ such that $p_a \MustP n_a$. This means that $\co{a}.p_a \MustP n$. But this would
imply that  $\co{a}.p_a \MustP \Unit$, which is impossible since 
$\Unit \NMust \co{a}.p_a$. 
So what we have shown is that $n_a \not\in \Usable[\peer]$ for every $a$ in $\Act$ and therefore
$\sum_{a \in A} n_a \not\in \Usable[\peer]$. The result now follows from Lemma~\ref{lemma:notusable}. 

\item
The only other possibility is that it has the form 
$\sum_{A \in \calA} \tau.n_A \Soption{\Unit}$. 
For a contradiction suppose that for all $A \in \calA$ there exists
some $a_A \in A$ such that $p_A \MustP n_{a_A}$. This means that 
$n \MustP p$, where $p$ is the peer 
$\sum_{A \in \calA} \co{a_A}.p_a$. 
But this contradicts the fact that $n \testleqP \Unit$ since 
$\Unit \NMust p$; the peer $\Unit$ cannot induce $p$ into a successful state. 

So we have established that there is some $A \in \calA$ such that 
$n_a \not\in \Usable[\peer]$ for every $a \in A$. Using 
 Lemma~\ref{lemma:notusable} and \Ename{(DZ1)} we can derive
$n_A \eqleqP \Cnil \Soption{\Unit}$, from which the result follows,
since $n \eqleqP \tau.n_A$. \qed
\end{enumerate}

\begin{thm}[Completeness: peers]\label{thm:comp.peers}
  In \CCSf, $p \testplusP q$ implies $p \eqleqP q$. 
\end{thm}
\proof
The proof follows the same structure as that of Theorem~\ref{thm:comp.clients},
but there are more details to be considered. Here 
let $n,\ m$ be the \pnfs for $p,\,q$ respectively; 
the  proof proceeds by induction on the combined size
of $n,\ m$, and an analysis of their possible structure,
as given in  Definition~\ref{def:pnfs}. Because of
Lemma~\ref{lemma:notusable} we may also assume
that $n \in \Usable[\peer]$. We also leave the uninteresting case
when it has the form  ${\tau^\infty} \Soption{\Unit}$ to the reader. 

\begin{enumerate}[label=\({\alph*}]
\item $n =  ( \sum_{A \in \calA} \tau.n_A ) \Soption{\Unit} $ and 
$m = m_B$ for some $B \subseteq \Acttick$.  This is precisely the case to
which Lemma~\ref{lemma:tau.to.stable.peers} applies. Let $A \in \calA, N$
and $B_0$ be as given in the statement of the lemma; because of  Lemma~\ref{lemma:unitcase}
we can assume that $B_0$ is not empty. Let~$A_0$ be~$A \backslash\sset{\ok}$. 
Our aim is to show
\begin{align}\label{eq:todo}
  n_{A_0} \eqleqP m_{B_0}
\end{align}
from which the required result will follow. This is a consequence of the
following:
\begin{itemize}
\item if $\Unit$ is a summand of $n$ then 
it must also be a summand of $m_B$; to see this consider the 
peer $\Unit \extc \tau^\infty$. 
\item 
$m_{B_0} \eqleqP m_B$; for  if $\ok \in B$ then  by
condition (2)(ii) of Definition~\ref{def:pnfs} $m_b$ must be of the form
$\Unit \extc m'_b$ for every $b \in B_0$, and because  $B_0$ is not empty we can apply
an instance of \Ename{(P2P2)} to one summand $b.m_b$ of $m_{B_0}$, to 
obtain $m_{B_0} \eqleqP m_B$.

\end{itemize}
So let us concentrate on establishing (\ref{eq:todo}). This relies on the following
set decompositions:
\begin{eqnarray*}
  A_0  &=& (A_0 \cap N)  \ \cupdot\ 
                (A_0\backslash N) \cap B_0  \ \cupdot\  (A_0\backslash N) \backslash B_0\\
  B_0  &=& (A_0 \cap N)  \ \cupdot\ (A_0 \backslash N) \cap B_0  
\end{eqnarray*}
The argument now proceeds in much the same way as in the corresponding case, (b), of
Theorem~\ref{thm:comp.clients}:
\begin{eqnarray*}
  n_{A_0}  &=& \sum_{a \in (A_0 \cap N)} a.n_a  \;\extc  \sum_{a \in (A_0 \backslash N) \cap B_0} a.n_a
              \;\extc  \sum_{a \in A_0 \backslash N \backslash B_0} a.n_a    &\\
       &\eqleqP& \sum_{a \in (A_0 \cap N)} a.m_a  \;\extc  \sum_{a \in (A_0 \backslash N) \cap B_0} a.n_a
              \;\extc  \sum_{a \in A_0 \backslash N \backslash B_0} a.n_a    &   ($\star$)\\
       &\eqleqP& \sum_{a \in (A_0 \cap N)} a.m_a  \;\extc  \sum_{a \in (A_0 \backslash N) \cap B_0} a.n_a &   
                          Lemma~\ref{lemma:notusable}, \Ename{(Zb)}\\
       &\eqleqP& \sum_{a \in (A_0 \cap N)} a.m_a  \;\extc  \sum_{a \in (A_0 \backslash N) \cap B_0} a.m_a &   
                          Lemma~\ref{lemma:notusable}, \Ename{(DZ1)} \\
       &=& m_{B_0}
\end{eqnarray*}
The step $(\star)$ uses  induction. From part
(2) of Lemma~\ref{lemma:tau.to.stable.peers} we know $\tau.n_a \testleqP m_a$ for every 
$a \in A_0 \cap N$. Lemma~\ref{lemma:plus} and induction give $\tau.n_a \eqleqP m_a$.
There are now two cases.  If $n_a \notok$ then an application of  \Ename{(S1a)}
gives  $a.n_a \eqleqP a.m_a$.  However if $n_a \ok$ this equation can not be used. 
However we can achieve the same conclusion as follows:
\begin{eqnarray*}
  a.n_a  &\eqleqP& a.(\Unit  \extc \tau.n_a)    & \Ename{(DP3)}\\
         &\eqleqP& a.(\Unit  \extc m_a)        & Induction\\
         &=&  a.m_a
\end{eqnarray*}
The last line follows from part (3) of Lemma~\ref{lemma:tau.to.stable.peers}.

\item 
Suppose 
 $n$ is as in  the previous case but that 
 $m$ is   $( \sum_{B \in \calB} \tau.m_B ) \Soption{ \Unit} $. 
Here we proceed as in case (c) of Theorem~\ref{thm:comp.clients}. 
Regardless of the presence or absence of the optional units, one can show
that $n \testleqP m_B$ for every $B \in \calB$. Therefore by part (a) we have
$n \eqleqP m_B$. 

Now suppose $n \notok$, that is $n$ does not contain $\Unit$ as a summand. 
Then using the derived \Ename{(D1)} we have
\begin{align}\label{eq:no.yes}
  n  &= \tau.n \ =\sum_{B \in \calB} \tau.n \qquad \eqleqP \sum_{B \in \calB} \tau.m_B
\end{align}
If $m$ also does not contain the summand $\Unit$ we are finished. But if it does,
we know that $\calB$ is non-empty and that each $B \in \calB$ contains $\ok$. 
Pick one such $B_0$, and applying \Ename{(P2P)} we obtain 
$\tau.m_{B_0} \eqleqP \tau.m_{B_0} + \Unit$. Using this in (\ref{eq:no.yes}) above
we obtain the required $n \eqleqP \sum_{B \in \calB} m_B + \Unit$.

Finally if both  $n$ and $m$ have $\Unit$ as a summand a simple variation on 
the argument (\ref{eq:no.yes}) above suffices.

\item
Now suppose that $n$ has the form $n_A$ for some $A \subseteq \Acttick$.
Reasoning as in the corresponding case of Theorem~\ref{thm:comp.clients}
we see that the only possible form for $m$ is $m_B$ for some $B \subseteq \Acttick$. 
Now  we use the fact that $\sum_{A \in \sset{A}} \tau.n_A \testplusP n_A \testleqP m_B$
to apply Lemma~\ref{lemma:tau.to.stable.peers}. This gives that 
$B_0 \subseteq A,\, A \cap N \subseteq B_0$, where $B_0,\, N$ are as described in 
that lemma. Now we can repeat the argument used in case (b) to show that 
$n_{A_0} \eqleqP m_{B_0}$ where again $A_0$ denotes $A \backslash \sset{\ok}$. 
Again a simple case analysis on whether $\ok$ is in either of $A,\, B$, as used also in case
(b), will allow us to conclude that $n_A \eqleqP m_B$. 
\end{enumerate}

\qed

\section{Conclusions}
\label{sec:conclusions}

Much of the recent work on behavioural preorders for processes 
has been carried out using formalisms for contracts for web-services,
proposed first in \cite{DBLP:conf/wsfm/CarpinetiCLP06}.
Spurred on by the recasting of the standard must preorder from
\cite{NH1984} as a server-preorder between contracts, these ideas 
have been developed further in
\cite{LP07,DBLP:journals/toplas/CastagnaGP09,DBLP:conf/ppdp/Barbanerad10,DBLP:journals/tcs/Padovani10}.

In these publications the standard refinements are referred to as {\em
  subcontracts} or {\em sub-server} relations and
\cite{LP07,DBLP:journals/toplas/CastagnaGP09,DBLP:journals/tcs/Padovani10,DBLP:conf/ppdp/Barbanerad10}
contain a range of alternative characterisations.  For example in
\cite{LP07,DBLP:journals/toplas/CastagnaGP09} the characterisations
are coinductive and essentially rely on traces and ready sets;
in \cite{DBLP:conf/ppdp/Barbanerad10} the characterisation is
coinductive and syntax-oriented.

To the best of our knowledge, the first paper to use a preorder for
clients is \cite{DBLP:conf/ppdp/Barbanerad10}. 
But  their setting is
much more restricted; they use so-called \emph{session behaviours} which correspond
to a much smaller class of processes than our language \CCS. As there are fewer contexts, their sub-server preorder differs from our server preorder: $a_1.\Unit \preceq_s a_1.\Unit \extc a_2.\Unit $, whereas $a_1.\Unit \; \not \testleqS \; a_1.\Unit \extc a_2.\Unit$.

The refinements in the papers mentioned above depend on a {\em
  compliance} relation, rather than must testing; this is also why
  in~\cite{DBLP:conf/ppdp/Barbanerad10} the peer preorder $\preceq:$ 
coincides with the intersection
  of the client and the server preorders; this is not the case for the
  must preorders (Example~\ref{ex:testleqSC-not-in-testleqS} can be
  tailored to the setting of session behaviours).  Moreover, in a
general infinite branching and non-deterministic LTS the refinements
in the above papers differ from the preorder $\testleqS$. The
subcontract relation of \cite{LP07} turns out to be not comparable
with $\testleqS$, whereas the strong subcontract $\sqsubseteq$ of
\cite{DBLP:journals/tcs/Padovani10} is strictly contained in
$\testleqS$, as the LTS there is convergent and finite branching. The 
comparison of $\testleqS$ with the refinement preorder of
\cite{DBLP:journals/toplas/CastagnaGP09} is complicated by their use
of a non-standard LTS.
A thorough comparison of the client and server refinements
given by the compliance and the $\Must$ testing can be found in \cite{DBLP:conf/sefm/BernardiH13}.

In~\cite{DBLP:conf/wsfm/BugliesiMPR09} a symmetric refinement due to the compliance, $\sqsubseteq^\mathsf{ds}$, is studied; it differs from 
our peer preorder ($\testleqSC \; \not \subseteq \; \sqsubseteq^\mathsf{ds}$), and its characterisation does not mention usability. This is because of the restrictions of the LTS in~\cite{DBLP:conf/wsfm/BugliesiMPR09}.
In more general settings the usability of contracts/services is
crucial; \cite{DBLP:conf/coordination/Padovani11c} talks of {\em
  viability}, while \cite{DBLP:journals/jlp/MooijSV10} talks of {\em
  controllability}.

Also subcontracts/subtyping for peers inspired by the should/fair-testing
of \cite{RV05} have been proposed in
\cite{DBLP:conf/sfm/BravettiZ09,DBLP:conf/wsfm/BugliesiMPR09,DBLP:conf/coordination/Padovani11c}. 
In \cite{DBLP:conf/sfm/BravettiZ09} the fair-testing preorder is used as
proof method for relating contracts, but no characterisation of 
their refinement preorder is given. A sound but incomplete
characterisation is given in~\cite{DBLP:conf/wsfm/BugliesiMPR09}.
The focus of \cite{DBLP:conf/coordination/Padovani11c} is on
multi-party {\em session types} which, roughly speaking, cannot express all the behaviours of our language \CCS. 
In view of the restricted form of session types,
they can give a syntax-oriented characterisation of their 
subtyping relation, $\leqslant$; this is in general incomparable
with our $\testleqSC$.

\paragraph{Future work:}
The most obvious open question about our two new refinement preorders
$\testleqC$ and $\testleqSC$ is the development of algorithms for
finite-state systems. The ability to check efficiently whether a
process is \emph{usable} will play an important role.

Another interesting question would be to characterise in some
equational manner the refinement preorders $\testleqC,\, \testleqSC$
themselves rather than their associated pre-congruences $\testplusC$
and $\testplusSC$. In the resulting equational theory we would have to
restrict in some way the form of reasoning allowed under the external
choice operator $- \extc -$, but the extra inequations needed in such 
a proof system might be simpler.

\leaveout{
\MHc{
One significant restriction on the refinement preorders investigated in
this paper is that they focus very much on \emph{binary interaction} 
between clients and servers, or between pairs of  peers. A more general scenario
would allow \emph{$n$-party interactions}. For example one might define 
a preorder $r_1 \testleqnC r_2$,  for any $n \geq 0$, by requiring 
\begin{quote}
   $p \Must^n r_2 \Par t_1 \Par \ldots \Par t_n$ whenever 
    $p \Must^n r_1 \Par t_1 \Par \ldots \Par t_n$
\end{quote}
for all servers $p$, and collections of partner clients $t_1$. 
Here intuitively the clients being compared, $r_1$ and $r_2$, 
are part of a larger environment in which numerous other  
independent clients are seeking satisfaction from the server $p$,
cite{bravetti1,bravetti2}. 
A formalisation of this idea would require a a precise definition
of the predicate  $p \Must^n r^1 \Par r^2 \Par \ldots \Par r^2$.
Should we require all clients $r^i$ to independently report success as
in \cite{bravetti1},
or only one of them? Variations of the peer-preorder along these lines 
are also possible. However because $\testleqArb$, where $\star$ is either
$\clt$ or $\peer$, are not in general preserved by $\Par$ it is likely that
this will give rise to a large number of different behavioural preorders. 
}
\GBc{
The peer preorder is amenable to a natural generalisation, that
accounts for $n$-peers rather than simply $2$.
We write 
$$
p \Par q_1 \Par \ldots \Par q_n \,\, \textsf{must succeed}
$$
if in all the maximal computations of the composition above 
every peer reaches a happy state, possibly at different times.
Then $p_1 \testleq[n] p_2$ if 
$ p_1 \Par q_1 \Par \ldots \Par q_n \textsf{must succeed}$ implies
$ p_2 \Par q_1 \Par \ldots \Par q_n \textsf{must succeed}$.

GB: Let me claim that $ p_1 \testleqSC p_2$ implies $p_1 \testleq[3]
p_2$. As of now, I do not see how the example we use to prove that
preorders are not preserved by $\extc$ could impact on the claim.
}
}

A further interesting question is the possible use of the parallel operator between clients and peers, either by allowing multi-party interactions as in \cite{DBLP:conf/sfm/BravettiZ09,DBLP:conf/wsfm/BugliesiMPR09}, or by deciding on how a parallel combination of clients should report success.

We have also confined our attention to refinement preorders based on must
testing. But one can also define client and peer preorders based on the
standard \emph{may testing} of \cite{NH1984}. We believe that these
refinement preorders can be completely characterised using a modified
notion of \emph{trace}, which takes into account the \emph{usability}
of residuals.
Other variations on client and peer preorders are worth
investigating: a ``synchronous'' formulation of
$\testleqSC$ where a computation\leaveout{in (\ref{eq:comp})} is successful
only if the peers report success {\em at the
  same time}; the client preorders for fair settings
\cite{DBLP:conf/coordination/Padovani11c,DBLP:conf/sfm/BravettiZ09},
or the ones based on the compliance \MHc{of} \cite{DBLP:journals/tcs/Padovani10}.

\paragraph{Acknowledgements}
The first author would like to acknowledge Vasileios Koutavas, 
for his help in unravelling the client preorder.
The paper has also benefited from comprehensive reviews by anonymous authors, 
which are greatly appreciated.
\appendix

\section{Justifying the derived equations}
\label{app:derived.laws}

\Ename{(D1)}:  The proof is by induction on $n$. For $i = 1$, the result follows by \Ename{(S1b)},
\Ename{(S4)} and Idempotency of $+$. Assume it is true for $k$; that is $\tau.z = z$, where $z$ abbreviates
$\sum_{1 \leq i \leq k} \tau.x_i$. Then 
\begin{eqnarray*}
  \tau.z \extc \tau.x_{k+1}  &=& \tau.(\tau.z \extc \tau.x_{k+1})  &\Ename{(S1a)} \\
                            &=&  \tau.(z \extc  \tau.x_{k+1})     &Induction
\end{eqnarray*}

\bigskip
\noindent
\Ename{(D2)}: 
\begin{eqnarray*}
  \ntickx \extc \tau.(\ntickx \extc y)  &=& \tau.(\ntickx \extc \ntickx \extc y) 
                                        \extc \tau.(\ntickx \extc y)  &\Ename{(S2)}\\
                                        &=& \tau.(\ntickx \extc y)    &Idempotency
\end{eqnarray*}

\bigskip
\noindent
\Ename{(D3)}: One direction is immediate from \Ename{(S5)}. Here is the converse:
\begin{eqnarray*}
  \mu.x \extc {\tau^\infty}  &\leq& \mu.x \extc \tau.{\tau^\infty}    &\Ename{(S5)} \\
                      &\leq& \tau.( \mu.x \extc {\tau^\infty}) \extc \tau.{\tau^\infty} &\Ename{(S2)} \\
                      &\leq& {\tau^\infty}    &\Ename{(S4)}
\end{eqnarray*}

\bigskip
\noindent
\Ename{(D4a)}: 
\begin{eqnarray*}
  \tau.\ntickx + \tau.y   &=&\tau.(\tau.\ntickx \extc  \tau.y)  &\Ename{(S1a)}\\
  &=& \tau.(\tau.\ntickx \extc \tau.(\ntickx \extc \tau.y))  &\Ename{(S2)}\\
  &=& \tau.(\tau.\ntickx \extc \tau.(\tau.(\ntickx \extc y) \extc
  \tau.y))         &\Ename{(S2)}\\
 &=& \tau.(\tau.\ntickx \extc \tau.(\ntickx \extc y) \extc \tau.y)
 &\Ename{(D1)}\\
 &=& \tau.\ntickx \extc \tau.(\ntickx \extc y) \extc \tau.y        &\Ename{(D1)}\\
\end{eqnarray*}

\bigskip
\noindent
\Ename{(D5a)}:
\begin{eqnarray*}
  \tau.x \extc \tau.(x \extc \nticky \extc z)   
&=&   \tau.x \extc  \tau.(x \extc \nticky \extc z)    
          \extc \tau.(x \extc \nticky \extc z \extc \tau.x) &\Ename{(S2)}\\
&=&   \tau.x \extc  \tau.(x \extc \nticky \extc z)    
          \extc \tau.(x \extc \nticky \extc z \extc \tau.(x \extc
          \nticky) )&\Ename{(S2)}\\
&=&   \tau.x \extc  \tau.(x \extc \nticky )    
          \extc \tau.(x \extc \nticky \extc z )&\Ename{(S2)}\\
\end{eqnarray*}

\bigskip
\noindent
\Ename{(DP1)}: One direction is straightforward from \Ename{(P2P2)}. Conversely:
\begin{eqnarray*}
  x                   &\leq&x \extc \Unit  &\Ename{(P2P1)}\\
  \Unit \extc \mu.x      &=& \Unit \extc \mu.(x \extc \Unit)  &Pre-congruence\\
                      &\leq& \Unit \extc \mu.(\tau.x \extc \Unit)  &\Ename{(P2P3)}, Idempotency\\
                      &\leq& \Unit \extc \mu.(\tau.x \extc \Unit)  &\Ename{(S4)}, Idempotency
\end{eqnarray*}

\bigskip
\noindent
\Ename{(DP2)}: This is a generalisation of (\ref{eq:P2P3}) above. 
It is proved by induction on $n$. 
The case when  $n =1 $ has already been discussed in
(\ref{eq:unit.tau}) on page~\pageref{eq:unit.tau}.
For the inductive case let $r$ denote $\sum_{1 \leq i \leq k}{\tau.x_i}$. 
By \Ename{(D1)} $r = \tau.\tau.r$ can be derived. Then
\begin{eqnarray*}
  \tau.(\Unit \extc \tau.x_{k+1} + r) &=&  \tau.(\Unit \extc \tau.x_{k+1} + \tau.(\tau.r) )\\
            &=& \tau(\Unit \extc  x_{(k+1)}) \extc  \tau(\Unit \extc  \tau.z) 
                      & (\ref{eq:P2P3}) above\\
      &=& \tau(\Unit \extc  x_{(k+1)}) \extc  \sum_{1 \leq k} \tau.(\Unit \extc x_i)
                      & Induction\\
\end{eqnarray*}

\bigskip
\noindent
\Ename{(DP3)}:
\begin{eqnarray*}
  \mu.x &\leq& \mu.(\Unit \extc x)  &\Ename{P2P1}, Pre-congruence \\
        &\leq& \mu.(\Unit \extc \tau.x) &\Ename{P2P3}, Idempotency
\end{eqnarray*}

\bigskip
\noindent
\Ename{(D4b)}: 
\begin{eqnarray*}
  \tau.(\ntickx \extc \Unit)  \extc \tau.(\nticky \extc \Unit  )   
&=&\tau.(  \tau.(\ntickx \extc \Unit)  \extc \tau.(\nticky \extc \Unit  )  )  &Lemma~\ref{lemma.tau.tau}\\
  &=&  \tau.(  \tau.(\ntickx \extc \Unit)  \extc \tau.( \ntickx \extc
  \Unit \extc 
  \tau.(\nticky \extc \Unit  )  )  )\\ 
  & & &\Ename{(S2)}, Idempotency\\
  &=&  \tau.(  \tau.(\ntickx \extc \Unit)  \extc \tau.( \Unit \extc 
  \tau.(\ntickx \extc \nticky \extc \Unit  ) \extc \tau.(\nticky \extc
  \Unit))  )\\ 
& & &\Ename{(S2)}, Idempotency\\
  &=&  \tau.(  \tau.(\ntickx \extc \Unit)  \extc 
\tau.( \tau.(\ntickx \extc \nticky \extc \Unit  ) \extc \tau.(\nticky \extc
  \Unit))  )  &\Ename{(DP2)}\\
&=&\tau.(\ntickx \extc \Unit)  \extc 
\tau.(\ntickx \extc \nticky \extc \Unit  ) \extc \tau.(\nticky \extc
  \Unit)  &\Ename{(D1)} twice
\end{eqnarray*}

\noindent
\Ename{(D5b)}: 
\begin{eqnarray*}
  \tau.x \extc \tau.(x \extc (\nticky  \extc \Unit) \extc z)   
&=&   \tau.x 
           \extc 
           \tau.(x \extc (\nticky \extc \Unit) \extc z ) 
           \extc 
           \tau.(x \extc (\nticky \extc \Unit) \extc z \extc \tau.x) 
            &\Ename{(S2)}\\
&=&    \tau.x 
            \extc 
           \tau.(x \extc (\nticky \extc \Unit) \extc z ) 
           \extc  
            \tau.(x \extc (\nticky \extc \Unit) \extc z \extc
                      \tau.(x  \extc \nticky)) 
\\ &&&\Ename{(S2)}\\
&=&    \tau.x 
            \extc 
           \tau.(x \extc (\nticky \extc \Unit) \extc z ) 
             \extc  
              \tau.(x \extc (\nticky \extc \Unit) \extc z 
          \extc  \tau.(x  \extc \nticky \extc \Unit) ) 
\\ &&&\Ename{(DP1)}\\
&=&    \tau.x \extc \tau.(x \extc (\nticky \extc \Unit))    
          \extc \tau.(x \extc (\nticky \extc \Unit) \extc z ) 
&\Ename{(S2)}\\
\end{eqnarray*}

\bibliographystyle{alpha}
\bibliography{Mutualt}

\end{document}